\documentclass[12pt,amsmath,amssymb,longbibliography,nofootinbib,eprint]{revtex4-1}
\usepackage{epigraph}
\usepackage{graphicx}
\usepackage{epstopdf}
\usepackage{braket}
\usepackage{bm}
\usepackage{numprint}
\usepackage{seqsplit}
\usepackage{float}
\usepackage[T1,T2A]{fontenc}
\usepackage[utf8]{inputenc}
\usepackage{hyperref}

\allowdisplaybreaks
\begin{document}
\preprint{V.M.}
\title{Market Dynamics.
  On A Muse Of Cash Flow And Liquidity Deficit.}
\author{Vladislav Gennadievich \surname{Malyshkin}} 
\email{malyshki@ton.ioffe.ru}
\affiliation{Ioffe Institute, Politekhnicheskaya 26, St Petersburg, 194021, Russia}

\date{August, 25, 2016}

\begin{abstract}
\begin{verbatim}
$Id: AMuseOfCashFlowAndLiquidityDeficit.tex,v 1.575 2019/03/28 06:02:45 mal Exp $
\end{verbatim}
A first attempt at obtaining market--directional information
from a non--stationary solution of the dynamic equation
``future price tends to the value that maximizes the number of shares traded per unit time''
\cite{2015arXiv151005510G} is presented.
We demonstrate that the concept of price impact
is poorly applicable to market dynamics. Instead, we consider
the execution flow $I=dV/dt$ operator with the ``impact from the future'' term
providing information about not--yet--executed trades.
The ``impact from the future" on $I$
can be directly estimated from the already--executed trades,
the directional information on price is then obtained from 
the experimentally observed fact that the
$I$ and $p$ operators have the same eigenfunctions
(the exact result in the dynamic impact approximation $p=p(I)$).
The condition for ``no information about the future''
is found and directional prediction quality is discussed.
This work makes a substantial contribution
toward solving the ultimate market dynamics problem:
find evidence of existence (or proof of non--existence)
of an automated trading machine which consistently makes positive P\&L
on a free market as an autonomous agent (aka the existence of the market dynamics equation).
\href{http://www.ioffe.ru/LNEPS/malyshkin/AMuseOfCashFlowAndLiquidityDeficit.zip}{The software}
with a reference implementation of the theory is provided.
\end{abstract}

\maketitle

\section{\label{intro}Introduction}
Market Dynamics is the central concept of modern economic study.
An ultimate form of the study to be an evidence 
of existence (or a proof of non--existence) of an automated trading machine,
consistently making positive P\&L (with a given value of risk)
trading on a free market as an autonomous agent.
In our previous study\cite{2016arXiv160204423G,2016arXiv160305313G}
we have shown experimentally that supply and demand match
each other down to milliseconds time scale, thus
their disbalance cannot be a source of market dynamics.
Moreover, supply and demand cannot be measured
or estimated from the data
even after transaction execution\cite{2016arXiv160204423G}.
In the modern world all available data is typically
represented in a form of recorded transactions,
where money, financial instruments,
goods, etc. change hands. In each such transaction
there are  two matched
parties (e.g. ``A'' sold $x$ goods to ``B'' and received $y$ dollars for that)
what means that in recorded data supply and demand are \textsl{matched}.
The disbalance of supply/demand cannot (even in principle!)
be measured from a sequence of transactions,
as any transaction assume the parties to match.
An example of information source, that is not a sequence of transactions,
is the Limit Order Book. However, using Limit Order Book
as a source of information
about Supply and Demand
is fruitless\cite{2016arXiv160305313G} since at least 2008--2010
and exchange trading is now little different from dark pool trading.
(We tried to consider the Limit Order Book
both: as not a sequence of transaction, and
as a sequence of add/\{cancel$|$execute\} transactions,
but without much success;
most typical limit order book pattern is: added order
spend almost no time in the order book,
it either get almost immediately executed or canceled.
The ratio observed is that more than 90\% of orders
being at best price level at some time end up being canceled\cite{nasdaqord,2015arXiv151005510G}. This is due to exchange fee structure, because
add/cancel order ``round trip'' cost (almost) no money and carry little risk
for market participants.)
This make us to conclude that the disbalance of supply and demand
is not a practically applicable concept,
because it cannot
be measured  from recorded transactions.

For practical applications we need a concept
that can be estimated from a sequence of transactions.
In \cite{2016arXiv160204423G,2015arXiv151005510G}
a concept of execution flow ($I=dV/dt$ a number of shares
traded in unit time, a number of dollars paid in unit time, etc.)
was introduced and practical approach to its calculation (based on Radon--Nikodym
derivatives and their generalization)
was developed.

An application of this approach in quasistationary case
was demonstrated in \cite{2016arXiv160204423G},
where we have shown that asset price is much more
sensitive to execution rate $I=dV/dt$, rather than to trading volume $V$,
and dynamic impact (sensitivity to $I$) was introduced
as a practical alternative to regular impact\cite{wiki:marketimpact} (sensitivity to $V$)\footnote{
Also see later developed\cite{MalMuseScalp} concept of
constrained optimization $I\xrightarrow[{\psi}]{\quad }\max$
subject to the constraint
$\Braket{\psi|C|\psi}=0$,
considered for a number of operators $\|C\|$.
This allows us,
within the framework of a single formalism
of constrained optimization,
take into account
 the
 driving force of the market  $I\to\max$,
 and the reaction, via the operator $\|C\|$,
of the market participants on it.
}.
In this paper we make one more step
forward, demonstrating an application of this approach
in a non--stationary case.
First, we show that price impact,
the central subject of many studies,
is poorly applicable to market dynamics.
A practical alternative to it is
an impact from the future on $I$,
that can be estimated from past sample.
Then we are trying to obtain
directional information on price from a knowledge of future $I$,
with the goal to obtain trading strategy with a positive P\&L.
There is a fundamental philosophical question\cite{BoikoTimeMachine2009}
about positive P\&L provided by an automated trading machine:
Assume one created a ``Real Time Machine'',
but looking only very few moments ahead in the future. How to prove
that a given ``Time Machine'' works? Attach it to an exchange
and show the P\&L! In this sense any dynamic
equation (Newton, Maxwell,
Schr\"{o}dinger) can be considered as some kind of ``Time Machine''.
Moreover, any intelligence can be considered as
a ``future prediction system'' \cite{hawkins2007intelligence},
thus, when applied to the market,
the P\&L can be considered
as an ``intelligence criteria'' of an automated trading machine.
There is a very deep difference between an intelligent agent
and statistical approach. For an intelligent agent
a single observation is enough to make a prediction.
For any statistical approach a large number of observations is required
to make any kind of inference. In \cite{2016arXiv160305313G}
we emphasized the inapplicability of any statistical approach
to exchange trading and the importance of the
\textbf{\textsl{dynamical}} approach,
a practical alternative to a statistical one.

The dynamic equation we introduced\cite{2015arXiv151005510G}
``future price tends to the value that maximizes the number of shares traded per unit time''
in this direct form requires to know ``future'' prices and flows,
and can be easily solved only in quasistationary case\cite{2016arXiv160204423G}.
In a non--stationary case the best result of 
our previous study\cite{2015arXiv151005510G}
was ``maximizing the number of shares traded per unit
time on past observations sample'', but with a limited success.
The concept of market dynamics
in its ultimate form requires to determine future market movement
from past observations sample.
In this paper a substantial progress is made toward this goal.
In Section \ref{FuturePsi}
an estimation (\ref{dI}) of the impact from the future on $I$ is made,
allowing
(from experimentally observed\cite{2016arXiv160204423G} fact that
$I$ and $p$ operators to have the same eigenfunctions,
at least for the states with high $I$)
to obtain  price directional answer.
This dynamic equation solution is equivalent to
some trending model,
but have an automatic selection of the relevant time scale,
a critically important feature of any automated trading system\cite{2015arXiv151005510G}.

In Ref. \cite{2015arXiv151005510G},
as a first application of the dynamic equation,
the concept of liquidity deficit trading
was introduced:
open a position on  low $I_0$ ($I_0$ is defined in Eq. (\ref{I0def})),
close already opened position on high $I_0$,
as the only way to build a strategy,
resilient to catastrophic  P\&L loss.
In Ref. \cite{2015arXiv151005510G} market directional information
was not obtained, thus only volatility trading was available for
practical implementation.
In this new study we
made a substantial progress in dynamic equation application:
to obtain market directional information
from the dynamic equation.

\href{http://www.ioffe.ru/LNEPS/malyshkin/AMuseOfCashFlowAndLiquidityDeficit.zip}{Computer code}
 with a reference implementation of the theory 
is presented in the Appendix \ref{appendix1}.

\section{\label{basis}Basis Selection}

To operate with introduced in\cite{2015arXiv151005510G}
concepts
we need to convert market observable timeserie variables
(time, execution price, shares traded)
to a set of distribution moments. The three bases, performing time averaging with the exponential
weight, are the most convenient for
market dynamics study.
Laguerre basis:
\begin{eqnarray}
  x&=&t/\tau \\
  x_0&=&0 \\
\Braket{Q_k f}&=&\int\limits_{-\infty}^{x_0} Q_k(x) f(t) \exp(x) dx
\label{flaguerre} \\
d\mu&=&\exp(x) dx \label{muflaguerre} \\
\mathrm{supp}(\mu(x)) &=& x \in [-\infty , x_0] \label{lagurrerange} \\
D(Q_k(x)) &=& \frac{dQ_k(x)}{dx}+\frac{Q_k(x)}{2} \label{DpsiLag} \\
\Braket{Q_k f}&=&\sum_{i} Q_k(-\frac{t_{now}-t_i}{\tau})  \exp(-\frac{t_{now}-t_i}{\tau}) f(t_i) \frac{t_i-t_{i-1}}{\tau}
\label{sampleLag}
\end{eqnarray}
Shifted Legendre basis:
\begin{eqnarray}
  x&=&\exp(t/\tau) \\
  x_0&=&1 \\
  \Braket{Q_k f}&=&\int\limits_{-\infty}^{0} Q_k(x) f(t) \exp(t/\tau) dt/\tau 
  = \int\limits_{0}^{x_0} Q_k(x) f(t) dx \label{slegendre} \\
  d\mu&=& \exp(t/\tau) dt/\tau   = dx \label{muslegendre}\\
  \mathrm{supp}(\mu(x)) &=& x \in [0 , x_0] \label{shiftedlegendrerange} \\
  D(Q_k(x)) &=& x\frac{dQ_k(x)}{dx}+\frac{Q_k(x)}{2} \label{DpsiLegendre} \\
  \Braket{Q_k f}&=&\sum_{i} Q_k(\exp(-\frac{t_{now}-t_i}{\tau}))  \exp(-\frac{t_{now}-t_i}{\tau}) f(t_i) \frac{t_i-t_{i-1}}{\tau}
  \label{sampleLeg}
\end{eqnarray}
Price Basis
\begin{eqnarray}
  x&=&p \\
  \Braket{Q_k f}&=&\int\limits_{-\infty}^{0} Q_k(p(t)) f(t) \exp(t/\tau) dt/\tau  \label{pspace}\\
  d\mu&=&\exp(t/\tau) dt/\tau  \label{mupspace} \\
  \mathrm{supp}(\mu(p(t))) &=& t \in [-\infty , 0]  \\
  \Braket{Q_k f}&=&\sum_{i} Q_k(p(t_i))  \exp(-\frac{t_{now}-t_i}{\tau}) f(t_i) \frac{t_i-t_{i-1}}{\tau}
\label{samplePrice}
\end{eqnarray}
$Q_k(x)$ is a polynomial of $k$--th order (e.g. monomials $\{1;x;x^2;x^3;\dots\}$),
but from numerical stability point\cite{2015arXiv151005510G}
for (\ref{muflaguerre}) a good choice
is the selection  $Q_k(x)=L_k(-x)$, with $L_k(x)$ Laguerre polynomials,
and for (\ref{muslegendre}) a good choice
is the selection $Q_k(x)=P_k(2x-1)$, with $P_k(x)$
Legendre polynomials.
This choice make the basis orthogonal in $d\mu$ measure:
$\int_0^{\infty}L_j(x)L_k(x)\exp(-x)dx=\delta_{jk}$
and $\int_0^{1}P_j(2x-1)P_k(2x-1)dx=\frac{1}{2k+1}\delta_{jk}$,
what drastically increase the numerical stability of calculations.
However, all results are invariant with respect to
polynomials selection.
The specific choice affects only numerical stability of calculations,
thus should be discussed separately\cite{laurie1979computation,beckermann1996numerical,2015arXiv151005510G,2015arXiv151101887G}.
Proper basis selection\cite{2015arXiv151101887G} allows us to
have the numerically stable results
even for two--dimensional basis with 100 basis functions in each dimension,
i.e. with 10000 basis functions total
for 64bit double precision computer arithmetic.

The Eqs. (\ref{sampleLag}), (\ref{sampleLeg}) and (\ref{samplePrice})
show how to calculate the $\Braket{Q_k f}$ moments from a timeserie sample $f(t_i)$.
To simplify working with averages introduce quantum mechanic
\href{https://en.wikipedia.org/wiki/Bra%E2%80%93ket_notation}{bra--ket notation}\cite{wiki:braketnotation} $\Bra{}$ and $\Ket{}$:
\begin{eqnarray}
  \Braket{Q_k f}&=&\int d\mu Q_k(x) f(t) \\
  \Braket{Q_j | f |Q_k }&=&\int d\mu Q_j(x) Q_k(x) f(t)
  \label{braket}
\end{eqnarray}
where the integral $\int d\mu$ in (\ref{braket})
is calculated directly from a timeserie according to
(\ref{sampleLag}), (\ref{sampleLeg}) or (\ref{samplePrice})
depending on basis used.
Familiar values can be easily presented with these definitions.
Price exponential moving average: put price at time $t_i$ as
the $f(t_i)$, then $\overline{p}_{\tau}=\Braket{Q_0 p}/\Braket{Q_0}$ is required moving average.
From all the considerations above one can easily see that
bra--ket  $\Bra{}$ and $\Ket{}$ notations
from quantum mechanic
are nothing more, than a ``glorified moving average'',
and think of $\Braket{Q_k|f|Q_j}$ as taking a moving average
with two basis functions product:
$\int d\mu Q_k(x(t)) f(t) Q_j(x(t))$.
Different $d\mu$ measures can be defined in a similar way.
However the measures (\ref{muflaguerre}) and  (\ref{muslegendre})
are special\cite{laurie2015privatecomm},
in a sense they allow to calculate the $\Braket{Q_k df/dt}$
moments from the $\Braket{Q_k f}$
moments using integration by parts.
The following condition also holds:
\begin{equation}
  Q_j(x_0) Q_k(x_0) =
  \Braket{ Q_j(x) D\left(Q_k(x)\right)}+
  \Braket{ D\left(Q_j(x)\right) Q_k(x)}
  \label{DQQ}
\end{equation}
Infinitesimal time--shift
 linear operator $D(\psi(x))$ from (\ref{DpsiLag}) and (\ref{DpsiLegendre}),
is different from plain differentiation because
exponent differentiation in (\ref{muflaguerre}) and  (\ref{muslegendre})
give an extra term.
The selection of  basis functions as a function of price
$Q_k(p(t))$ in (\ref{mupspace})
is extremely convenient in the quasistationary case\cite{2016arXiv160204423G}
but does not possess such a simple infinitesimal time--shift
transform.

\subsection{\label{demoIPbasis} $I=dV/dt$ as Radon Nikodym Derivative of Lebesgue Measures.}
\begin{figure}
\includegraphics[width=18cm]{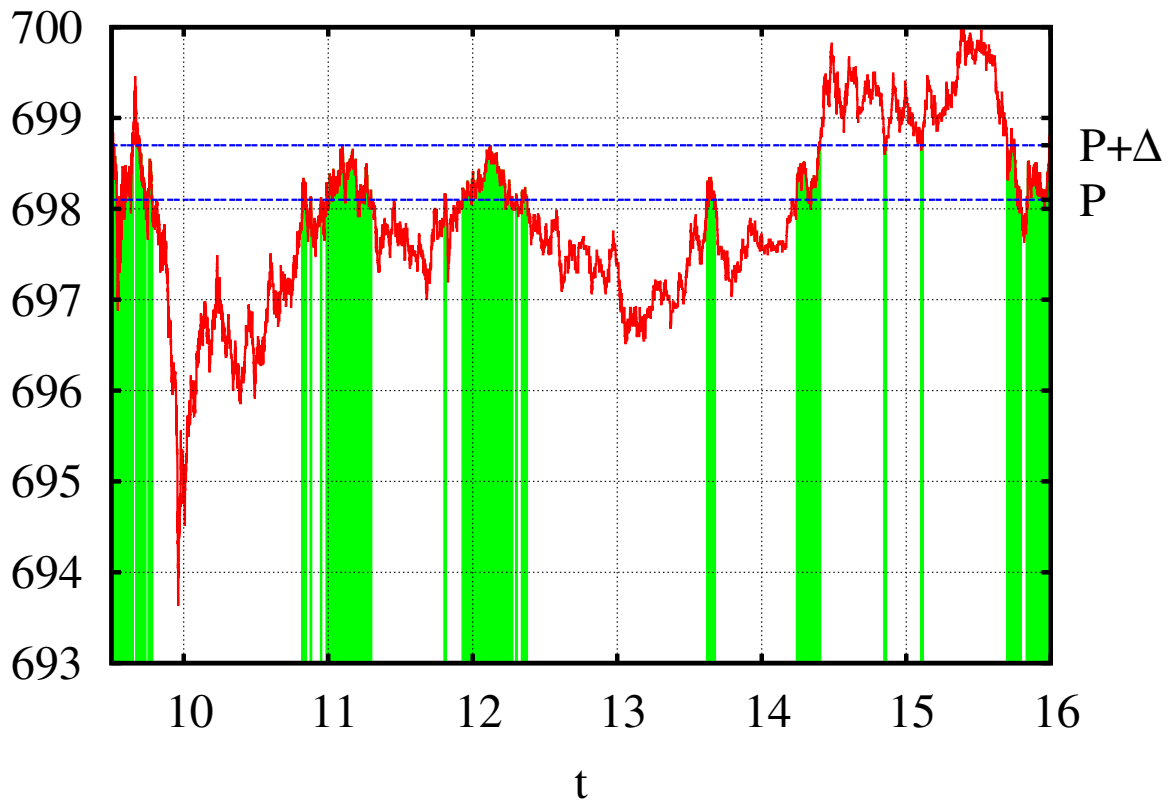}
\caption{\label{figPLebeg}
  The AAPL stock price on September, 20, 2012.
  Demonstration of Lebesgue Integral concept:
  time spent or volume traded with price inside $[P : P+\Delta P]$  interval.
}
\end{figure}
In this subsection we demonstrate price basis convenience
for execution flow calculation in the quasistationary case
and it's relation to Radon--Nikodym derivatives,
the main technique of our \cite{2016arXiv160204423G,ArxivMalyshkinLebesgue} papers.
The idea is to split price range on a number of $\Delta P$ intervals,
then, for each interval
calculate:
\begin{itemize}
\item time spent
\item volume traded
\end{itemize}
of timeserie observations when the price is inside the $[P : P+\Delta P]$ interval,
 see Fig. \ref{figPLebeg} for illustration.
These calculations give us two Lebesgue  measures:
$\Delta t = \mu_t(P) \Delta P$ and $\Delta V = \mu_V(P) \Delta P$.
These measures give time spend and volume traded
when the price is inside the range $[P : P+\Delta P]$.
By itself these two Lebesgue  measures
are very similar to each other and are nothing more than
a ``glorified price--volume distributions'',
both having distribution maximum near price median,
see Fig. 3 (top) of Ref. \cite{2016arXiv160204423G}.
But when one take a ratio of these two measures, it gives
trades execution flow $I(P)=\mu_V(P)/\mu_t(P)$,
with singularities near price tipping points,
see Fig. 3 (center) of Ref. \cite{2016arXiv160204423G}.
The execution rate, the central concept of our theory,
$I(P)=\mu_V(P)/\mu_t(P)$ can be considered
as Radon--Nikodym derivative of two Lebesgue measures $\mu_t(P)\Delta P$ and $\mu_V(P)\Delta P$.
For numerical calculations the described above histogram--like procedure
works well only if discretization scale $\Delta P$ is properly chosen,
what is a non--issue for manual analysis, but can be a real problem
for an automated system.
From numerical perspective there is a much
better way to calculate Radon--Nikodym derivative
of two measures, a calculation from distribution moments,
see the formula (\ref{frn}) below,  the answer in the form of Nevai operator\cite{nevai}.
Given sufficient number of moments (what may be a problem
to calculate numerically, unless a stable basis is chosen\cite{2015arXiv151005510G})
the (\ref{frn}) is a superior numerical
estimator of Radon--Nikodym derivatives.

\section{\label{probabilitystate}Wavefunction}
Introduce a wavefunction $\psi(x)$ to be a linear combination
of basis function $Q_k(x)$ (here $n$ is time--space dimension,
typically $n$ take some value between 4 and 20).
\begin{eqnarray}
  \psi(x)&=&\sum\limits_{k=0}^{n-1} \alpha_k Q_k(x) \label{psiintr}
\end{eqnarray}
Then any observable (or calculable) market--related value 
$f_{\psi}$, corresponding to a probability density  $\psi^2(x)$
can be calculated as:
\begin{eqnarray}
  f_{\psi}&=&\frac{\Braket{\psi|f|\psi}}{\Braket{\psi|\psi}}
  \label{faver} \\
  f_{\psi}&=&\frac{\sum\limits_{j,k=0}^{n-1}\alpha_j\Braket{Q_j|f|Q_k}\alpha_k}{\sum\limits_{j,k=0}^{n-1}\alpha_j\Braket{Q_j|Q_k}\alpha_k}
  \label{faverexpand}
\end{eqnarray}
The (\ref{faver}) is plain ratio of two moving averages,
but the weight is not just a regular decaying exponent according to
(\ref{muflaguerre}) or  (\ref{muslegendre}),
but exponent, multiplied by the $\psi^2(x)$,
thus the $\psi^2(x)$ define how to average a timeserie sample $f(t_i)$.
The (\ref{faverexpand}) is (\ref{faver}) with parentheses
expanded according to (\ref{psiintr}).
This way any $\psi(x)$ function is defined by $n$ coefficients $\alpha_k$,
and the value of any observable variable, corresponding to this $\psi(x)$ state
is a ratio of 
two quadratic forms (built on $\alpha_k$ coefficients)
of dimension $n$, an estimator
of stable form\cite{malha}. The representation of an observable
in a form of two quadratic forms ratio (\ref{faverexpand})
is conceptually different from the representation
of an observable in a form of linear superposition of basis functions.
In (\ref{faverexpand}) a wavefunction $\psi(x)$
is represented as a linear superposition of basis functions,
the $\psi^2(x)d\mu$ define probability density,
then  $f_{\psi}$ is  calculated as $f(t_i)$ averaged with
this probability density\cite{2016arXiv161107386V}.
This approach allows do decouple
 variables determining market dynamics
and variables determined by market dynamics,
what is critically important for any market dynamics study.

\subsection{\label{intexample}Interpolation Example}
Given the definitions above,
let us show some familiar answers.
Let $f(t)$ be some function, obtain $\beta_k$, such as the interpolation
$A_{LS}(y(t))=\sum_{k=0}^{n-1}\beta_k Q_k(y(t))$, minimize
least squares norm: $\Braket{\left(f(x(t))-\sum_{k=0}^{n-1}\beta_k Q_k(x(t))\right)^2}\to\min$.
Taking the derivatives of the norm on $\beta_k$ obtain the solution:
\begin{eqnarray}
  A_{LS}(y)&=&\sum\limits_{j,k=0}^{n-1}Q_j(y)(G^{-1})_{jk}\Braket{f Q_k}
  \label{leastsquares}
\end{eqnarray}
Here $G^{-1}$ is the inverse to Gramm matrix $G_{jk}=\Braket{Q_j | Q_k}$
and the (\ref{leastsquares}) is a regular least squares solution,
a polynomial of $n-1$ order, where
the coefficients are obtained as the solution of a linear
system with Gramm matrix.

A much more interesting case is to obtain probability density
$\psi^2_y(x)d\mu$,
which is localized at given $y$, then calculate $A_{RN}(y)=\frac{\int f(x) \psi^2_y(x)d\mu}{\int \psi^2_y(x)d\mu}$,
using 
probability density with interpolated $\psi_y(x)$.
There are several forms\cite{2015arXiv151005510G} of such localized $\psi_y(x)$,
the simplest one give (\ref{frn}),  Nevai operator\cite{nevai}:
\begin{eqnarray}
  \psi_y(x)&=&\sum\limits_{j,k=0}^{n-1}Q_j(y)(G^{-1})_{jk}Q_k(x)
  \label{psileastsquares} \\
  A_{RN}(y)&=&\frac{\sum\limits_{j,k,l,m=0}^{n-1}Q_j(y) (G^{-1})_{jk} \Braket{Q_k|f|Q_l} (G^{-1})_{lm} Q_m(y)}
  {\sum\limits_{j,k=0}^{n-1}Q_j(y)(G^{-1})_{jk}Q_k(y)}
  \label{frn}
\end{eqnarray}
The (\ref{psileastsquares}) is interpolated localized wavefunction (localized at $y$,
compare it to $A_{LS}$ interpolation (\ref{leastsquares})),
then this localized at $y$ probability density is put to (\ref{faverexpand})
to obtain (\ref{frn}), that is now considered as
Radon--Nikodym
interpolation of $f$ at $y$.
In contrast with the least squares answer (\ref{leastsquares})
(which is a linear combination of basis functions),
the (\ref{frn}) is a ratio of two quadratic forms of basis functions,
a ratio of two polynomials $2n-2$ order each in case of polynomial basis.
The (\ref{frn}) is used for numerical estimation
of $\frac{d\nu}{d\mu}=\frac{f(x)d\mu}{d\mu}$, considered as Radon--Nikodym
derivative. The (\ref{frn}) answer
(basis--invariant answers (\ref{leastsquares}) and (\ref{frn})
take very simple form\cite{2015arXiv151005510G,2016arXiv161107386V}
in the basis of eigenfunctions of operator,
generated by the $f$),
is typically the most convenient
one among other available,
because it requires only one measure to be positive. Other
answers\cite{BarrySimon,2015arXiv151005510G}
require both measures to be positive.
Radon--Nikodym interpolation (\ref{frn}) has several
critically important advantages\cite{2015arXiv151005510G,2015arXiv151101887G,malyshkin2015norm}
compared to the least squares
interpolation (\ref{leastsquares}):
stability of interpolation, there is no divergence
outside of interpolation interval,
oscillations near interval edges are very much suppressed,
even in multi--dimensional case\cite{2015arXiv151101887G}.
These advantages come from the very fact,
that probability density is interpolated first,
then the result is obtained by averaging with this, always positive, interpolated
probability.

\subsection{\label{wfpractical}Probability States}
Considered in subsection \ref{intexample}
localized wavefunction give a simple example,
illustrating the power of the technique.
However, much more interesting results
can be obtained considering not only localized states
such as (\ref{psileastsquares}),
but arbitrary $\psi(x)$. This allows us to decouple
observable variables
and probability state.

As we emphasized in\cite{2015arXiv151005510G}
system dynamics cannot be obtained from price.
  The price is secondary and typically fluctuates
  few percent a day in contrast with the liquidity flow, that
  fluctuates in orders of magnitude. (This also allows to estimate
  maximal workable time scale for an automated trading machine: the scale on which execution flow
  fluctuates at least in an order of magnitude. Minimal time scale
  is typically determined by  available market liquidity\cite{2016arXiv160305313G}).
The main idea is to obtain the state $\psi$ from
the variables, determining the dynamics (e.g. execution flow $I=dV/dt$,
execution flow changes $dI/dt$, etc.)
and then use obtained state to
determine the values of interest (e.g. price, price change, or P\&L).
A critically important feature of this approach
is that both:
the variables determining the dynamics
and  the variables determined by the dynamics
can be directly calculated from recorded data,
what is drastically different from Supply--Demand
approach, where the disbalance of it cannot be calculated from recorded
transactions data, because
in all recorded transactions Supply and Demand are matched.

\section{\label{pimpact}Price Impact}
Price impact \cite{2009PhRvE..80f6102M,gatheral2013dynamical,2014arXiv1412.0141D}
is typically considered as path--dependent impact
of executed shares number on asset price.
However the price can be affected by a number of other factors
and, moreover, an impact defined in such a way may diverge or even do not exist.
In a style of previous section, define price impact
as price change in a given $\psi(x)$ state.
With the approach we develop in this paper
price impact is calculated in two steps.
First, find the state
of interest  $\psi(x)$  (e.g. corresponding to a large $I$ or $dI/dt$, etc.).
Second calculate price change corresponding to the $\psi(x)$ found
on the first step.
We define price change, corresponding to the $\psi(x)$,
as generalized price impact in the $\psi$ state: $\Delta_{\psi}P$.
The selection of $\psi(x)$ will be discussed in the next section.
In this section we only demonstrate how to calculate price impact for a given $\psi(x)$.
There are two practical answers:

1. The moments $\Braket{Q_k dp/dt}$ can be directly
calculated from a sample using (\ref{sampleLag}), (\ref{sampleLeg}) or (\ref{samplePrice})
with the replacement
of the factor $f(t_i) (t_i-t_{i-1})/\tau$
by the factor $(p(t_i)-p(t_{i-1}))/\tau$.
After the calculation of $\Braket{Q_k dp/dt}$ moments
the $\Delta_{\psi}P$ can be obtained directly:
\begin{eqnarray}
  \Delta_{\psi}P&=&\frac{\Braket{\psi| \frac{dp}{dt} |\psi }}
        {\Braket{\psi|\psi}}=\frac{\sum\limits_{j,k=0}^{n-1}\alpha_j\Braket{Q_j|\frac{dp}{dt}|Q_k}\alpha_k}{\sum\limits_{j,k=0}^{n-1}\alpha_j\Braket{Q_j|Q_k}\alpha_k}
        \label{DPDirect}
\end{eqnarray}
The (\ref{DPDirect}) give an answer calculated directly from sample.

2. In some situations the moments $\Braket{Q_k dp/dt}$
are not convenient to use or not available
and only $\Braket{Q_k p I}$ sampled moments are available.
Then calculate the price $p_{\psi}$, corresponding to the $\psi(x)$ state,
and variate $\psi(x)$ using infinitesimal time--shift operator $D(\psi)$ from (\ref{DpsiLag}) or
(\ref{DpsiLegendre}) depending on the basis used.
\begin{eqnarray}
  p_{\psi}&=&\frac{\Braket{\psi|pI|\psi}}{\Braket{\psi|I|\psi}} \label{pIcalc}\\
  \Delta_{\psi}P&=&-2\left(\frac{\Braket{D(\psi)|pI|\psi}}{\Braket{\psi|I|\psi}}
  - \frac{\Braket{\psi|pI|\psi}}{\Braket{\psi|I|\psi}}
  \frac{\Braket{D(\psi)|I|\psi}}{\Braket{\psi|I|\psi}}\right)
  \label{DPDirectRQ}
\end{eqnarray}
The (\ref{DPDirectRQ}) is the first order variation of Rayleigh quotient (\ref{pIcalc}),
the second order variation of Rayleigh quotient can be also calculated,
see the (\ref{psivar}) below with $\delta \psi=-D(\psi)$,
but note that that $D(D(\psi))$ terms need to be added to (\ref{rqD2})
in general case.

The (\ref{DPDirect}) and (\ref{DPDirectRQ})
may or may not give similar answer,
because they treat the boundary $x=x_0$ (time is ``now'') differently.
Substantial difference in between (\ref{DPDirect}) and (\ref{DPDirectRQ})
typically indicates a large contribution of the boundary,
and is a signal of possible discrepancy in generalized price impact estimation.
But, as we emphasized earlier\cite{2015arXiv151005510G},
in practical applications
other than price, dynamics--related attributes (e.g. P\&L or $I$) should
be considered instead.

\section{\label{psistate}Wavefunction States Important For Market Dynamics}
Localized $\psi$ state, considered in the subsection \ref{intexample},
is of interest for interpolation problem only.
For dynamic problem other $\psi$ to be considered.
There is a number of interesting situations to consider,
but consider the two forms of $\psi$,
the most promising for market dynamics
and for generalized price impact calculation.

\subsection{\label{psiImax}$\psi$ Corresponding to Maximal $I$}
We have already emphasized\cite{2016arXiv160204423G} the importance of the states, corresponding
to maximal $I$.
The problem of maximizing $I$ on ``past'' sample\cite{2015arXiv151005510G}
can be reduced to a generalized eigenvalue problem (\ref{GEVI}).
\begin{eqnarray}
  \frac{\Braket{\psi|I|\psi}}{\Braket{\psi|\psi}}&\to&\max
  \label{Imax} \\
  \sum\limits_{k=0}^{n-1} \Braket{Q_j|I|Q_k} \alpha^{[i]}_k &=&
  \lambda_I^{[i]} \sum\limits_{k=0}^{n-1} \Braket{Q_j|Q_k} \alpha^{[i]}_k
  \label{GEVI} \\
  \psi^{[i]}_I(x)&=&\sum\limits_{k=0}^{n-1} \alpha^{[i]}_k Q_k(x)
  \label{psiIsol}
\end{eqnarray}
Generalized eigenvalue problem (\ref{GEVI}) provide
$n$ solutions ($i=[0\dots n-1]$),
each $i$ corresponds to the (eigenvalue,eigenfunction) pair
$(\lambda_I^{[i]},\psi^{[i]}_I(x))$.
The state $\psi^{[IH]}_I(x)$, corresponding to the maximal $\lambda_I$,
is a first good candidate for generalized price impact calculation.

\subsection{\label{psidImax}$\psi$ Corresponding to Maximal $dI/dt$}
The state, corresponding to maximal $dI/dt$ can be also of interest
for market dynamics.
In contrast with the $\Braket{Q_j|dp/dt|Q_k}$, $\Braket{Q_j|I|Q_k}$ and $\Braket{Q_j|pI|Q_k}$ matrices
the matrix $\Braket{Q_j|dI/dt|Q_k}$  cannot be directly calculated from sample.
However, in a presence of an infinitesimal time--shift
operator (\ref{DQQ}) this matrix can be calculated by applying integration
by parts:
\begin{eqnarray}
  \Braket{Q_j\left|\frac{dI}{dt}\right|Q_k} &=& I^{f}Q_j(x_0)Q_k(x_0)- \Braket{D(Q_j)|I|Q_k}-\Braket{Q_j|I|D(Q_k)}
  \label{dIMatr}
\end{eqnarray}
Edge $x=x_0$ value $I^{f}$
is unknown in general case.
We have tried various values for $I^{f}$,
but for simplicity of calculation let us put $I^{f}=0$ in this section
(see the Section \ref{FuturePsi} below for the case $I^{f}=\lambda_I^{[IH]}$).
The $I^{f}=0$ means that the trading ``now'' is expected to stop at this price.
Then the $\Braket{Q_j|dI/dt|Q_k}$ matrix can be
obtained from (\ref{dIMatr}) and generalized eigenvalue problem
can be written in a usual way:
\begin{eqnarray}
  \frac{\Braket{\psi\left|\frac{dI}{dt}\right|\psi}}{\Braket{\psi|\psi}}&\to&\max
  \label{dImax} \\
  \sum\limits_{k=0}^{n-1} \Braket{Q_j\left|\frac{dI}{dt}\right|Q_k} \alpha^{[i]}_k &=&
  \lambda_{dI}^{[i]} \sum\limits_{k=0}^{n-1} \Braket{Q_j|Q_k} \alpha^{[i]}_k
  \label{GEVdI} \\
  \psi^{[i]}_{dI}(x)&=&\sum\limits_{k=0}^{n-1} \alpha^{[i]}_k Q_k(x)
\end{eqnarray}
Generalized eigenvalue problem (\ref{GEVdI}) provide
$n$ solutions ($i=[0\dots n-1]$),
each $i$ corresponds to the (eigenvalue,eigenfunction) pair
$(\lambda_{dI}^{[i]},\psi^{[i]}_{dI}(x))$.
The state $\psi^{[dIH]}_{dI}(x)$, corresponding to the maximal $\lambda_{dI}$,
is a second good candidate for generalized price impact calculation.

\subsection{\label{psiloc}$\psi$ Localized at $x_0$}
Localized at $x_0$ (the state ``time is now'') the wavefunction $\psi_0(x)$
is of ``interpolatory'' type 
and does not provide any valuable information
about market dynamics but is useful in some applications.
Take (\ref{psileastsquares}) and put $y=x_0$ to obtain the $\psi_0(x)$.
In \cite{2016arXiv160204423G,2015arXiv151005510G}, just for convenience,
we used normalized  $\psi_0(x)$:
\begin{eqnarray}
  \psi_0(x)&=&\frac{\sum\limits_{j,k=0}^{n-1}Q_j(x_0)(G^{-1})_{jk}Q_k(x)}
      {\sqrt{\sum\limits_{j,k=0}^{n-1}Q_j(x_0)(G^{-1})_{jk}Q_k(x_0)}}
      \label{psix0} \\
     1&=& \Braket{\psi_0|\psi_0}
\end{eqnarray}
The (\ref{psix0}) is plain normalized (\ref{psileastsquares}),
normalization factor  cancels in the numerator and
in the denominator of (\ref{faver})
when calculating an observable.

\section{\label{PImpactnumeric}Demonstration Of Generalized Price Impact Calculation}
\begin{figure}
  \includegraphics[width=16cm]{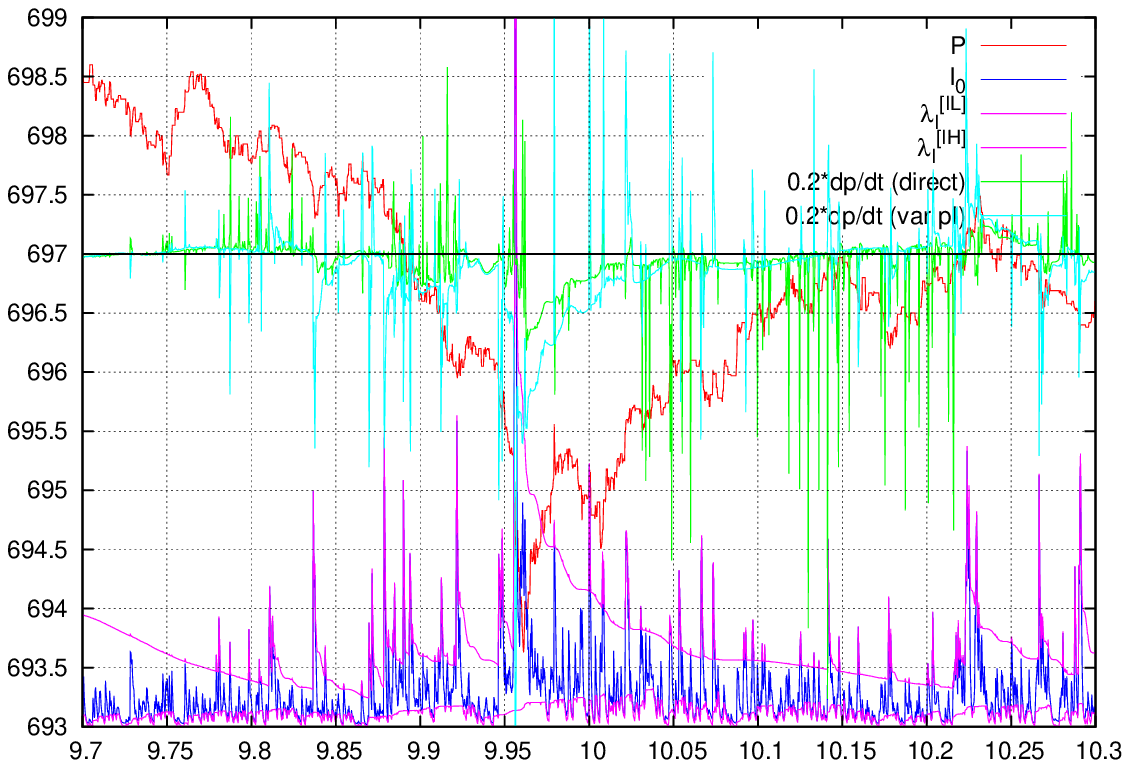}
  \includegraphics[width=16cm]{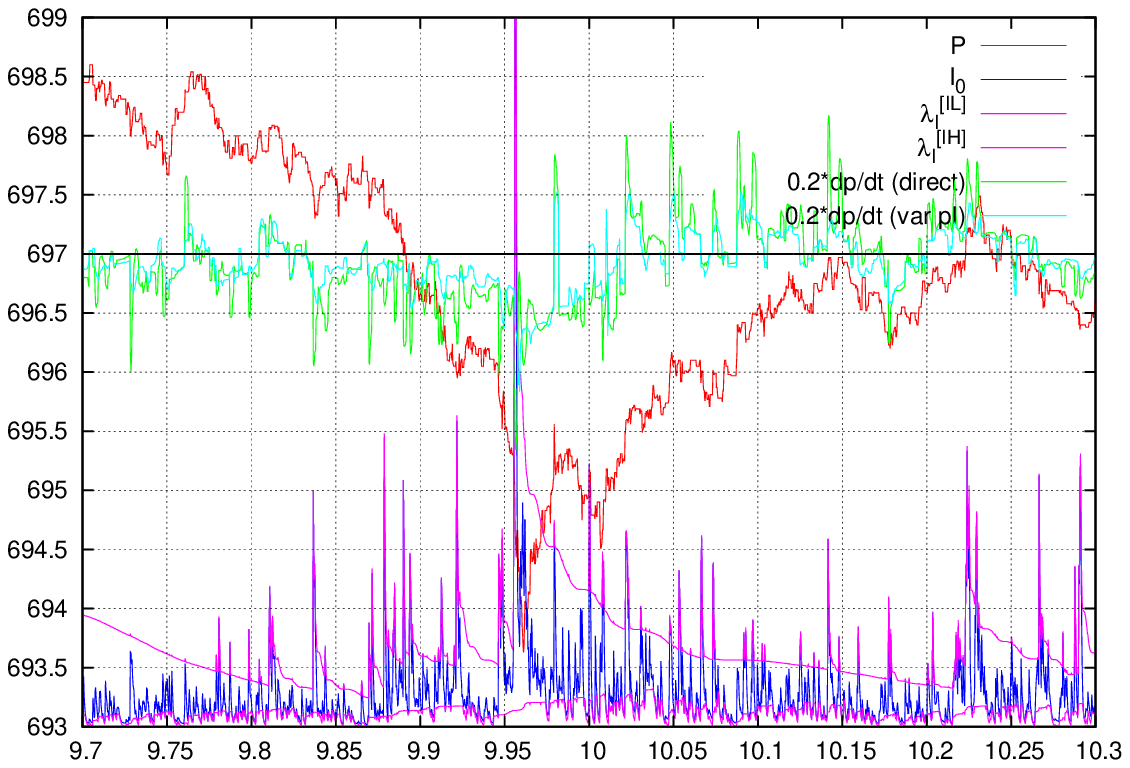}
\caption{\label{figdpdtdirect}
  The AAPL stock price on September, 20, 2012.
  Calculated in Shifted Legendre basis with $n=7$ and $\tau$=128sec.
  Calculations are performed using $dp$ moments ($direct$, Eq. (\ref{DPDirect}))
  and $pI$ moments ($var\, pI$, Eq. (\ref{DPDirectRQ})).
  Top:   Generalized Price Impact on $\psi$ state, corresponding
  to the maximal $I$, (\ref{Imax}).
  Bottom:   Generalized Price Impact on $\psi$ state, corresponding
  to the maximal $dI/dt$, (\ref{dImax}).
}
\end{figure}
In this section we calculate generalized price impact
on $\psi$ states discussed in the previous section.
In Fig. \ref{figdpdtdirect}
price change, corresponding to the state of maximal $I$ from (\ref{Imax})
subsection \ref{psiImax} and $dI/dt$ from (\ref{dImax})
subsection \ref{psidImax} are presented.
In these figures
\begin{eqnarray}
I_0=\Braket{\psi_0|I|\psi_0}
\label{I0def}
\end{eqnarray}
is the ``$I$ now'', calculated  with the $\psi_0$ from (\ref{psix0}),
the $\lambda_I^{[IH]}=\Braket{\psi^{[IH]}_I|I|\psi^{[IH]}_I}$, max $I$ solution of (\ref{GEVI}),
and $\lambda_I^{[IL]}=\Braket{\psi^{[IL]}_I|I|\psi^{[IL]}_I}$, the one corresponding to the minimal $\lambda_I$ of (\ref{GEVI}).
The $dp/dt (direct)$ is calculated using (\ref{DPDirect})
and $dp/dt (var\, pI)$ is calculated using (\ref{DPDirectRQ}).
From these charts it is clear that:
\begin{itemize}
\item
  Boundary $dp/dt$ contribution much exceed  non--boundary contribution,
  especially for large $I_0$;
  large $dp/dt$ typically corresponds to the boundary, i.e. large trading
  have just started ($\psi^{[IH]}_I(x)$ state is close to $\psi_0(x)$).
\item The Eqs. (\ref{DPDirect}) and (\ref{DPDirectRQ})
  give similar answers only when the boundary contribution is small.
\item The $dp/dt$ is typically much larger in
  the $\psi^{[IH]}_I(x)$ state, than in the $\psi^{[IH]}_{dI}(x)$ state.
\end{itemize}
This make us to conclude that:
\begin{enumerate}
  \item
 The eigenfunctions of $I$ operator (\ref{GEVI})
are  more important to market dynamics than the eigenfunctions
of $dI/dt$ operator (\ref{GEVdI}).
\item The concept of price impact
is poorly applicable to market dynamics, because of large contribution
of the boundary $x=x_0$. Because future ($x\ge x_0$) prediction
is the goal of any market dynamics study the attributes with
large boundary contribution (e.g. $dp/dt$)
are poorly applicable\cite{2015arXiv151005510G}.
\item Any consideration
  of infinitesimal time shifts 
  (e.g. price impact in (\ref{DPDirect}) or (\ref{DPDirectRQ}) form)
is poorly applicable to market dynamics.
A multi--state consideration (e.g. two different $\psi$ for enter and exit,
not infinitesimal variation of some $\psi$)
may be required.
\item At large $I_0$ the price has a singularity,
  same as in the quasistationary case\cite{2016arXiv160204423G}.
  In this paper we do not use a ``boundary condition $\psi(x_0)=0$''
  as we did in \cite{2015arXiv151005510G}, so we always
  have $\lambda_I^{[IL]}\le I_0 \le \lambda_I^{[IH]}$, see Fig. \ref{figdpdtdirect}.
  Bounded to $[0\dots 1]$  projections
  \begin{eqnarray}
    w^{[IL]}_I&=&\Braket{\psi_0|\psi^{[IL]}_I}^2 \label{wL} \\
    w^{[IH]}_I&=&\Braket{\psi_0|\psi^{[IH]}_I}^2 \label{wH}
  \end{eqnarray}
  $w^{[IL]}$ and $w^{[IH]}_I$
  are good indicators of ``low'' and ``high'' value of $I_0$ (also see 
  Eq. (\ref{skewnesslikeS0}) below for an alternative criteria).
  For a decision about ``low'' or ``high'' value of an 
  attribute, the
  estimation of wavefunction projection to the state of interest is a superior
  approach to any classical one with  a norm (i.e. $L^2$ or any other)
  and a threshold\cite{malyshkin2015norm}.
\item This confirms our approach\cite{2015arXiv151005510G}
to make a transition from price dynamics to execution flow and P\&L dynamics.
This to be considered next.
\end{enumerate}

\section{\label{FuturePsi} Impact From The Future.}
While the quasistationary case\cite{2016arXiv160204423G}
of dynamic equation is easy,
in a non--stationary case there are several fundamental questions
to be answered before considering any practical application.
We start with the ``infinitesimal future'' problem:
knowing the last price value, what information about
future price change can be obtained.

\subsection{\label{openQuestions}
  Open Questions (With Possible Answers)}
\begin{itemize}
\item {\bf What ``practically useful observable'' can be directly predicted
  from the dynamic equation\cite{2015arXiv151005510G}:
  ``Future price tends to the value that maximizes the number of shares traded per unit time''?}
  Future value of $I_0$ can be predicted. The (\ref{I0def}) gives ``current''
  value of $I_0$, it is calculated on already executed trades.
  Future value of $I_0$ (to be calculated on yet unexecuted trades)
  can be estimated as $\lambda_I^{[IH]}$, the very important  fact is 
  that future $I_0$ estimator $\lambda_I^{[IH]}$ is calculated on already executed trades!
  If trading ``now'' is slow ($I_0$ from (\ref{I0def}) is small),
  this means that at current price buyers and sellers do not match well and
  asset price has to move. Asset price is expected to move due to
  an increase in
  the
  ``future'' $I_0$, caused by the ``future execution''.  
  In this sense
  the more slow the market now is, the more dramatic market move to be expected in the future.
  The ``past most dramatic $I$'', the $\lambda_I^{[IH]}$, can be used
  as a reasonably good estimator (\ref{iofuture})
  of the ``future dramatic $I$'':
  \begin{eqnarray}  
  I_0^{f}&=&\lambda_I^{[IH]} \label{iofuture} \\
  dI &=& I_0^{f} -I_0  \label{dI}  \\
  dI &\ge &0 \label{dIge0}
  \end{eqnarray}
  Note, that similar ideology is often applied by market practitioners
  to asset prices or their standard deviations.
  This is incorrect.
  Experimental observations\cite{2016arXiv160204423G}
  show: this ideology can be applied {\em only} to execution flow $I=dV/dt$,
  not to the trading volume, asset price standard deviation
  or any other observable.
\item {\bf Given the role of the execution flow $I$, what is a criteria
  of presence (or absense) information about the ``future'' in the ``past data''?}
  If current $I_0$ from (\ref{I0def})
  is close to $\lambda_I^{[IH]}$, this means that we already have
  a ``very dramatic market'' and there is no much information
  about the future of this market.
   This is \textbf{\textsl{the condition of no information about the future}}:
  \begin{eqnarray}
    dI&=&0
    \label{dIeq0}
  \end{eqnarray}
But the most intriguing task would be to obtain
  directional information on price.
  \textbf{\textsl{The condition of no directional information
      about the future}}:
   \begin{eqnarray}
     \Ket{I^{f} |\psi_0}&=&\lambda \Ket{\psi_0}
    \label{dIfuturedir}
  \end{eqnarray}
is more restrictive than (\ref{dIeq0}).
  If the state ``time is now'', the $\psi_0(x)$ from (\ref{psix0}),
  is an eigenfunction of $\|I^{f}\|$ operator (\ref{Ifuture}),
  then past dynamics of $I$ has no information about the future
  (also note, that if  $\psi_0(x)$  is $\|I^{f}\|$
  eigenfunction, then it is $\|I\|$ eigenfuction either).
  The (\ref{dIeq0}) is a special
  case of (\ref{dIfuturedir}).
  Imagine extremely high volume was traded
  at $x=x_0$. Then the (\ref{GEVI}) solution, corresponding
  to $\lambda_I^{[IH]}$ is exactly the $\psi_0(x)$, and all other
  eigenfunctions ($i\ne IH$) have $\psi^{[i]}_I(x_0)=0$,
  what immediately give the (\ref{dIeq0}).
  Another example of (\ref{dIfuturedir}) condition
  is the case when execution occurred only ``now'' ($x=x_0$)
  and in the moments of $\psi_0(x)$ roots,
  that are the nodes of Gauss--Radau quadrature 
  built on the measure $(x_0-x)d\mu$, see Ref. \cite{2015arXiv151005510G}
  and computer code for calculating
  Gauss-type quadratures\cite{polynomialcode}.
  One more example is, for an arbitrary $\|\widetilde{I}\|$, to consider
  $\|I\|=\|\widetilde{I}\| - \frac{\Ket{\widetilde{I}|\psi_0}\Bra{\psi_0|\widetilde{I}}}{\Braket{\psi_0|\widetilde{I}|\psi_0}}$,
  then this  $\|I\|$ give the (\ref{dIfuturedir}) $\|I^{f}\|$.
  There is one more very important situation, when information about the
  future cannot be obtained: assume we have a trading
  without execution flow fluctuations, $I=const$, then
  $\|I\|$ operator is degenerated (all eigenvalues are the same: $\lambda^{[i]}_{I}=I=const$), what immediately lead to
  both (\ref{dIeq0}) and (\ref{dIfuturedir})  being satisfied.

\item {\bf While the $I=dV/dt$ dynamics is more or less understood,
  how can it  be converted to a price dynamics?}
  This is the most difficult problem.
The relation between $p$ and $I$ is the fundamential
question of market dynamics.
We started this discussion in \cite{2016arXiv160204423G},
and have shown experimentally, that execution flow affect
price much stronger (dynamic impact), than traded volume (regular impact).
We also noticed there, that $p$ and $I$ often reach an extremum
in the same $\psi$ state, i.e. their operators have the same eigenfunctions.
Introduce
\textbf{\textsl{dynamic impact approximation}}
assuming asset price is affected only by the execution flow $I$,
not by the volume traded:
\begin{eqnarray}
  p&=&p(I) \label{dynimpactdef}
\end{eqnarray}
If (\ref{dynimpactdef}) holds
then
$p$ and $I$ have the same  tipping points,
the behaviour we experimentally observed in Ref.\cite{2016arXiv160204423G}.
More generally, if price is only a function of $I$
then corresponding $\|p\|$  and $\|I\|$ operators to have
the same eigenfunctions,
the behaviour we observed\cite{2016arXiv160204423G}
for the states with high $I$.
We already estimated (\ref{iofuture})
future value of $I_0$
as $\lambda_I^{[IH]}$ and can build
$\|I^{f}\|$ operator (\ref{Ifuture}),
having  $dI$  contribution  ``from the future'' (\ref{dI}).
Then future value of price
can be estimated 
considering the $\|p^mI^{f}\|$ operator (\ref{pImfutureOp}),
on eigenstates  already found for $\|I^{f}\|$ operator (\ref{GEVIf}).
The price is secondary to the liquidity flow,
but their common
eigenfunctions allows to use future value of $I$ to calculate future value of $p$.
\end{itemize}

\subsection{\label{openQuestionsWithout}
  Open Questions (Without Answers)}
\begin{itemize}
\item  {\bf What is the role of infinitesimal time--shift operator,
  available in some bases, e.g.  (\ref{DpsiLag}) and (\ref{DpsiLegendre})?}
  It is very seductive to use infinitesimal time--shift operator
  to define a
  Lagrange functional (combining price volatility and execution rate),
  build an action $\cal S$ (like other dynamic theories do),
  then try to minimize  $\cal S$
  to build a theory combining both trend following (due to execution flow)
  and price reverse (due to price volatility)\cite{2015arXiv151005510G}.
  Despite all our effort we failed with this plan.
  Even first order infinitesimal time--shift give
  the results similar to price impact
  of Section \ref{PImpactnumeric} above.
  Typical for other dynamic theories second order infinitesimal time--shifts
  give
  an answer with even larger boundary $x=x_0$ contribution, thus having little predictive power. This make us to conclude that
  infinitesimal time--shifts are not very perspective
  for market dynamics and  finite variations to be considered instead.
\item  {\bf What is the role of $dp/dt$  in the dynamic equation,
  especially, whether price volatility can be expressed through the $(dp/dt)^2$
  term \cite{2015arXiv151005510G}?}
   As we already emphaised several times above ``the price is secondary to liquidity flow'', the $dp/dt$ spikes are just a consequency of liquidity fluctuations, the charts of  Section \ref{PImpactnumeric} above seems to prove this.
   But this statement results in
   ``future price does not depened on past prices'',
   what make our theory too provocative,
   e.g. it predicts that 
   all theories of  ``trend following'' or ``reverse to the mean''
   based only on price trends  are invalid.
 \item  {\bf What is the role of basis minimal and maximal time scale
   (how to determine $n$ and $\tau$)?}
   If we assume that the $\Braket{Q_j |f| Q_k}$ matrix has all the
  information about $f$, then we can easily calculate
   the values, that cannot be directly calculated
   from a sample\cite{2015arXiv151005510G}.
   For example
   price volatility matrix in the form $(dp/dt)^2$,
   that cannot be calculated directly from sample,
   can be expressed through calculatable directly from sample
   $dp/dt$ matrix using $f=g=dp/dt$:
\begin{eqnarray}
  \Braket{Q_j |f g | Q_k}&=&
  \sum\limits_{l,m=0}^{n-1}\Braket{Q_j |f| Q_l}
  \left(G^{-1}\right)_{lm}
  \Braket{Q_m |g| Q_k}
  \label{fg}
\end{eqnarray}
   Numerical experiment have shown
   this approach is not a very successful one.
   One can also try to compare the $\|pI\|$ matrix calculated directly
    $\Braket{Q_j |pI| Q_k}$
   and Hermitian part of  (\ref{fg}) calculated with $f=p$ and $g=I$.
   The $\tau$ determines a ``base'' time scale,
   $n$ determines the time--scale variation. While this approach
   is a great advance from ``moving average''--type of approaches
   with a single predefined time--scale (corresponds to $n=1$),
   now we automatically select
   the state out of $n$ eigenfunctions with their own time--scales
   (in practice $n\le 15$), we still do not have
   a formal way to select proper $n$ and $\tau$.
\end{itemize}

\subsection{\label{FuturePsiOp} Impact From The Future Operator.}
As we stated above maximal (\ref{GEVI}) eigenvalue, the $\lambda_I^{[IH]}$,
can serve as an estimator of future $I_0$.
Then  execution flow operator
with an impact from the future is:
 \begin{eqnarray}  
  \|I^{f}\|&=&\|I\| + \Ket{\psi_0}dI\Bra{\psi_0} \label{Ifuture}
  \end{eqnarray}
The term $\Ket{\psi_0}dI\Bra{\psi_0}$  is proportional to
the execution flow of not yet executed trades $dI$ from (\ref{dI});
we now have $\Braket{\psi_0 | I^{f} | \psi_0}=I_0^{f}$ and $\Braket{\psi_0 | I | \psi_0}=I_0$.
To find future equilibrium wavefunction, according to  dynamic equation,
eigenvalues problem for $\|I^{f}\|$ operator needs to be solved
\begin{eqnarray}
  \Ket{I^{f}|\psi_{I^{f}}^{[i]}}&=&\lambda_{I^{f}}^{[i]}\Ket{\psi_{I^{f}}^{[i]}}
  \label{GEVIf} 
\end{eqnarray}
the Eq. (\ref{GEVIf}) is the same as the Eq. (\ref{GEVI}), but
with the $\|I^{f}\|$ operator from (\ref{Ifuture})
instead of $\|I\|$ operator in  (\ref{GEVI}).
Eigenvalue selection in (\ref{GEVI}) was easy, it was the
state with the maximal $\lambda_{I}^{[i]}$, according to our dynamic equation
(\ref{Imax}), from where we received the (\ref{iofuture}).
But for (\ref{GEVIf}) the answer is not so trivial.
As we demonstrated in \cite{2016arXiv160204423G},
asset price is much more
sensitive to execution rate $I=dV/dt$, rather than to trading volume $V$,
thus in dynamic impact approximation (\ref{dynimpactdef})
the contribution of  $\Ket{\psi^{[i]}_{I^{f}}}$ state
to future price changes
is proportional to the flow of not yet executed trades $\Braket{\psi^{[i]}_{I^{f}}|\psi_0}^2dI$.
For this reason we are going to keep all eigenfunctions of (\ref{GEVIf}) problem.
The $\Ket{\psi^{[i]}_{I^{f}}}$ is $\|I^{f}\|$ operator
eigenfunction (\ref{GEVIf}),
thus first order variation (\ref{VarpsiFuture})
is equal to zero for arbitrary  $\Ket{\delta \psi}$.
\begin{eqnarray}
\frac{1}{2}\delta \frac{
  \Braket{\psi^{[i]}_{I^{f}}|I^{f}|\psi^{[i]}_{I^{f}}}}
       {\Braket{\psi^{[i]}_{I^{f}}|\psi^{[i]}_{I^{f}}}}=
       \Braket{\psi^{[i]}_{I^{f}}|I^{f}|\delta\psi}-
       \Braket{\psi^{[i]}_{I^{f}}|I^{f}|\psi^{[i]}_{I^{f}}}
       \Braket{\psi^{[i]}_{I^{f}}|\delta\psi}
       &=&0
       \label{VarpsiFuture}
\end{eqnarray}
The $\|p^mI^{f}\|$ operator
(for practical applications it is more convenient to
consider operator $p^mI$ instead of $p^m$)
with an impact from the future is:
\begin{eqnarray}
  \|p^mI^{f}\|&=&\|p^mI\| + \Ket{\psi_0}P^{fm}dI\Bra{\psi_0} \label{pImfutureOp}\\
  P^{fm}&=&\left(P^{last}\right)^m
  \label{Pfmestlast}
\end{eqnarray}
The term $\Ket{\psi_0}P^{fm}dI\Bra{\psi_0}$ for $m=1$ is proportional to
execution capital flow of  not yet executed trades
at unknown future price $P^{f1}$ with known future execution rate contribution
$dI$ from (\ref{dI}).
``The last price as $P^{f}$ estimator (\ref{Pfmestlast})''
is the simplest estimation,
meaning the best estimation of future price is current value.
In equilibrium the $\|p\|$ and $\|I^{f}\|$ to have
the same eigenfunctions $\Ket{\psi^{[i]}_{I^{f}}}$,
at least for the states with a high $\lambda^{[i]}_{I^{f}}$,
so the most promissing idea is to consider $\|p^mI^{f}\|$ operator
on eigenstates of  $\|I^{f}\|$ and $\|\frac{d}{dt}I^{f}\|$.

\subsection{\label{naive} Equilibrium Price in Na\"{\i}ve Dynamic Impact Approximation}
In pure dynamic impact approximation
formal answer for future equilibrium price can be obtained.
This answer is not a very practical,
so we would call it \textsl{Na\"{\i}ve Dynamic Impact Approximation},
but it is worth
considering to compare it with the answer
from our previous work\cite{2015arXiv151005510G}.
\begin{figure}
  \includegraphics[width=16cm]{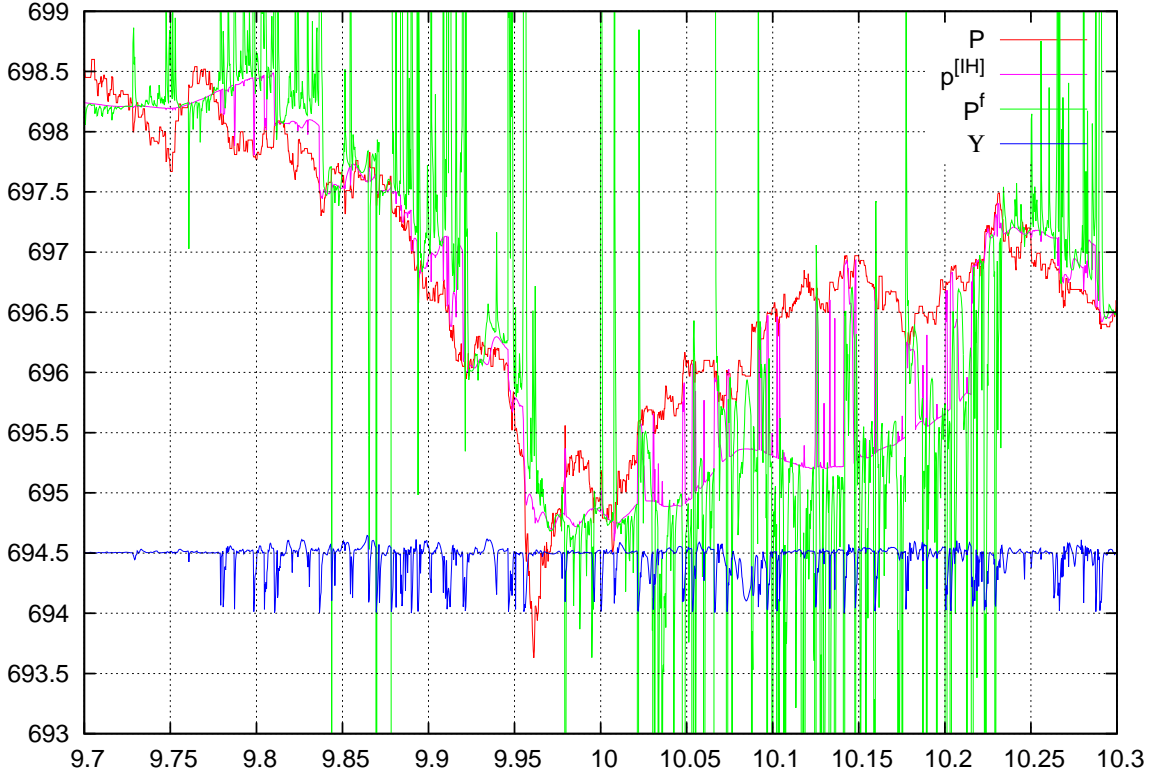}
\caption{\label{pfpih}
  The AAPL stock price on September, 20, 2012.
  $P^{[IH]}$ (\ref{PIH}) (pink), $P^{f}$ (\ref{PfNative}) (green),
  and $\Upsilon$ (\ref{degeneray}) (shifted to 694 level to fit the chart).
 Calculated in Shifted Legendre basis with $n=7$ and $\tau$=128sec.
}
\end{figure}

Future equilibrium price $P^{f}$ enter
impact from the future operator (\ref{pImfutureOp})
from which $P_0$ is calculated as:
\begin{equation}
  P_0=  \frac{\Braket{\psi_0|pI^{f}|\psi_0}}{\Braket{\psi_0|I^{f}|\psi_0}}
  \label{p0Exa}
\end{equation}
Now, assume $\|pI^{f}\|$ and $\|I^{f}\|$
are diagonal in the same basis, the solution of (\ref{GEVIf}).
Expanding $\Ket{\psi_0}=\sum_{i=0}^{n-1}\Braket{\psi_0|\psi_{I^{f}}^{[i]}}\Ket{\psi_{I^{f}}^{[i]}}$
and assuming all off diagonal ($i\ne j$) matrix $\|pI^{f}\|$
elements are zero: $\Braket{\psi_{I^{f}}^{[i]}|pI|\psi_{I^{f}}^{[j]}}=0$,
same as we have for $\|I^{f}\|$ in (\ref{GEVIf}).
Then the $P_0$ can be estimated only from diagonal elements of $\|pI^{f}\|$:
\begin{equation}
  P_0=
  \sum\limits_{i=0}^{n-1}\frac{\Braket{\psi^{[i]}_{I^{f}}|pI^{f}|\psi^{[i]}_{I^{f}}}}{\lambda_{I^{f}}^{[i]}}\Braket{\psi_0|\psi^{[i]}_{I^{f}}}^2
  \label{p0Dyn}
\end{equation}
Then (\ref{p0Exa}) and (\ref{p0Dyn}) with (\ref{pImfutureOp}) give the solution for $P^{f}$:
\begin{eqnarray}
  \frac{\Braket{\psi_0|pI^{f}|\psi_0}}{I_0^{f}}&=&
  \sum\limits_{i=0}^{n-1}\frac{\Braket{\psi^{[i]}_{I^{f}}|pI^{f}|\psi^{[i]}_{I^{f}}}}{\lambda_{I^{f}}^{[i]}}\Braket{\psi_0|\psi^{[i]}_{I^{f}}}^2
  \\
  \Upsilon&=&1-\sum\limits_{i=0}^{n-1}\Braket{\psi_0|\psi^{[i]}_{I^{f}}}^4\frac{I_0^{f}}{\lambda_{I^{f}}^{[i]}} \label{degeneray} \\
  P^{f}&=&\frac{1}{\Upsilon dI}\left(
  -\Braket{\psi_0|pI|\psi_0}+\sum\limits_{i=0}^{n-1}\Braket{\psi^{[i]}_{I^{f}}|pI|\psi^{[i]}_{I^{f}}}\Braket{\psi_0|\psi^{[i]}_{I^{f}}}^2\frac{I_0^{f}}{\lambda_{I^{f}}^{[i]}}
  \right)
  \label{PfNative}
\end{eqnarray}
Conceptually (but not practically) the (\ref{PfNative})  directional answer is a giant step forward
from our previous work\cite{2015arXiv151005510G},
where the best directional  estimator  was
the difference between last price and
the price $P^{[IH]}$,
corresponding to the state of maximal $I$ {\em on past sample},
the (\ref{pIcalc}) calculated on $\Ket{\psi^{[IH]}_I}$ state (\ref{psiIsol}):
\begin{eqnarray}
  P^{[IH]}&=&\frac{\Braket{\psi^{[IH]}_I|pI|\psi^{[IH]}_I}}{\lambda_I^{[IH]}}
  \label{PIH}
\end{eqnarray}
This Ref. \cite{2015arXiv151005510G} answer is
asset  price averaged on past sample with
always positive weight $\left(\psi^{[IH]}_I(x)\right)^2 I d\mu(x)$;
no explicit information about the future
is used in this averaging.
The (\ref{PfNative}) answer is very different:
it directly incorporates information  about
not yet executed trades from the future
using $dI$ and $\Ket{\psi^{[i]}_{I^{f}}}$
obtained from (\ref{iofuture}) assumption about $I_0^{f}$.
The $\Upsilon$ from (\ref{degeneray})
formally define the degree of degeneracy,
how much directional information can be obtained from the sample,
it is zero when $\Ket{\psi_0}$ is (\ref{GEVIf}) eigenvector, condition (\ref{dIfuturedir}).
Future volatility prediction is easy,
for example (\ref{wL}) and (\ref{wH}) projections
can be used to estimate whether current $I_0$ (\ref{I0def})
is ``low'' or ``high'', then use (\ref{dI}).
Future directional prediction is much more complicated,
the (\ref{PfNative}) is the simplest (na\"{\i}ve)
directional answer that can be obtained.
In Fig. \ref{pfpih} the $P^{[IH]}$ (\ref{PIH}),
$P^{f}$ (\ref{PfNative}), and $\Upsilon$ (\ref{degeneray})
are presented. The degeneracy $\Upsilon$
typically has a value $1/2$, but going to $0$
at times of high $I_0$, what correspond to (\ref{dIfuturedir}) condition.
In \cite{2015arXiv151005510G} the difference
between last price  and $P^{[IH]}$ was used as a directional estimator.
If $P^{f}$ is used instead, the result, as one see
from Fig. \ref{pfpih} is very similar (sign does not change),
but, as expected,  $P^{f}$ is not close to last price
at high $I_0$.
The (\ref{PfNative}) is
asset  price averaged on past sample,
but, in contrast with (\ref{PIH}),
with the weight, which \textsl{is not} always positive. This
lead to a divergence in $P^{f}$  (especially at low $\Upsilon$ and/or small $dI$). This divergence typically does not change the $P^{last}-P^{f}$ sign.
Overall the (\ref{PfNative})
seems to be  a marginal improvement over our old answer (\ref{PIH}),
this is why we call (\ref{PfNative}) \textsl{na\"{\i}ve answer}.
For computer implementation see the \texttt{\seqsplit{PnLdIDSk.Pf\_from\_pt\_true\_pi}}
for $P^f$ and
\texttt{\seqsplit{PnLdIDSk.deg\_from\_pt\_true\_pi}} for $\Upsilon$.
Computer code structure is described in appendix \ref{codestr}.

\section{\label{Thr}Selection of  Time--Scale,
  Then Determine Price Distribution Asymmetry From Quadrature.
  Trend--Following vs. Reverse to the Mean
}
Equilibrium price estimation, let it be (\ref{PfNative})
of previous section 
or (\ref{PIH})
of our previous work\cite{2015arXiv151005510G},
and using the difference between $P^{last}$ and calculated
price as directional indicator,
typically does not give a satisfactory results,
as price is secondary concept to market dynamics.
The characteristics, describing the P\&L
distribution should be considered instead.

Let us start with the simplest problem of price distribution.
As we discussed in Section \ref{probabilitystate}
a measure
is defined by a wavefunction $\psi(x)$,
the measure is $\psi^2(x)d\mu$,
then price moments $\pi_m$, $m=0,1,2,3$ are:
\begin{eqnarray}
  \pi_m&=&\Braket{\psi|p^mI|\psi}
  \label{pimdef}
\end{eqnarray}
(similar expression  without $I$ can be used $\Braket{\psi|p^m|\psi}$,
but (\ref{pimdef}) choice is better in applications).
The (\ref{pimdef}) expression selects the time scale
based on $\psi(x)$ choice. This way (via $\psi(x)$)
the (\ref{dI}) information
about future $I$ can be incorporated.
Different $\psi(x)$ choices
are considered below.
For now assume, that some $\psi(x)$ is  chosen and
the goal is 
 to estimate price distribution
 on the measure generated by this $\psi(x)$ .
The standard approach is to consider price
average, standard deviation and skewness.
In the Appendix C of Ref. \cite{2015arXiv151005510G}
modified skewness estimator was introduced.
The $\pi_m$ moments describe how the price
is distributed at times of the support of the  measure.
The skewness of the distribution
is typically used for estimation of future price direction.
However, a much better, than a regular skewness, answer can be obtained.
The idea is to build two--point Gauss quadrature out of $\pi_m$, $m=0,1,2,3$
moments then consider quadrature weights asymmetry
(single--point Gauss quadrature require two moments $\pi_0$ and $\pi_1$
to calculate
and give price average as the node:
$p_1=\pi_1/\pi_0$, the weight $w_1=\pi_0$).
It is very important, that besides weights, two--point quadrature nodes
can be used to determine threshold levels.
The two nodes $\lambda_p^{[s]}$
are generalized eigenvalue problem solution:
\begin{eqnarray}
    \left(
  \begin{array}{ll}
    \pi_1 &  \pi_2 \\
   \pi_2 & \pi_3
  \end{array}
  \right)
  \left(
  \begin{array}{l}
    \alpha_{0}^{[s]}\\
    \alpha_{1}^{[s]}
  \end{array}
    \right)
    &=&
    \lambda_{p}^{[s]}
    \left(
 \begin{array}{ll}
    \pi_0 &  \pi_1 \\
   \pi_1 & \pi_2
  \end{array}
  \right)
  \left(
  \begin{array}{l}
    \alpha_{0}^{[s]}\\
    \alpha_{1}^{[s]}
  \end{array}
  \right)
  \label{evp3quadr} \\
  p_{\{1,2\}}&=&\lambda_{p}^{[\{1,2\}]} \label{evp3nodes} \\
  w_{\{1,2\}}&=&\frac{1}{\left(\alpha_{0}^{[\{1,2\}]}+\lambda_{p}^{[\{1,2\}]}\alpha_{1}^{[\{1,2\}]}\right)^2}
  \label{evp3weights} \\
  \Gamma&=&\frac{w_1-w_2}{w_1+w_2} =
\frac{2\overline{p}-p_1-p_2}{p_1-p_2}
  \label{skewness}
\end{eqnarray}
The quadrature nodes $p_{\{1,2\}}$ are the eigenvalues (\ref{evp3nodes})
(we assume $p_1<p_2$),
and the quadrature weights $w_{\{1,2\}}$ are expresses via
the eigenfunction (\ref{evp3weights}), for numerical
calculation see the class \texttt{\seqsplit{com/polytechnik/utils/Skewness.java}}.
Note that defined in (\ref{skewness}) skewness $\Gamma$
 is similar in concept to
the ``signed volume'' (the difference between market--sell matched limit--buy
and market--buy matched limit--sell orders).
As we emphasized earlier\cite{2016arXiv160305313G},
regular signed volume concept is not a practical one.
Important, that (\ref{skewness}) definition
allows us to obtain volume difference from trades
history only, no matching type knowledge is required.
See alternative formulas for (\ref{skewness})
in the Appendix C of Ref. \cite{2015arXiv151005510G}
to obtain (\ref{evp3nodes}) and (\ref{evp3weights})
by minimizing over the $p_{\{1,2\}}$ nodes the expression:
\begin{eqnarray}
  L^4volatility&=&\Braket{\psi|(p-p_1)^2(p-p_2)^2I|\psi}\to\min \label{p4min}
\end{eqnarray}
The (\ref{p4min}) is the definition of $L^4$ volatility,
minimization of which give the $p_{\{1,2\}}$ nodes (\ref{evp3nodes}).
Compare it to well known ``minimizing volatility as standard deviation
over the $\overline{p}$'':
\begin{eqnarray}
  L^2volatility&=&\Braket{\psi|(p-\overline{p})^2I|\psi}\to\min
 \label{p2min}
\end{eqnarray} 
that gives the (\ref{paverW}) expression for the average price $\overline{p}$
(single node Gauss quadrature) and to kurtosis
calculation as $\Braket{\psi|(p-\overline{p})^4I|\psi}$.
For two variables $p$ and $r$ a $L^4covariation$,
correlating (\ref{evp3quadr}) eigenfunction (they are proportional to Lagrange interpolating polynomials) for $p$ and $r$ quadratures
can be introduced,
see Appendix \ref{twovarcorrela} below for calculations.

Two point Gauss quadrature give
exact integration answer for integration of a polynomial
of degree 3 or less ($n$ point quadrature is exact
for a polynomial of degree $2n-1$ or less).
Familiar average, standard deviation
and skewness  can be expressed by averaging
at $p_1$ with the weight $w_1$ and at $p_2$ with the weight $w_2$:
\begin{eqnarray}
  \pi_0&=&w_1+w_2\\
  \pi_1&=&p_1 w_1+p_2 w_2\\
  \pi_2&=&p^2_1 w_1+p_2^2 w_2\\
  \pi_3&=&p_1^3 w_1+p_2^3 w_2\\
  \overline{p}&=&\frac{\pi_1}{\pi_0}=p_1\frac{ w_1}{w_1+w_2}+p_2\frac{w_2}{w_1+w_2} \label{paverW} \\
  \overline{\left(p-\overline{p}\right)^2}&=&
  (p_1-\overline{p})^2\frac{ w_1}{w_1+w_2}+(p_2-\overline{p})^2\frac{w_2}{w_1+w_2} \\
 \overline{\left(p-\overline{p}\right)^3}&=&
  (p_1-\overline{p})^3\frac{ w_1}{w_1+w_2}+(p_2-\overline{p})^3\frac{w_2}{w_1+w_2}
\end{eqnarray}
The distribution itself can now be considered
as two--mode distribution:
trading at $p_1$ with the weight $w_1$ and trading at $p_2$ with the weight $w_2$.
This gives huge advantage: an opportunity to implement ``follow the trend'' type of strategy.
For a single--point Gauss quadrature the only node
is price average $\overline{p}$
and only strategy available
is ``reverse--to-the--average'' type of strategy
(average price as an attractor).
For two--point Gauss quadrature one can
implement a ``follow the trend'' type of strategy
(average price as a repeller, $p_{\{1,2\}}$ as the attractors),
in a most simplistic way it is: ``Open Short when $p_1<P^{last}<\overline{p}$;
Open Long when $\overline{p}<P^{last}<p_2$; combine with weights asymmetry''.
The two new price levels: $p_1$ and $p_2$
allow to have a completely new look
to trend--following trading:
if $\pi_m$ are moving--average  moments,
then the $p_1$ and $p_2$ are much better thresholds
than often used $\overline{p}\pm\sigma$,
because they include the skewness of price distribution,
the thresholds are now
different for up and down moves, according to the distribution skewness.
This approach is much more generic, than this simple demonstration.
The key components of it are:
\begin{itemize}
\item Find the $\psi$ of interest.
  Several choices of $\psi$
  are considered below. As we emphasized above the most
  interesting $\psi$ is the one maximizing the $\|I^{f}\|$ operator
  according to the dynamic equation. However, other
  $\psi$ choices can be also considered, at least for the
  purpose of the  demonstration of the technique.
\item Given $\psi$ obtain the measure $\psi^2(x)d\mu$
  to calculate price moments $\pi_m$ from  (\ref{pimdef})
  Then Gauss quadrature
  nodes $p_{\{1,2\}}$ and weights $w_{\{1,2\}}$ to be obtained.
  This quadrature determines the distribution of price in the $\psi$ state.
  One can try to obtain some directional information on price
  from this distribution (e.g. skewness estimation (\ref{skewness})).
  Note, that when using (\ref{pImfutureOp}) operators,
  with an impact from the future term,
  future price $P^{f}$ is required to calculate the moments,
  ``the last price as $P^{f}$ estimator (\ref{Pfmestlast})''
  is a very crude approximation.
  While future price $P^{f}$ is unknown,
  all the calculations above can be reperated using $P^{f}$
  as a parameter, see Appendix \ref{quadrParam} below
  where the dependendce of $\Gamma(P^{f})$ on $P^{f}$
  is obtained (\ref{spkewnessPf}).

\item In addition, some
  other value $r$ (e.g. market index, etc.)
  can be considered and cross--correlation of
  Appendix \ref{twovarcorrela} below can be performed.

\end{itemize}

\section{\label{quadexample}Demonstration
  of price--distribution estimation from two--point Gauss quadrature
  built for a measure of interest}
  
Let us demonstrate the technique
of building two--point Gauss quadrature out of $\pi_m$ moments (\ref{pimdef}) 
calculated for a number of $\psi$
choices.

\subsection{\label{movavermoments}Measure: Moving Average and Moving Average --Like }
\begin{figure}
  \includegraphics[width=16cm]{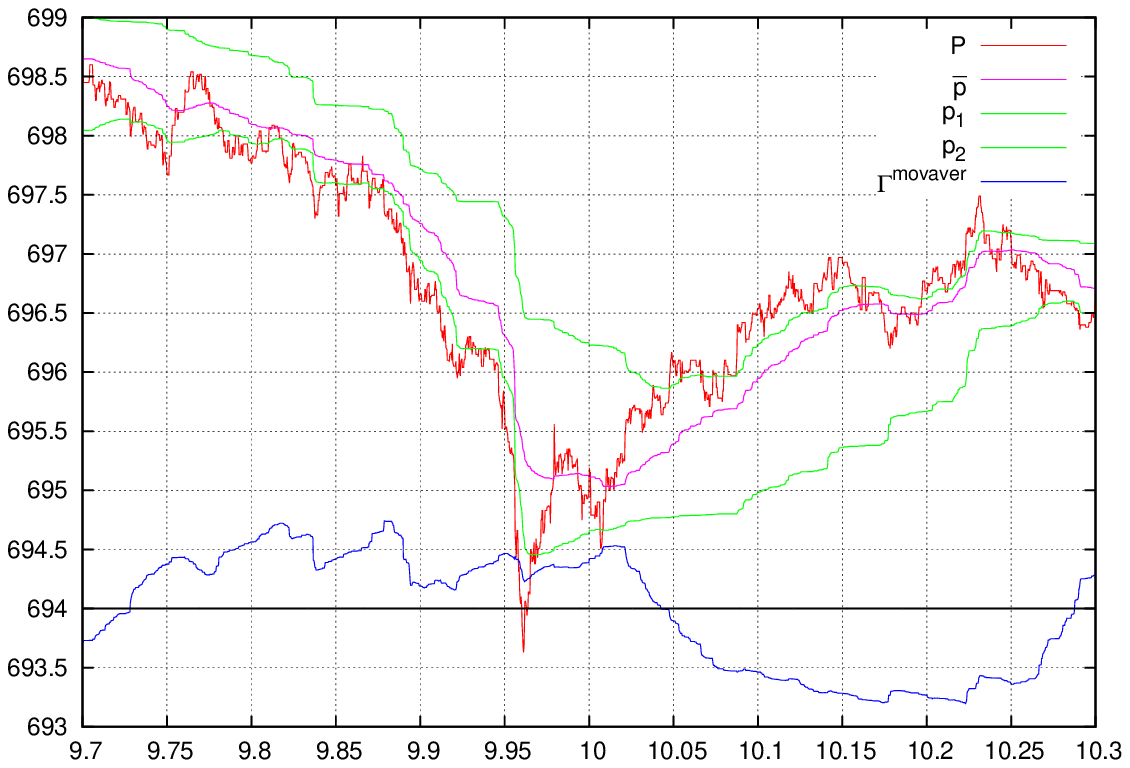}
  \includegraphics[width=16cm]{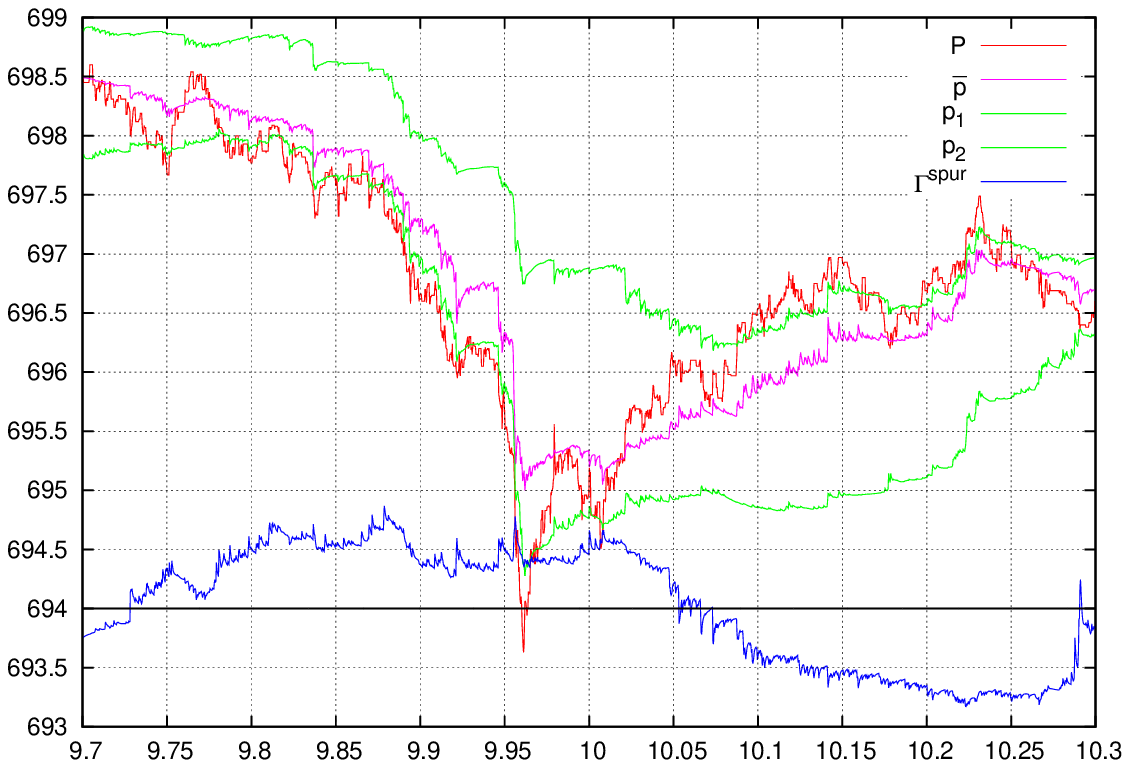}
\caption{\label{movaverp3}
  The AAPL stock price on September, 20, 2012.
  Top: Demonstration of Gauss quadrature calculation with
  moving average  (\ref{pimmovaver}) moments,
  $\overline{p}=\pi_1/\pi_0$ -- exponential moving average with $\tau$=128sec,
  $p_1$, $p_2$ -- quadrature nodes calculated according to (\ref{evp3quadr}),
  and modified skewness (\ref{skewness}) $\Gamma$ (shifted to 694 level to fit the chart).
  Bottom: same thing with (\ref{pimspurnodI}) mixed state moments.
}
\end{figure}
The most simple example is moving average--type of measure
(corresponds to $\psi(x)=1$, also assume here, that there is no impact from the future: $dI=0$).
Calculate the moments:
\begin{eqnarray}
  \pi_m&=&\Braket{p^mI}
  \label{pimmovaver}
\end{eqnarray}
  Then
$\overline{p}_{\tau}=\pi_1/\pi_0=p_1\frac{w_1}{w_1+w_2}+p_2\frac{w_2}{w_1+w_2}$
is regular exponential moving average.
Gauss quadrature nodes $p_{\{1,2\}}$ and weights $w_{\{1,2\}}$
are calculated according to (\ref{evp3quadr}),
and $\Gamma$ from (\ref{skewness}).
These values are presented in Fig. \ref{movaverp3}.
Even in this non--practical example (because of fixed time--scale $\tau$)
we clearly see an asymmetry between $\overline{p}_{\tau}$ and $p_{\{1,2\}}$.
Median estimator $(p_{1}+p_{2})/2$ is equal to average $\overline{p}_{\tau}$
only in the case of zero skewness. We also see good skewness correlation
with price trend, but, as for any model with
a fixed time--scale, there is fixed time delay between
price trend change and skewness change. However, the asymmetry
between $\overline{p}$ and $p_{\{1,2\}}$ is a remarkable
feature that may be incorporated to a trading model,
because three levels now allow to implement a ``follow--the--trend''
type of strategy.

There is a characteristics, that is very similar to exponential moving average,
but described by
a density--matrix state, it cannot be reduced to a
state of some $\Ket{\psi}$.
In its simplistic form the $\pi_m$ moments are matrix spur:
\begin{eqnarray}
  \pi_m&=&\sum\limits_{i=0}^{n-1}
  \Braket{\psi^{[i]}_{I}|p^mI|\psi^{[i]}_{I}}
  \label{pimspurnodI}
\end{eqnarray}
These are different from (\ref{pimspur}) in Section \ref{MeasureSpur} below
 in absence of the impact from the future term, $dI=0$.
(Note, that (\ref{pimspurnodI}) is invariant with respect to basis
 transform, also see\cite{2015arXiv151005510G} Appendix E
 of the expression in a non--orthogonal basis:
 $\pi_m=\sum\limits_{j,k=0}^{n-1}(G^{-1})_{jk} \Braket{Q_k|p^mI|Q_j}$).
The result is presented in Fig. \ref{movaverp3} bottom.
It is very similar to moving average result, as expected.
These two kind of ``moving average'':
with (\ref{pimmovaver}) ``pure state'' and (\ref{pimspurnodI}) ``mixed state''
moments, demonstrate wavefunction and density--matrix  approaches.
In this section we specifically chose the situation,
when both approaches give
very similar result.

\subsection{\label{PsiFutureI}
 Measure: The Period of Maximal Future $I$}
\begin{figure}
  \includegraphics[width=16cm]{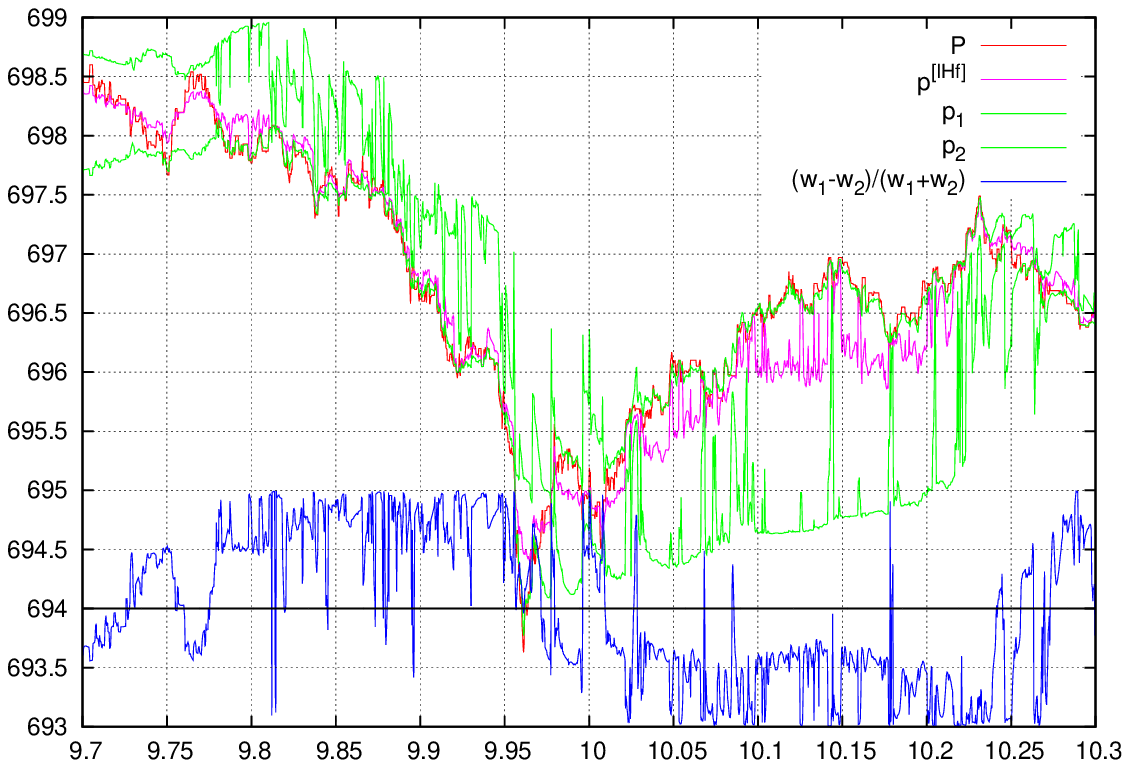}
  \includegraphics[width=16cm]{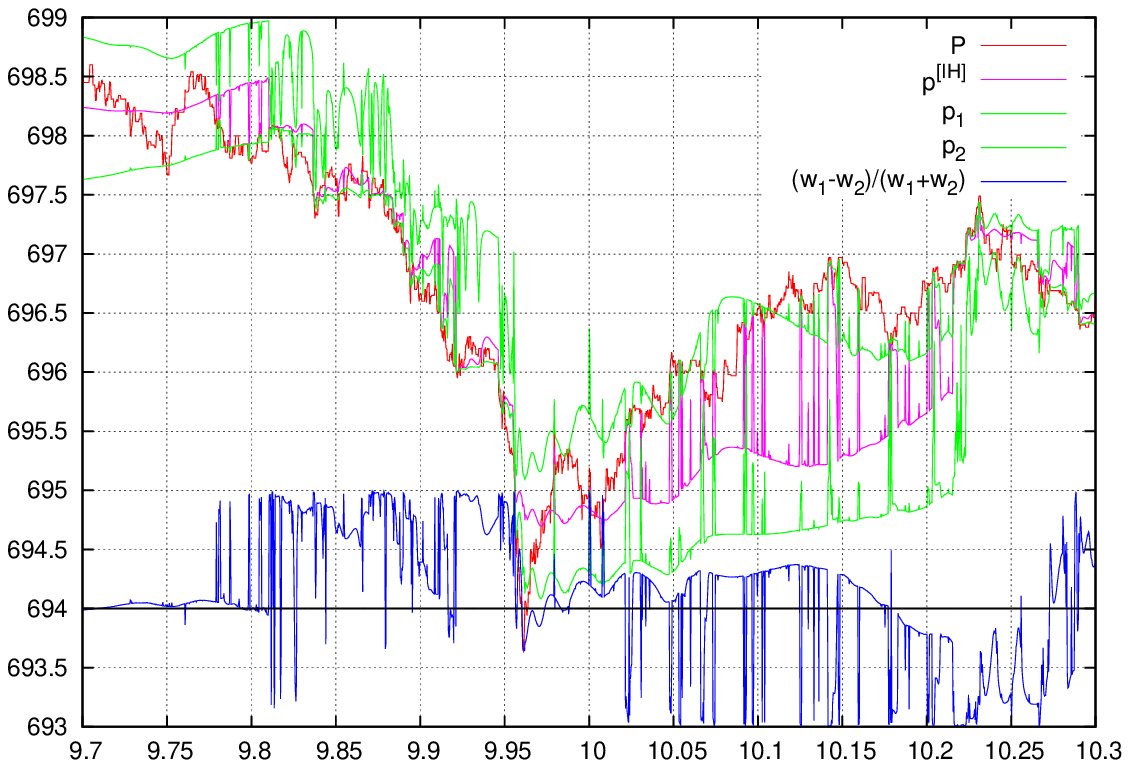}
\caption{\label{psiIfp3}
  The AAPL stock price on September, 20, 2012.
  Demonstration of Gauss quadrature calculation with
  the state (moments $\pi_m$ from (\ref{pimfuture})),corresponding to maximal $\|I^{f}\|$ (top)
  and (moments $\pi_m$ from (\ref{pimpast})), maximal $\|I\|$ (bottom).
  The prices and skewness are presented
  as in Fig. \ref{movaverp3} above.
}
\end{figure}

Consider the periods of maximal future $I$.
The ``future'' time scale is determined by
the future state $\Ket{\psi^{[IH]}_{I^{f}}}$, the eigenfunction
of (\ref{Ifuture}) operator, the (\ref{GEVIf})
solution, corresponding to maximal eigenvalue $\lambda_{I^{f}}^{[IH]}$.
The  $\|p^mI^{f}\|$ operators
and $\pi_m$ moments for $m=0,1,2,3$ are:
\begin{eqnarray}
  \pi_m&=&\Braket{\psi^{[IH]}_{I^{f}}|p^mI^{f}|\psi^{[IH]}_{I^{f}}}
  \label{pimfuture}
\end{eqnarray}
To practically calculate the $\pi_m$ ---
the value of $dI$ is known (\ref{dI})
and last price $P^{last}$ can be used as $P^{fm}$ estimator (\ref{Pfmestlast}).
The result is presented in Fig. \ref{psiIfp3} top.

Then compare the results with the $\Ket{\psi^{[IH]}_{I}}$
choice for $\psi(x)$, not having an impact from the future contribution,
when the moments
\begin{eqnarray}
  \pi_m&=&\Braket{\psi^{[IH]}_{I}|p^mI|\psi^{[IH]}_{I}}
  \label{pimpast}
\end{eqnarray}
are calculated
in the $\Ket{\psi^{[IH]}_{I}}$ state, the (\ref{GEVI}) solution
(without an impact from the future term
$P^f$ estimator is not required).
The result is presented in Fig. \ref{psiIfp3} bottom.
One can see the importance of the impact from the future term,
however in this simplistic form price skewness
has some issues as market directional indicator.

\subsection{\label{PsiFutureIPeq}
  Measure: The Period of Maximal Future $I$ with equilibrium $P^{f}$ estimator}
\begin{figure}
  \includegraphics[width=16cm]{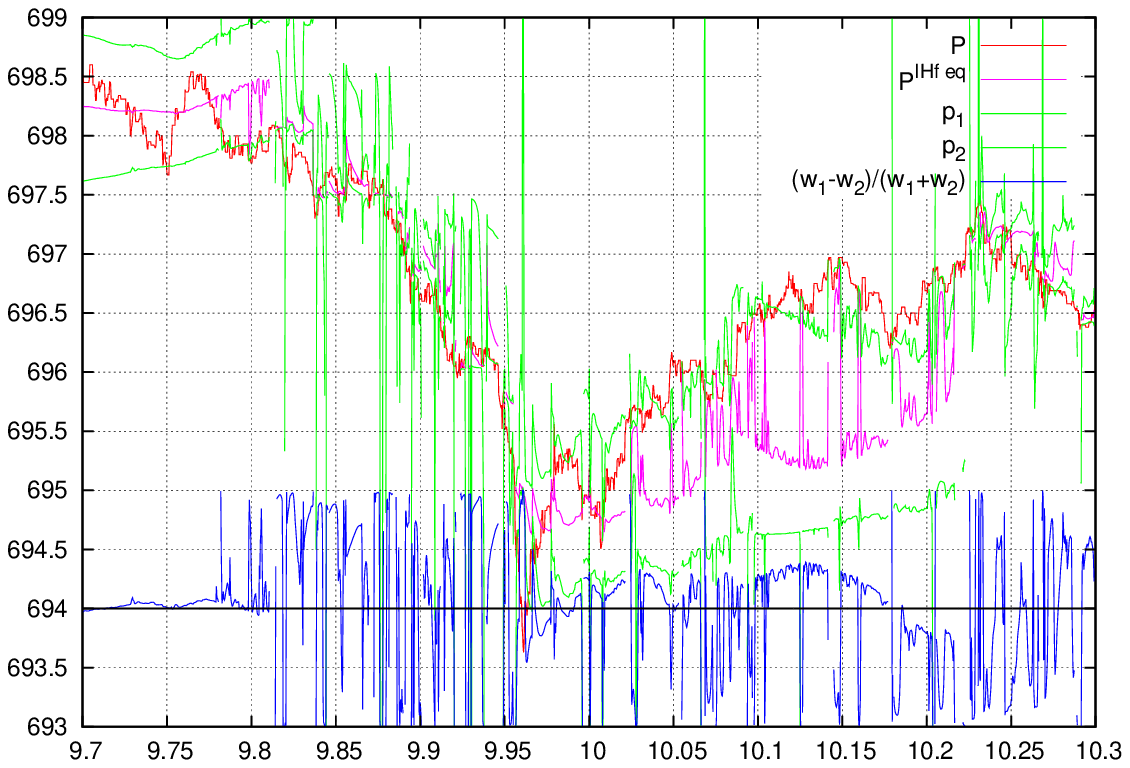}
\caption{\label{psiIfp3Peq}
  The AAPL stock price on September, 20, 2012.
  Demonstration of Gauss quadrature calculation with
  the state (moments $\pi_m$ from (\ref{pimEq})),
  corresponding to maximal $\|I^{f}\|$ 
  The prices and skewness are presented
  as in Fig. \ref{movaverp3} above.
}
\end{figure}

While (\ref{pimfuture}) moments from previous
section are very promising they have one conceptional weakness:
using $P^{last}$ as $P^f$ estimator (\ref{Pfmestlast}).
Consider $\|p^mI^{f}\|$ operator (\ref{pImfutureOp}) with an impact from the future.
The idea is to modify (\ref{Pfmestlast}) estimator
to obtain some ``equilibrium'' value of $P^{fm}$.

As we discussed  in Section \ref{FuturePsiOp}
the $\|I^{f}\|$ and the $\|p^mI^{f}\|$ operators to have the same eigenfunctions,
thus first order variation 
should be equal to zero for arbitrary  $\Ket{\delta \psi}$,
same as for
$\|I^{f}\|$ in (\ref{VarpsiFuture}):
\begin{eqnarray}  
   \Braket{\psi^{[i]}_{I^{f}}|p^mI^{f}|\delta\psi}-
       \Braket{\psi^{[i]}_{I^{f}}|p^mI^{f}|\psi^{[i]}_{I^{f}}}
       \Braket{\psi^{[i]}_{I^{f}}|\delta\psi}
       &=&0
       \label{VarpmIf0}
\end{eqnarray}
The (\ref{VarpsiFuture}) holds for arbitrary $\Ket{\delta \psi}$,
but for variations (\ref{VarpmIf0}) only a single parameter $P^{fm}$ is available,
thus zero--sensitivity condition can be 
satisfied only for a single $\Ket{\delta \psi}$,
besides trivial $\Ket{\delta \psi}=\Ket{\psi^{[i]}_{I^{f}}}$.
There are a number of options for $\Ket{\delta \psi}$ variation to consider:
\begin{itemize}
\item $\Ket{D(\psi^{[i]}_{I^{f}})}:$ Zero price impact (\ref{DPDirectRQ})
  (zero sensitivity to infinitesimal time--shift).
\item $\Ket{\psi_0}:$ Zero sensitivity to
 $\Ket{\psi^{[i]}_{I^{f}}}\to\Ket{\psi_0}$ transition.
\item $\Ket{\psi^{[IH]}_{I}}:$  Zero sensitivity to
$\Ket{\psi^{[i]}_{I^{f}}}\to\Ket{\psi^{[IH]}_{I}}$ transition.
\end{itemize}
among many others.

The $P^{[i]\, fm}dI$ estimation, corresponding to  (\ref{VarpmIf0})  equilibrium
of (\ref{pImfutureOp}) operator
on $\Ket{\psi^{[i]}_{I^{f}}}$ state with $\Ket{\delta\psi}$ variation is:
\begin{eqnarray}
  P^{[i]\, fm}dI&=&
  \frac{\Braket{\psi^{[i]}_{I^{f}}|p^mI|\psi^{[i]}_{I^{f}}}-\frac{\Braket{\psi^{[i]}_{I^{f}}|p^mI|\delta\psi}}{\Braket{\psi^{[i]}_{I^{f}}|\delta\psi}}    
  }{
    \frac{\Braket{\psi^{[i]}_{I^{f}}|\psi_0}\Braket{\psi_0|\delta\psi}}
    {\Braket{\psi^{[i]}_{I^{f}}|\delta\psi}}
    -
    \Braket{\psi^{[i]}_{I^{f}}|\psi_0}^2    
  } \label{pIeq} 
\end{eqnarray}
For the most interesting case $\Ket{\delta \psi}=\Ket{\psi_0}$ obtain:
\begin{eqnarray}
  P^{[i]\, fm}dI&=&
  \frac{\Braket{\psi^{[i]}_{I^{f}}|p^mI|\psi^{[i]}_{I^{f}}}-
    \frac{\Braket{\psi^{[i]}_{I^{f}}|p^mI|\psi_0}}{\Braket{\psi^{[i]}_{I^{f}}|\psi_0}}
  }{
    1-\Braket{\psi^{[i]}_{I^{f}}|\psi_0}^2
  } \label{pmIeq0}
\end{eqnarray}
Then for the state with the maximal $\lambda_{I^{f}}^{[i]}$ ($i=IH$):
\begin{eqnarray}
  \pi_m&=&
  \frac{\Braket{\psi^{[IH]}_{I^{f}}|p^mI|\psi^{[IH]}_{I^{f}}}
    -\Braket{\psi^{[IH]}_{I^{f}}|p^mI|\psi_0}\Braket{\psi^{[IH]}_{I^{f}}|\psi_0}}
  {
    1-\Braket{\psi^{[IH]}_{I^{f}}|\psi_0}^2
  }
  \label{pimEq}  
\end{eqnarray}
Obtained $\pi_m$ have a term $\Braket{\psi^{[IH]}_{I^{f}}|p^mI|\psi_0}$
added to have zero variation (\ref{VarpmIf0}).
In Fig. \ref{psiIfp3Peq} corresponding chart is presented.
First, what is clearly seen is that Gauss
quadrature does not always exist. This is because
(\ref{pmIeq0}) may not always give a positive standard deviation.
However, the formulae for the  first moment
is actually similar to na\"{\i}ve dynamic impact approximation
of Section \ref{naive} and demonstrate an approach
of searching a $\Ket{\delta \psi}$ to variate  (\ref{VarpmIf0}).
Despite all our effort we did not achieve
much success with this search of $\Ket{\delta \psi}$,
and now think that (\ref{VarpmIf0}) variation
can be a good option only for the first moment,
what can give only a equilibrium price (first moment).

\subsection{\label{PsiAfterFutureI}
  Measure: The Period After Maximal Future $I$}
\begin{figure}
  \includegraphics[width=16cm]{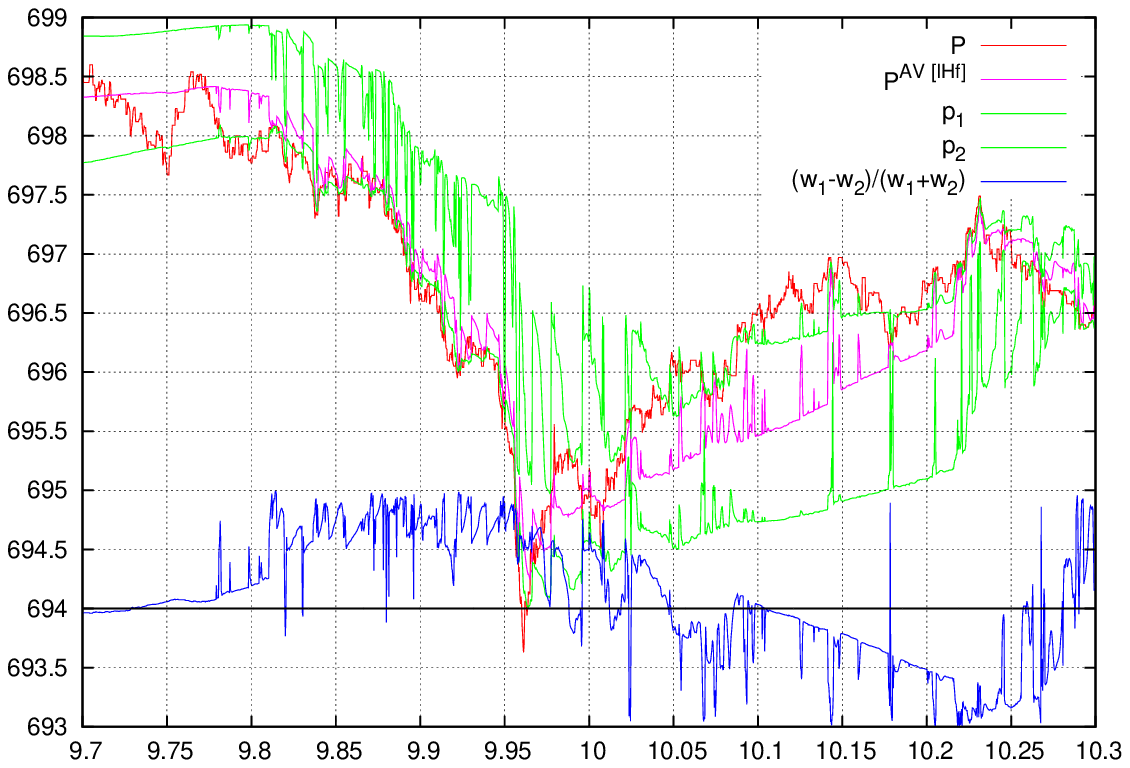}
  \includegraphics[width=16cm]{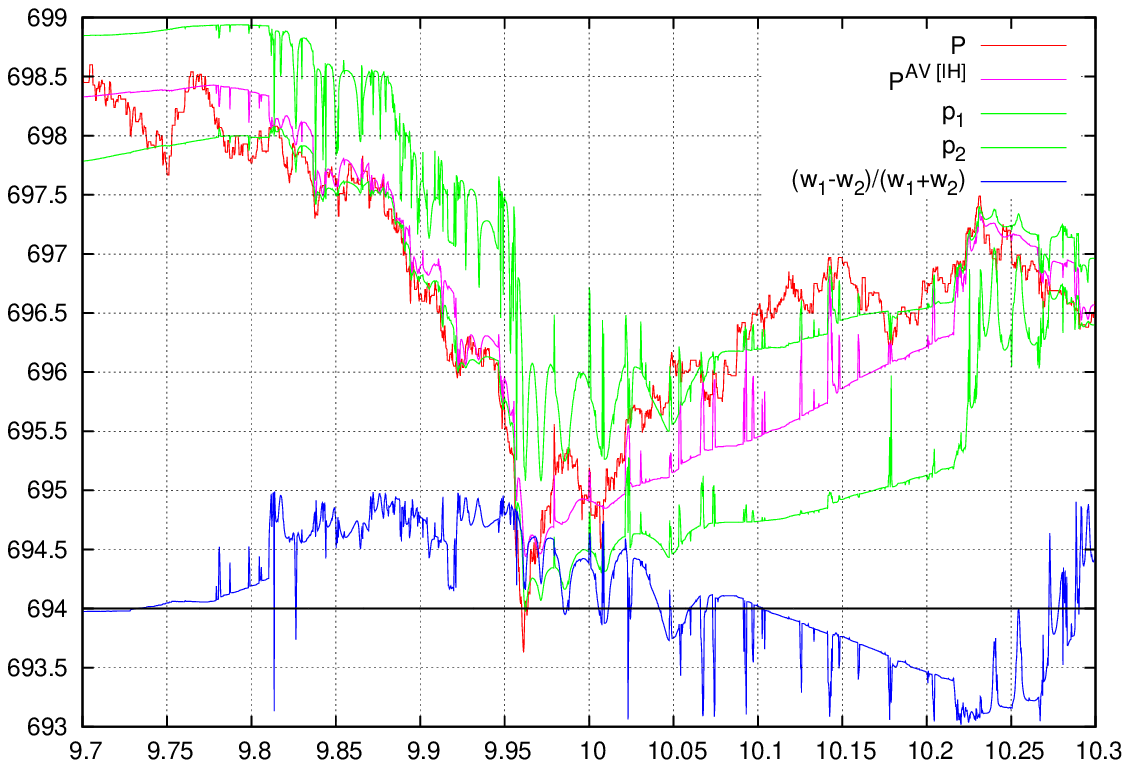}
  \caption{\label{psiIfVp3}
      The AAPL stock price on September, 20, 2012.
  Demonstration of Gauss quadrature calculation with
  the state (moments $\pi_m$ from (\ref{pimCdef})
  with the impact from the future),
  corresponding to the state of the maximal $\|I^{f}\|$ (top)
  and (moments $\pi_m$ from (\ref{pimCdef}), without an impact from the future),
  corresponding to the state of the maximal $\|I\|$ (bottom).
  The prices and skewness are presented
  as in Fig. \ref{movaverp3} above.
  }
\end{figure}
The $\pi_m$ choices (\ref{pimfuture}) and (\ref{pimpast})
are considering price
distribution during the spikes for the future and for the past $I$ respectively.
It is very interesting to consider the time period
after a spike in $I$.
Consider $V_{m}$ and $T_m$:
\begin{subequations}
  \label{VTmdef}
\begin{align}
  V_{m}(t)&=\int\limits_{t}^{t_{now}} p^mI dt^{\prime} =
  \int\limits_{t}^{t_{now}} p^m dV^{\prime}
  \label{cmdef}  \\
   T_{m}(t)&=\int\limits_{t}^{t_{now}} p^m dt^{\prime}
  \label{cmdefT}
\end{align}
\end{subequations}
Here $V_{0}(t)$ is traded volume,
$V_{1}(t)$ is traded capital,
$V_{1}(t)/V_{0}(t)$ is volume--weighted average price,
$T_{1}(t)/T_{0}(t)$ is time--weighted average price.
These values are calculated for the interval
between $t$ and $t_{now}$.
Then for a given $\psi(x)$
\begin{eqnarray}
  \pi_m&=&\Braket{\psi|V_{m}|\psi}
  \label{pimCdef}
\end{eqnarray}
Note, that for the measures allowing an integration by parts
(i.e. the ones with infinitesimal
time--shift operators such as
(\ref{DpsiLag}) or (\ref{DpsiLegendre}))
the (\ref{pimCdef}) can be interpreted 
as a transition from
an averaging with
the  $\psi^2(x)d\mu$ weight to an averaging with the $w_{\psi}(t)dt$ weight:
\begin{eqnarray}
  w_{\psi}(t)&=&\int\limits_{-\infty}^{t}\psi^2(x^{\prime})\frac{d\mu^{\prime}}{dt^{\prime}}dt^{\prime}
  \label{wfdef} \\
  \pi_m&=&\int\limits_{-\infty}^{t_{now}}p^m I w_{\psi}(t^{\prime})dt^{\prime}
  \label{pimw}
\end{eqnarray}
$w_{\psi}(t_{now})=1$ follows from the $\psi(x)$ normalizing.
For (\ref{muflaguerre}) and (\ref{muslegendre})
measures the (\ref{pimCdef})
can be calculated from 
the $\Braket{Q_kp^mI}$ matrix elements using an integration by parts.
For these measures 
Eqs. (\ref{pimCdef}) and (\ref{pimw})
are identical.

Consider a $\psi(x)$,
defining the spikes in $I$, the  $\Ket{\psi^{[IH]}_{I^{f}}}$
or $\Ket{\psi^{[IH]}_{I}}$ from the previous section.
Then (\ref{pimCdef}) moments
give very much a ``moving average with automated time--scale selection''
measure.
These averages are calculated for the period of time:
between the spike in $I$ and $t_{now}$.

The results are presented in Fig. \ref{psiIfVp3}.
They are worse than that of the previous sections, what probably manifest
the importance of the execution flow $I$ dynamics
over the volume $V$ dynamics.
This correspond to our earlier work \cite{2016arXiv160204423G},
where an importance of dynamic impact was emphasized experimentally.
See also the discussion below in Section \ref{Muse},
where the  $V$-- and $I$-- dynamics
are discussed from a different perspective.

\subsection{\label{MeasureSpur}Measure: Density matrix mixed state of pure $\Ket{\psi^{[i]}_{I^{f}}}$
  states.}
\begin{figure}
  \includegraphics[width=16cm]{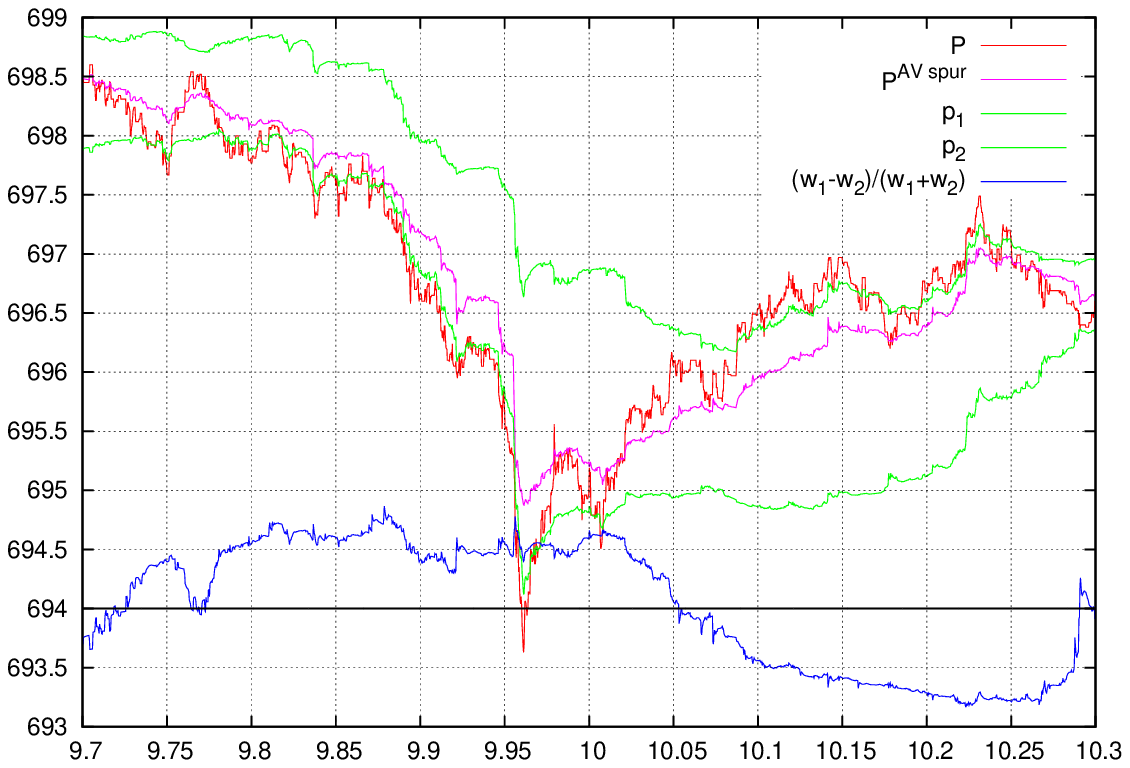}
  \includegraphics[width=16cm]{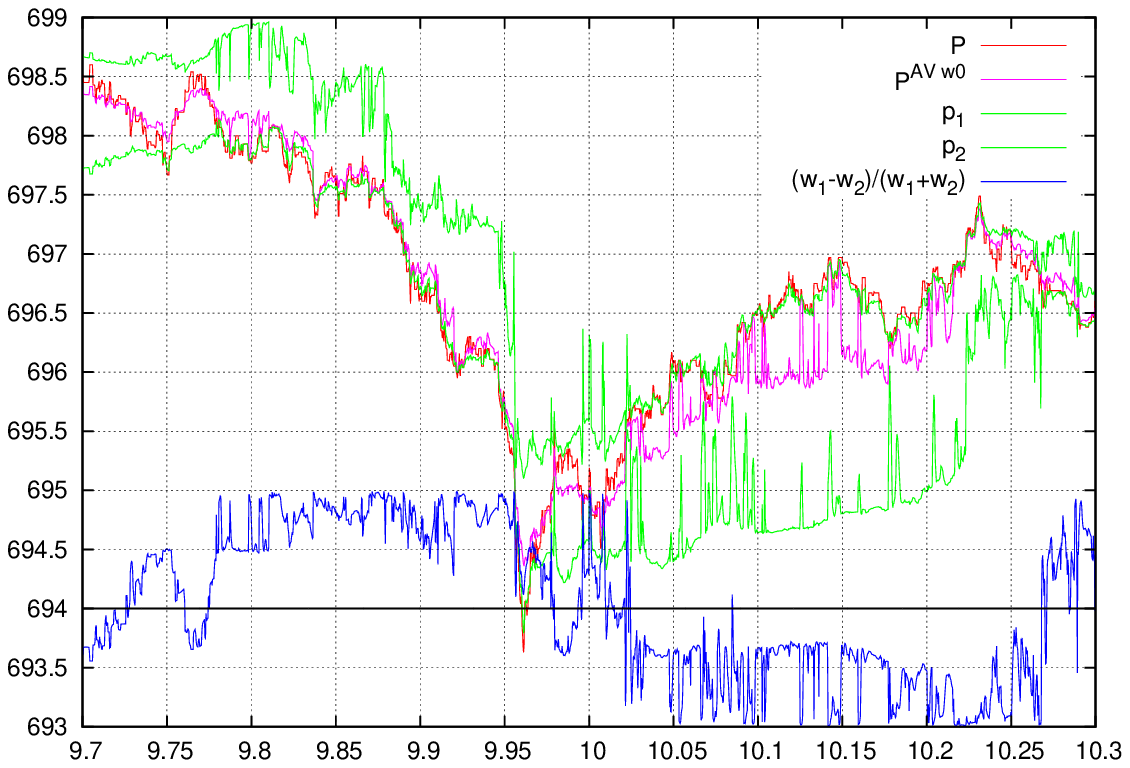}
  \caption{\label{psiIfVp3MS}
      The AAPL stock price on September, 20, 2012.
  Demonstration of Gauss quadrature calculation with
  the mixes states: moments $\pi_m$ from (\ref{pimspur}) (top)
  and  moments $\pi_m$ from (\ref{pimw0}) (bottom). 
  The prices and skewness are presented
  as in Fig. \ref{movaverp3} above.
  }
\end{figure}
As we discussed in Section \ref{FuturePsiOp} above,
in case of the impact from the future presence,
proper eigenstate selection is not a trivial question.
In the Sections   \ref{PsiFutureI} and \ref{PsiFutureIPeq}
the state, corresponding to the maximal $I$,
was considered.
There are several alternatives. Consider matrix--averages
(introduced in the Appendix E of Ref. \cite{2015arXiv151005510G},
see Ref. \cite{malyshkin2015norm} for quantum mechanics density matrix
mixed state relation):
\begin{eqnarray}
  \pi_m&=&\sum\limits_{i=0}^{n-1}
  \Braket{\psi^{[i]}_{I^{f}}|p^mI^{f}|\psi^{[i]}_{I^{f}}}
  \label{pimspur}
\end{eqnarray}
(in this section, when estimating the $\|p^mI^{f}\|$ matrix elements,
we assume (\ref{Pfmestlast})
$P^{fm}$ estimation for simplicity).
The (\ref{pimspur})  answer is very much a moving--average
type of answer (\ref{pimmovaver}),
it is basis--invariant (a unitary transform of
$\Ket{\psi^{[i]}_{I^{f}}}$ basis does not change the result)
and can be considered as a density--matrix mixed state\cite{malyshkin2015norm}
with equal contribution of each pure state.

Alternatively, a density--matrix mixed state with
$\Braket{\psi^{[i]}_{I^{f}}|\psi_0}^2$
contribution of a pure state $\Ket{\psi^{[i]}_{I^{f}}}$
can be considered:
\begin{eqnarray}
  \pi_m&=&\sum\limits_{i=0}^{n-1}
  \Braket{\psi^{[i]}_{I^{f}}|p^mI^{f}|\psi^{[i]}_{I^{f}}}\Braket{\psi^{[i]}_{I^{f}}|\psi_0}^2
  \label{pimw0}
\end{eqnarray}
The (\ref{pimw0}) result is not basis--invariant
and implicitly assume  dynamic impact approximation (\ref{dynimpactdef})
of $\|p\|$ and $\|I^{f}\|$ operators being
simultaneous diagonal in the $\Ket{\psi^{[i]}_{I^{f}}}$ basis.
The (\ref{pimw0}) is similar to (\ref{pimfuture}),
because $\|I^{f}\|$ state with the maximal $\Ket{\psi_0}$
projection is almost always the $\Ket{\psi^{[IH]}_{I^{f}}}$ state.
The results are presented in Fig. \ref{psiIfVp3MS}.
They are not much different from the Sections \ref{movavermoments}
and \ref{PsiFutureI} of above. This section demonstrate that
density matrix approach is  a viable  option
for the market dynamics, but, at this stage of development,
does not give much compared to wavefunction pure states.

\subsection{\label{PsiFutureIIter}
 Measure:  Combine maximal Future $I$ and minimal price volatility}
\begin{figure}
  \includegraphics[width=16cm]{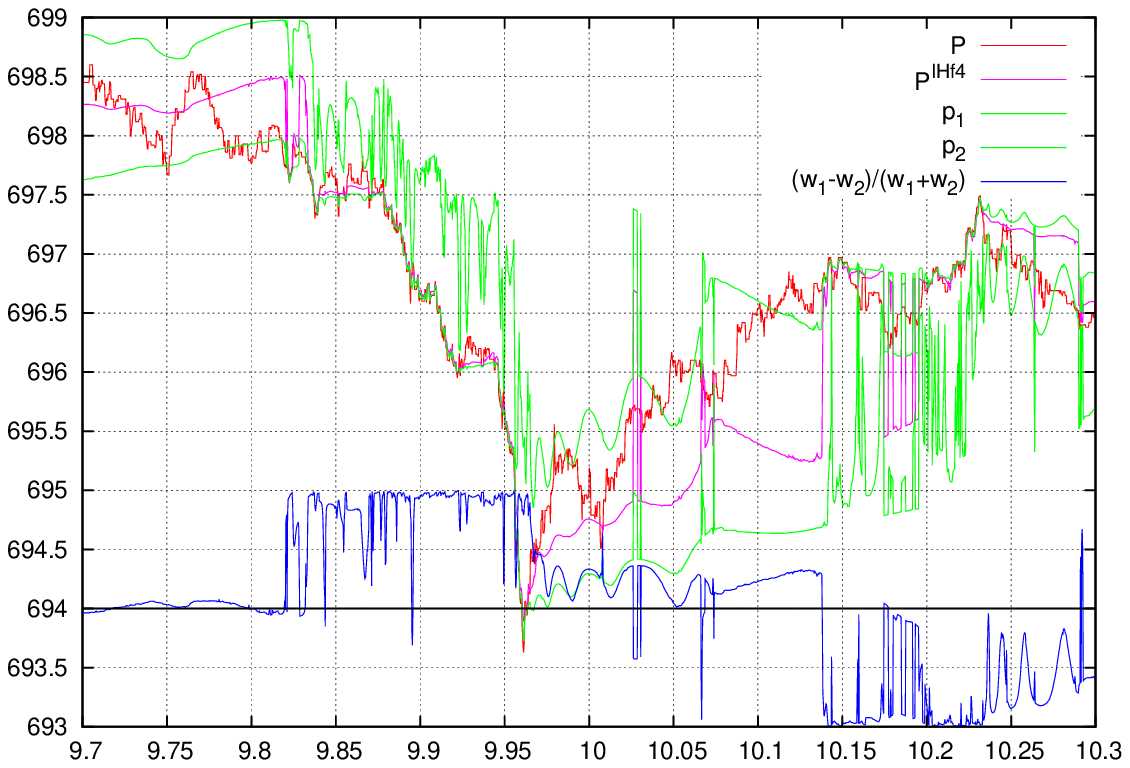}
  \includegraphics[width=16cm]{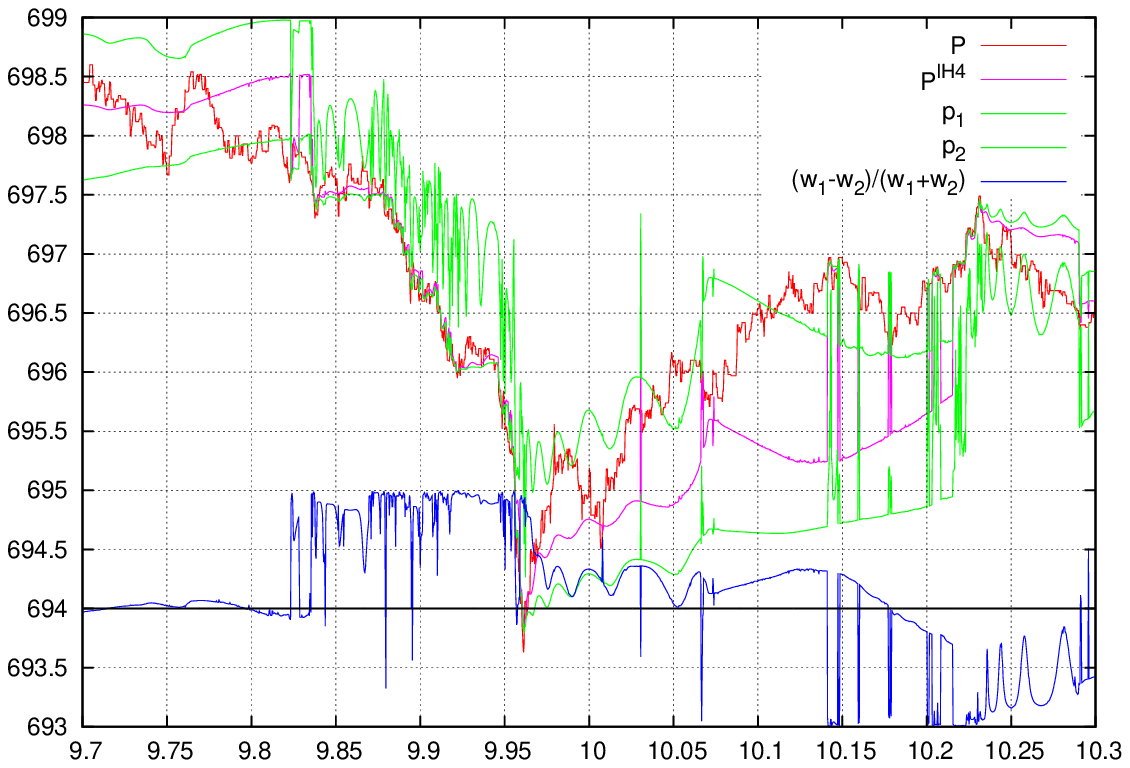}
\caption{\label{psiIfp3opt4}
  The AAPL stock price on September, 20, 2012.
  Demonstration of Gauss quadrature calculation with
  the state corresponding to
  $\max\limits_{\psi}\min\limits_{p_1,p_2}\Braket{\psi|(p-p_1)^2(p-p_2)^2I^{f}|\psi}$ (top)
  and $\max\limits_{\psi}\min\limits_{p_1,p_2}\Braket{\psi|(p-p_1)^2(p-p_2)^2I|\psi}$ (bottom).
  The prices and skewness are presented
  as in Fig. \ref{movaverp3} above.
}
\end{figure}
The approach of section \ref{PsiFutureI}
where $\psi$ corresponding to the maximum of $\Braket{\psi|I^{f}|\psi}/\Braket{\psi|\psi}\to\max$ was found on the first stage, then, for the $\psi$ found
the $p_{\{1,2\}}$ corresponding to the minimum of
$\Braket{\psi|(p-p_1)^2(p-p_2)^2I^{f}|\psi}\to\min$ are obtained (\ref{p4min}).
Consider a ``combined'' problem (despite it contradicts to the
ideology we develop):
\begin{eqnarray}
    &&\max\limits_{\psi}\min\limits_{p_1,p_2}\frac{\Braket{\psi|(p-p_1)^2(p-p_2)^2I|\psi}}
    {\Braket{\psi|\psi}}
     \label{p4opt}
\end{eqnarray}
The idea is to find a saddle point of (\ref{p4opt}),
the solution that has the maximum over $\psi$ and the minimum over $p_{\{1,2\}}$.
The results are presented in Fig. \ref{psiIfp3opt4}
(top: for $\|I^{f}\|$ operator with the (\ref{Pfmestlast}) price estimation,
bottom: for $\|I\|$ operator).
They are not very promising. This was one of
our many tries to built a functional, like an action $\cal S$
in other dynamic theories,
to search for a state of maximum  $I$
and minimum price volatility.
As with the other approaches of this type which we have tried,
this specific one was also not a very successful.
This make us to think that price volatility minimization
approach is probably not a very perspective direction.

\section{\label{Pcorrdirinfo}Market Directional Information and
  $P$ vs. $I$ Probability Correlation}
In Section \ref{quadexample}
we provided a few demonstrations of price skewness estimation technique,
consisting in constructing a measure,
building the $\pi_m=\Braket{p^mI}$ price moments on this measure
(either ``pure state'' (\ref{pimdef})
or ``mixed state'' of Section \ref{MeasureSpur}, depending on the measure used),
then a two--node Gauss quadrature
is built out of them
and price distribution skewness
is estimated as weight asymmetry (\ref{skewness}).
This approach has a built--in asymmetry of $P$ and $I$,
because the $\Braket{I^m}$ moments are difficult to calculate
at best or they are non--exist at worst.
It is very attractive
to introduce some \textsl{basis-invariant}
formulation of skewness concept,
obtain $P$ and $I$ skewness, and then actually try
to trade based on the skewnesses obtained. In the Appendix \ref{twoPvarcorrela}
a concept of probability correlation $\widetilde{\rho}(p,I)$
is introduced, but to trade we only need generalized skewness.
Assume we have an observable $s$,
for $m=0,1,2$ a basis $Q_m(x)$ (a polynomial of $m$--th order),
and inner product $\Braket{Q_j(x)|s|Q_k(x)}$ ($j,k=0,1$)
are defined in a way it can be calculated directly from sample.
Important, that now $x$ and $s$
are not the same variables, in
Section \ref{Thr}
for skewness calculation they were both equal to price.
Average $s$ can be obtained in a regular way:
\begin{eqnarray}
  \overline{s}&=&\frac{\Braket{sQ_0}}{\Braket{Q_0}}
  \label{saver} \\
  \widetilde{\Gamma}&=&\frac{2\overline{s}-s_{\min}-s_{\max}}
            {s_{\min}-s_{\max}}
            \label{skewnesslikeS}
\end{eqnarray}
To build $\widetilde{\Gamma}$, a similar to (\ref{skewness})
skewness--like estimator (like a difference between median and average),
we need $s_{\min}$ and $s_{\max}$
estimators of $s$.
These can be obtained solving optimization
problem:
\begin{eqnarray}
  \frac{
    \Braket{\Big[\alpha_0Q_0(x)+\alpha_1Q_1(x)\Big]^2s}
  }{
    \Braket{\Big[\alpha_0Q_0(x)+\alpha_1Q_1(x)\Big]^2}
    }&\to& \{\min ;  \max\}
\end{eqnarray}
After parenthesis expansion the problem is reduced
to $n=2$ generalized eigenvalue problem (\ref{d2EVproblem}),
the eigenvalues of which are quadratic equation roots.
The $\min$/$\max$ estimators of $s$
are equal to minimal/maximal eigenvalues $\lambda^{[0]}_s$
and $\lambda^{[1]}_s$ respectively, what allows us to obtain (\ref{skewnessLike})
skewness--like\footnote{
  The (\ref{skewnesslikeS}) is $\psi(x)=const$ state $\Ket{\psi_C}$
  weight asymmetry
  expansion over the states corresponding to min/max $s$:
  $\widetilde{\Gamma}=\Braket{\psi_C|\psi^{[0]}_s}^2-\Braket{\psi_C|\psi^{[1]}_s}^2$.
  Instead of $\overline{s}=\Braket{\psi_C|s|\psi_C}={\Braket{sQ_0}}\big/{\Braket{Q_0}}$ a different $s$ values can be used, e.g. $s_0=\Braket{\psi_0|s|\psi_0}$,
  corresponding to the state ``time is now'' $\Ket{\psi_0}$ from (\ref{psix0}):
  $\widetilde{\Gamma^{0}}=\Braket{\psi_0|\psi^{[0]}_s}^2-\Braket{\psi_0|\psi^{[1]}_s}^2=
  {\left(2s_0-s_{\min}-s_{\max}\right)}\big/{\left(s_{\min}-s_{\max}\right)}$,
  this ``skewness'', (\ref{skewnesslikeS0}) for $n=2$,
  describe $s_0$ asymmetry (compare it with $\widetilde{\Gamma}$, that describe
  $\overline{s}$ asymmetry).
} estimator $\widetilde{\Gamma}$
in (\ref{skewnesslikeS}).
If $s=x=p$, then we receive exactly the $\Gamma$ from (\ref{skewness}),
which requires total 4 moments:
$\Braket{1},\Braket{p},\Braket{p^2},\Braket{p^3}$ to calculate.
To calculate $\widetilde{\Gamma}$ it requires total 6 moments:
$\Braket{1},\Braket{x},\Braket{x^2},\Braket{s},\Braket{sx},\Braket{sx^2}$;
(for  $s=x=p$, there are only 4 independent among them).
See the file \texttt{\seqsplit{com/polytechnik/utils/Skewness.java:getGSkewness}} for
implementation example of numerical calculation of generalized skewness $\widetilde{\Gamma}$.
The most important property of $\widetilde{\Gamma}$
is that it can be readily applied to non--Gaussian variables, e.g. $I$.
In our previous study\cite{2016arXiv160305313G}
we emphasized the inapplicability of a
regular statistical characteristics (e.g. standard deviation) to market dynamics,
and, instead, spectral operators should be applied 
to sampled non--Gaussian data\cite{2016arXiv161107386V,liionizversiyaran}.
The (\ref{d2EVproblem}) generalized eigenvalue problem,
finding min/max $s$ estimates $\lambda^{[0]}_s$ and $\lambda^{[1]}_s$ 
from operator spectrum is the simplest application.

\subsection{\label{skewnessDemonstration} $I$ Skewness.
  A demonstration of skewness estimation for non--Gaussian distribution.}
\begin{figure}
\includegraphics[width=16cm]{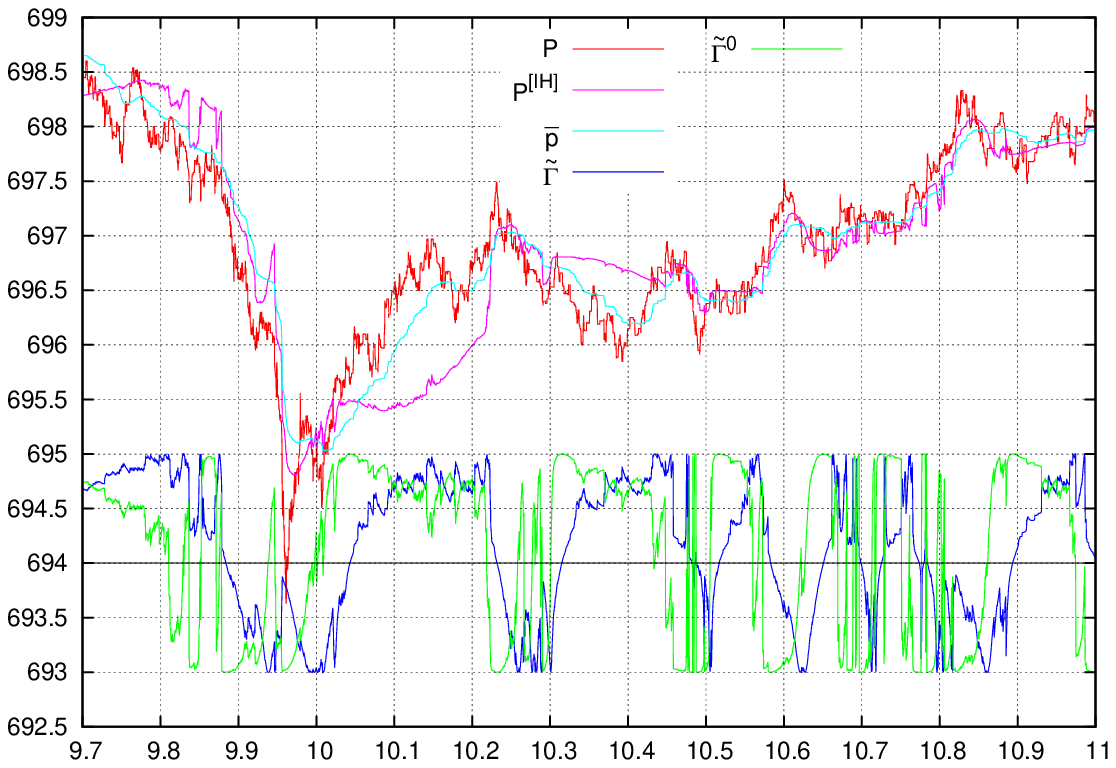}
\includegraphics[width=16cm]{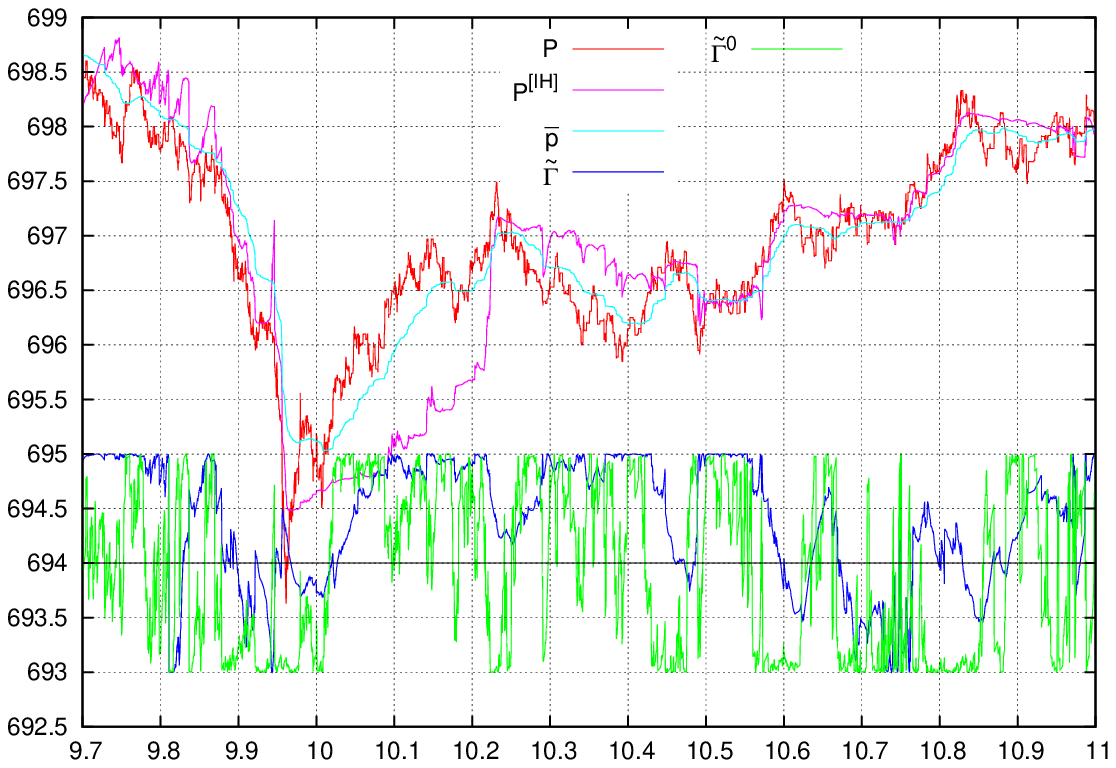}
\caption{\label{FigSimpleSkewness}
  Generalized skewness of $I$ calculated with $\widetilde{\Gamma}$
  (\ref{skewnesslikeS}) for $\tau=128$sec and $n=2$ (blue);
  same but with $\widetilde{\Gamma^{0}}$(green), (in (\ref{skewnesslikeS})
  the $\overline{I}$ is replaced with $\Braket{\psi_0|I|\psi_0}$ ).
  Price $P$, average price $\overline{p}$ and $P^{[IH]}$ for $n=2$
  are also presented.
  Top: for $t^k$ basis and (\ref{muflaguerre}) measure. Bottom: for $p^k$
  basis and (\ref{mupspace}) measure.
}
\end{figure}
Let us give a simple example of (\ref{skewnesslikeS})
skewness estimation application. Consider $s=I=dV/dt$ execution flow,
polynomial basis $Q_k(x)$, and
a measure (such as (\ref{muflaguerre}),  (\ref{muslegendre}),
or (\ref{mupspace})),
that can be calculated directly from sample:
(\ref{sampleLag}), (\ref{sampleLeg}) or (\ref{samplePrice}).
The problem: \textsl{to estimate $I$ skewness}.
``Classical'' approach, that requires $\Braket{1}$, $\Braket{I}$,
$\Braket{I^2}$, and $\Braket{I^3}$ moments
to calculate either traditional
$\Braket{\left(I-\overline{I}\right)^3}$ estimator,
or $\Gamma$ from (\ref{skewness})
is not applicable, because second $\Braket{I^2}$ and third
$\Braket{I^3}$ moments
are infinite (note that first moment $\Braket{I}$
has a meaning of the traded volume
and zeroth moment $\Braket{1}$ is a constant).

However the $\widetilde{\Gamma}$ skewness from (\ref{skewnesslikeS})
can be calculated directly. All six moments:
$\Braket{Q_0}$, $\Braket{Q_1}$, $\Braket{Q_2}$,
$\Braket{IQ_0}$, $\Braket{IQ_1}$, $\Braket{IQ_2}$
are finite, $2\times 2$ matrices
$\Braket{Q_j|I|Q_k}$ and $\Braket{Q_j|Q_k}$
obtained from these moments, eigenvalues problem (\ref{d2EVproblem})
solved by solving the quadratic equation $0=\det\| \Braket{Q_j|I|Q_k}-\lambda_{I}\Braket{Q_j|Q_k}\|$;
$\min I=\lambda_{I}^{[0]}$, $\max I=\lambda_{I}^{[1]}$
obtained, and $\widetilde{\Gamma}$ from (\ref{skewnesslikeS}) calculated.

In Fig. \ref{FigSimpleSkewness} we present the calculation of $I$ skewness
for two measures: (\ref{muflaguerre}) and (\ref{mupspace}).
Blue line:  $\widetilde{\Gamma}$ from (\ref{skewnesslikeS}),
the asymmetry of $\overline{I}$; green line:
the asymmetry of $I_0$, $\widetilde{\Gamma^{0}}=w^{[IL]}_I-w^{[IH]}_I=
{\left(2I_0-I_{\min}-I_{\max}\right)}\big/{\left(I_{\min}-I_{\max}\right)}$,
calculated using (\ref{wL}) and (\ref{wH}) with $n=2$.
Positive $I$ skewness correspond to liquidity deficit event (low $I$, slow market),
a signal to open a position (but to determine the
sing (long/short) of a position to open
is a much more problematic task).
Negative $I$ skewness
corresponds to the liquidity excess event (high $I$, fast market),
a signal to close already opened position. 
From these charts one can clearly see that both $\widetilde{\Gamma}$
and $\widetilde{\Gamma^{0}}$ can be a good indicator of
slow/fast markets, but the $\widetilde{\Gamma^{0}}$ skewness
is a  better indicator as it shows how the $I_0$ ($I$ now)
is related to past min/max $I$.
Note, that
calculated skewness of $I$ does not carry market directional price
information. Instead, $I$--skewness tells us about
when (at negative skewness of $I$) the position have to be closed to avoid
unexpected market move against position held,
otherwise just a single such a move can easily kill all the P\&L collected.
Directional information (whether to open long or short
position at positive $I$--skewness),
cannot be decided from $I$--skewness,
it to be decided from price or P\&L dynamics.

\subsection{\label{movavermomentsSK}Price Skewness.}
\begin{figure}
\includegraphics[width=16cm]{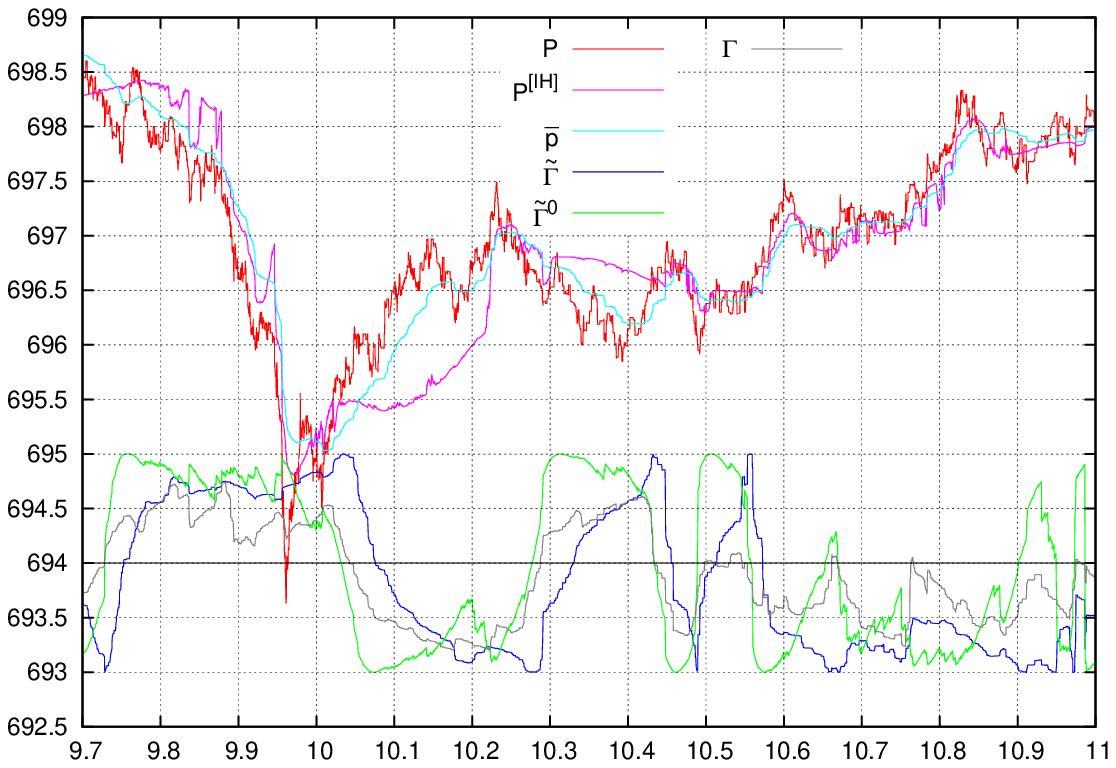}
\includegraphics[width=16cm]{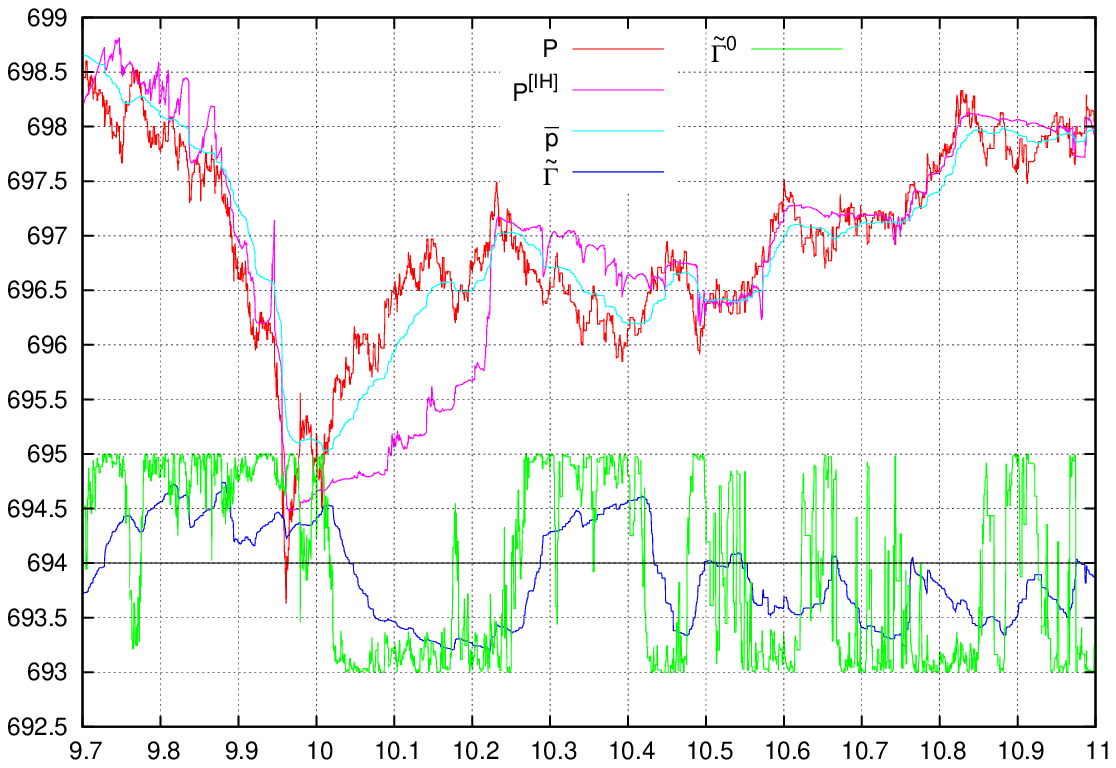}
\caption{\label{FigSimplePSkewness}
  Generalized skewness of price calculated with $\widetilde{\Gamma}$
  (\ref{skewnesslikeS}) for $\tau=128$sec and $n=2$ (blue);
  same but with $\widetilde{\Gamma^{0}}$(green), (in (\ref{skewnesslikeS})
  the $\overline{I}$ is replaced with $\Braket{\psi_0|I|\psi_0}$ ),
  and regular skewness $\Gamma$ from (\ref{skewness}) (gray).
  Price $P$, average price $\overline{p}$ and $P^{[IH]}$ for $n=2$
  are also presented.
  Top: for $t^k$ basis and (\ref{muflaguerre}) measure. Bottom: for $p^k$
  basis and (\ref{mupspace}) measure (in this basis $\Gamma=\widetilde{\Gamma}$,
  so regular price skewness (gray line) is not presented.
}
\end{figure}
In the previous section we have considered $I$
skewness, than generate ``position open/position close''
signals. However the direction (open long or open short)
cannot be determined from that.
Directional information to be determined from P\&L dynamics.
Consider the simplest case.

According to the arguments presented in Ref. \cite{2015arXiv151005510G}
price or price changes cannot be used for directional predictions,
and P\&L dynamics should be considered instead\cite{2016arXiv160305313G}.
P\&L dynamics 
 includes not only price dynamics, but also trader actions.
In Ref. \cite{2015arXiv151005510G}
(Section ``P\&L operator and trading strategy'')
we used probability states trying to analyze  P\&L dynamics,
but here let us start with a very simple problem:

Assume exchange trading take place,
and some speculator
knows the future for specific time interval (investment horizon)
from Oracle Precognition.
What trading strategy to be implemented
to maximize trading P\&L and minimize introduced impact to the markets?
The answer is trivial: for the investment horizon
calculate price median,
then
trade at exactly the same time moments when ``natural trading'' to occur
buying an asset when the price is below the median and
selling it when the price is above the median,
this is equivalent to frontrun the buyers at price below median
and to frontrun the sellers at price above median.
Why median price as a threshold? Only when price threshold
is equal  to the median, total position held
at the end of investment horizon will be zero.
If one use average price as a threshold
then, depending on distribution skewness,
speculator ends up with long or short position accumulated
(to maximize the P\&L speculator have to trade all the time)
at the end of investment horizon (what means taking market risk
because the future is assumed not to be
known outside of investment horizon).
In the simplest case price skewness,
that is proportional
to the difference between  median price (estimated
as midpoint $\frac{1}{2}\left[\lambda^{[0]}_P+\lambda^{[1]}_P\right]$)
and average price $\overline{p}$
can serve as directional price indicator.
Consider a simple demonstration:
\begin{itemize}
\item Select a measure to define inner product $\Braket{\cdot}$,
  that can be calculated directly from sample.
  \item
    Calculate price skewness $\widetilde{\Gamma}_{P}$ out of moments: $\Braket{IQ_0},\Braket{IQ_1},\Braket{IQ_2},\Braket{pIQ_0},\Braket{pIQ_1},\Braket{pIQ_2}$.
\end{itemize}
In Fig. \ref{FigSimplePSkewness} we present skewness calculation
in two bases: $t^k$ (\ref{sampleLag}) and $p^k$ (\ref{samplePrice})
(top and bottom respectively).
For $n=2$, we have $\Gamma$ (gray line), $\widetilde{\Gamma}$ (blue line),
and $\widetilde{\Gamma^0}$ (green line) calculated.
For $p^k$ basis $\widetilde{\Gamma}=\Gamma$
(and also equal to $\Gamma$ in Fig. \ref{movaverp3} top),
so gray line is not presented in this case.
The $\widetilde{\Gamma}$ define how close
average $p$ is to min/max estimated as $\lambda^{[0]}_P$, and $\lambda^{[1]}_P$
respectively. The $\widetilde{\Gamma^0}$ do the same for $p$
in $\Ket{\psi_0}$ state.
It is of interest to look in Fig. \ref{FigSimplePSkewness} top,
where one can see the difference between $\widetilde{\Gamma}$ and $\Gamma$ (gray and blue lines),
that sometimes occur near price tipping points.

\subsection{\label{FutureISK}Skewness of future $I$.}
\begin{figure}
  \includegraphics[width=16cm]{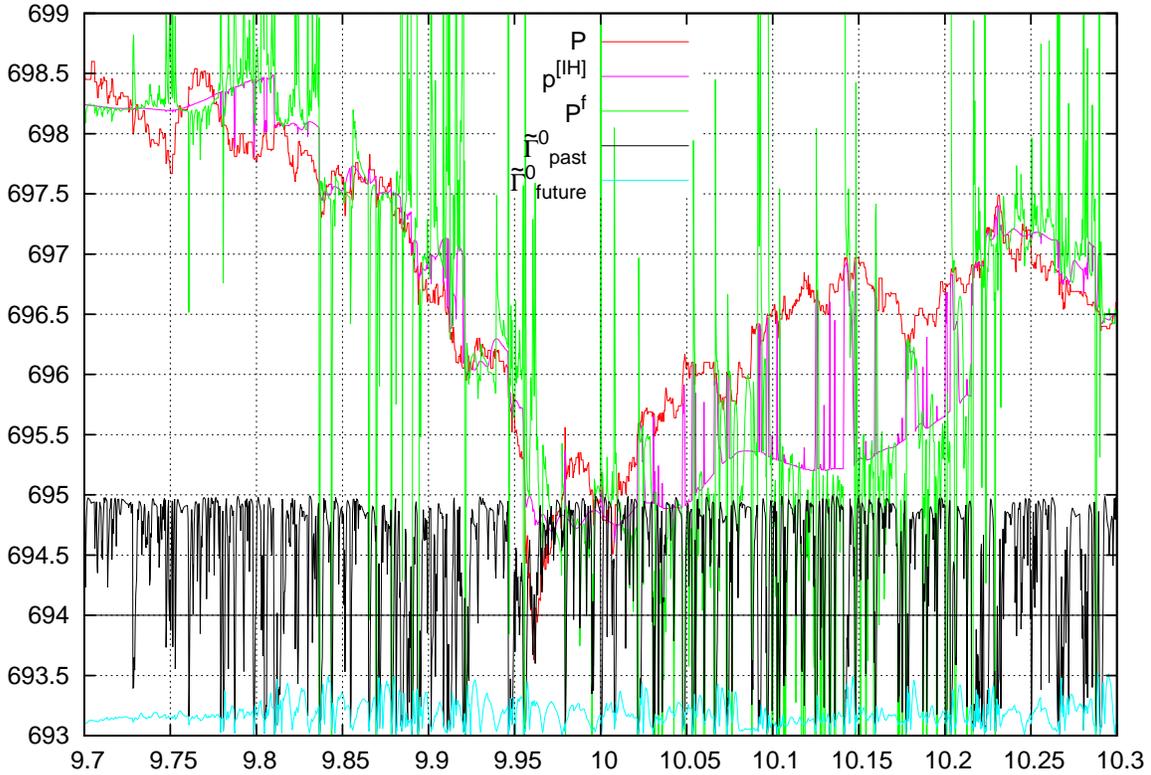}
\caption{\label{pfSkewnessFig}
  The AAPL stock price on September, 20, 2012.
  $P^{[IH]}$ (\ref{PIH}) (pink), $P^{f}$ (\ref{PfSkewF}) (green),
  skewness $\widetilde{\Gamma^0}_{\mathrm{past}}$ (black, for $\|I\|$)
  and skewness $\widetilde{\Gamma^0}_{\mathrm{future}}$ (blue, for $\|I^{f}\|$) are calculated according
  to (\ref{skewnesslikeS0}); 
  (data shifted to 694 level to fit the chart).
 Calculated in Shifted Legendre basis with $n=7$ and $\tau$=128sec.
}
\end{figure}
In Section \ref{Pcorrdirinfo} $\widetilde{\Gamma}$ concept (\ref{skewnesslikeS})
  was introduced and, for $n=2$, it can be rigorously defined
  (along with probability correlation concept)
in Appendix \ref{twoPvarcorrela}.
However a modified concept is convenient in applications.
Introduce $\widetilde{\Gamma^0}$ (the $s$ can be either price $p$ or execution flow $I=dV/dt$) :
\begin{eqnarray}
  s_0&=&\Braket{\psi_0|s|\psi_0}
  \label{s0aver} \\
  \widetilde{\Gamma^0}&=&\frac{2s_0-s_{\min}-s_{\max}}
            {s_{\min}-s_{\max}}
            \label{skewnesslikeS0}
\end{eqnarray}
$\widetilde{\Gamma^0}$
measure how $s_0$ ($s$ ``now'')
compares with $s_{\min}$ and $s_{\max}$ (min/max
eigenvalues of $\Ket{s|\psi}=\lambda\Ket{\psi}$ problem),
calculated on past observations. For $n=2$ we have $\widetilde{\Gamma^0}=\Braket{\psi_0|\psi^{[\min s]}}^2-\Braket{\psi_0|\psi^{[\max s]}}^2$,
(as we already mentioned this, regarding (\ref{wL}), (\ref{wH}) projections difference),
but for $n>2$ this is not the case. For $n>2$ the $\widetilde{\Gamma^0}$
is plain indicator of how $s_0$ fares with $s_{\min}$ and $s_{\max}$.
The (\ref{skewnesslikeS0}) answers
the major questions of our dynamic theory:
``whether the $I_0$ we currently observe is low or high''.
The $\widetilde{\Gamma^0}$ is bounded to $[-1\dots 1]$
interval. $\widetilde{\Gamma^0}$ value close to $1$ means we have liquidity
deficit event ($I_0$ is low), $\widetilde{\Gamma^0}$ value close to $-1$
means we have liquidity
excess event ($I_0$ is high).
Note, that $I$ is a non--Gaussian variable with
infinite second moment $\Braket{I^2}$,
so no approach utilizing a standard deviation of $I$ can be applied.

Because we do know future $\|I^{f}\|$ operator (\ref{Ifuture}),
the $\widetilde{\Gamma^0}^f$ can be calculated for it.
Now consider $\|pI^{f}\|$ operator (\ref{pImfutureOp}) with unknown $P^f$,
and assume it has \textsl{the same} skewness on the states of $\|I^{f}\|$ operator, then:
\begin{eqnarray}
  \widetilde{\Gamma^0}^f &=&
  \frac{2\Braket{\psi_0|pI^f|\psi_0}
    -\Braket{\psi_{I^{f}}^{[IL]}|pI^f|\psi_{I^{f}}^{[IL]}}
    -\Braket{\psi_{I^{f}}^{[IH]}|pI^f|\psi_{I^{f}}^{[IH]}}
  }{
    \Braket{\psi_{I^{f}}^{[IL]}|pI^f|\psi_{I^{f}}^{[IL]}}
    -\Braket{\psi_{I^{f}}^{[IH]}|pI^f|\psi_{I^{f}}^{[IH]}}
  }
  \\
  \Upsilon&=&2-\Braket{\psi_{I^{f}}^{[IL]}|\psi_0}^2-\Braket{\psi_{I^{f}}^{[IH]}|\psi_0}^2-\widetilde{\Gamma^0}^f\left[
        \Braket{\psi_{I^{f}}^{[IL]}|\psi_0}^2-\Braket{\psi_{I^{f}}^{[IH]}|\psi_0}^2
        \right] 
      \label{gegFSkewness} \\
      P^{f}dI \Upsilon &=&\widetilde{\Gamma^0}^f\left[
          \Braket{\psi_{I^{f}}^{[IL]}|pI|\psi_{I^{f}}^{[IL]}}
    -\Braket{\psi_{I^{f}}^{[IH]}|pI|\psi_{I^{f}}^{[IH]}}
    \right] \nonumber \\
        &-&\left[2\Braket{\psi_0|pI|\psi_0}
    -\Braket{\psi_{I^{f}}^{[IL]}|pI|\psi_{I^{f}}^{[IL]}}
    -\Braket{\psi_{I^{f}}^{[IH]}|pI|\psi_{I^{f}}^{[IH]}}\right]
      \label{PfSkewF}
\end{eqnarray}
The (\ref{PfSkewF}) is  $P^{f}$ that, for $\|pI^{f}\|$ operator (\ref{pImfutureOp}),
give the same skewness as the one for  $\|I^{f}\|$.
This answer is similar to na\"{\i}ve dynamic impact approximation
of Section \ref{naive} (compare (\ref{gegFSkewness}) with (\ref{degeneray}),
and (\ref{PfSkewF}) with (\ref{PfNative})).
The results are presented in Fig. \ref{pfSkewnessFig}.
As for na\"{\i}ve dynamic impact approximation,
the $P^{f}$ from (\ref{PfSkewF}) behave similar to $P^{[IH]}$ from (\ref{PIH}),
and have numerical instability for low $\Upsilon$. Future skewness $\widetilde{\Gamma^0}_{\mathrm{future}}$
(for $\|I^{f}\|$)
 is negative (the impact from the future
$dI$ (\ref{dI}) make it such). Past skewness $\widetilde{\Gamma^0}_{\mathrm{past}}$
(for $\|I\|$) is positive during liquidity deficit
and negative during liquidity excess. Trader should open
a position during positive $\widetilde{\Gamma^0}_{\mathrm{past}}$
and close it during negative $\widetilde{\Gamma^0}_{\mathrm{past}}$,
this is the only way to avoid catastrophic P\&L hit from an unexpected market move.

\section{\label{Muse}On A Muse of Cash Flow And Liquidity Deficit Existence }

We finally reached the point to decide  what
information can be obtained from historical (time, execution price, shares traded)
market observations
deploying introduced in\cite{2015arXiv151005510G} the dynamic equation:
``Future price tends to the value that maximizes the number of shares traded per unit time''.
While volatility trading is much easier
to implement algorithmically\cite{2015arXiv151005510G}, it is much more difficult
to implement practically, on exchange, because it requires
building some synthetic assets (such as Straddle \cite{wiki:Straddle})
using options (or other derivatives).
Compared to regular HFT equity trading accounts,
HFT derivative trading accounts are much more costly
and derivative markets often have insufficient available liquidity for
a practical trading strategy implementation.
In addition to that trading strategies including derivatives
are way more difficult to  backtest for the reasons of data availability and
insufficient liquidity.
In this section we are going to discuss
whether a much more ambitions goal, to obtain directional price information
(not only volatility!),
can be practically achieved with the dynamic equation.
Our study show, that there are two
pieces of information, required to obtain directional information:

 First. Price directional information of the past.
  A trivial information of this type
  is ``last price minus moving average'' currently is in common use.
  We obtained few more sources of this information,
  having the benefit of automatic time--scale selection.
  These are: $P^{[IH]}$ (price corresponding to max $I$ on past sample (\ref{PIH})),
  skewness of price on  max $I$ state of $\|p^mI^{f}\|$ operator
  with an impact from the future (Section \ref{PsiFutureI}),
  the skewness of price (or P\&L) of Section \ref{Pcorrdirinfo},
  and few other.

Second. Execution flow ($I=dV/dt$) directional information.
  Since Adam Smith\cite{wiki:adamsmith} and Karl Marx the volume of the
  trade is considered to be the key element of goods/money exchange process
  between buyers and sellers.
  The concept of \textsl{Velocity of money}\cite{wiki:velocityofmoney}, velocity of circulation,
  ($I=dV/dt$ is the velocity of shares, $pI$ is the velocity of money)
  while being widely recognized as an important macroeconomic concept,
  is not in use among both academics and
  exchange trading practitioners (at best they  use the volume, 
  assuming the consumption of shares is limited by the number of shares bought:
  ``\textsl{The tailor does not attempt to make his own shoes, but buys them of the
 shoemaker, page 350}''\cite{smith1827inquiry}).
Modern exchange trading currently exists of market participants,
that are simultaneously  buyers and sellers (modern ``shoemaker''
not only sells the shoes he made,
but also \textsl{buys} shoes to sell them later),
and, because of leveraged trading,  weakly sensitive
to the volume $V$ (regular impact\cite{wiki:marketimpact}) of the position.
As we have shown experimentally,
they are much more sensitive to the rate of trading $I=dV/dt$ (dynamic impact\cite{2016arXiv160204423G}).
The situation of market separation of $V$-- and $I$-- trading
can be currently observed in Electricity Market\cite{wiki:electricitymarket}
that is separated on \textsl{Energy} and \textsl{Power} markets
on legislative level. Our exchange experiments show that modern exchange trading
is actually a \textsl{Power}--like market. The reason why the
velocity of money was not actively used for exchange trading
is, from our opinion, the absences of mathematical technique
to estimate $I$ (execution flows are non--Gaussian).
Because Radon--Nikodym derivatives can be effectively applied
to non--Gaussian processes it is the proper tool for velocity of money
analysis.
Two indicators of $I$ are used in this paper. These are the projections (\ref{wL}) and (\ref{wH}) difference
that show whether current $I$ is ``low'' or ``high'', and
the skewness of $I$, the $\widetilde{\Gamma_I}$ from (\ref{skewnesslikeS})
(or more useful in practice $\widetilde{\Gamma^0}$ from (\ref{skewnesslikeS0})).
The skewness of $I$
can be estimated only from Radon--Nikodym approach, because
regular skewness estimators are not applicable
for the reason of infinite $\Braket{I^2}$ and $\Braket{I^3}$.

Practical trading to be this:
Determine price direction (e.g. from $P^{[IH]}$ (\ref{PIH})),
or the skewness of $P$, Section \ref{movavermomentsSK},  with some measure).
Then calculate $I$--skewness $\widetilde{\Gamma^0}$.
Open a position (according to price direction found)
when $\widetilde{\Gamma^0}$ is close to $1$,
close  already opened position (but do not take opposite position!)
when $\widetilde{\Gamma^0}$ is close to $-1$
to avoid catastrophic P\&L drain in case of
unexpected market move against position held.
Such a strategy do provide provide a P\&L,
and, important, is resilient to unexpected market hits.
In the next paper I will try to present
a demonstration of this strategy computer implementated.
Do not expect a big miracle, (even  a  ``small miracle'' of paper trading P\&L),
but avoiding big P\&L hits can also be considered as a miracle of some kind.

\begin{acknowledgments}
   Vladislav Malyshkin would like to thank Alexei Chekhlov at \href{http://systematicalpha.com/}{Systematic Alpha}
 for fruitful discussions on the link between liquidity deficit and execution flows,
 and Misha Boroditsky at \href{http://www.cantor.com/}{Cantor Fitzgerald}
 for his comments on trading systems' impact on financial markets.
\end{acknowledgments}

\appendix
\section{\label{tdist}Time--Distance Between $\psi$ States}
For two $\psi$ states from (\ref{psiIsol}),
already separated in $I$--space by the value of eigenvalue $\lambda_I$,
the separation in time space is often required.
For this a ``time--distance function'',
$d_{jk}$ between
the $\psi^{[j]}_I(x)$ and $\psi^{[k]}_I(x)$ states from (\ref{psiIsol})
is required.
The $d_{jk}$ is an antisymmetric matrix,
showing which state
$\psi^{[j]}_I(x)$ or $\psi^{[k]}_I(x)$ is later (in time) and which one is earlier.
\begin{eqnarray}
  d_{jk}&=&-d_{kj}
  \label{dantisymm}
\end{eqnarray}
There are several $d_{jk}$ choices, that can be applied to the task.
All of them can be obtained
from  two--point propagator--like expressions
with some antisymmetric $DI(x,y)$
\begin{eqnarray}
  DI(x,y)&=& -DI(y,x) \\
  d_{jk}&=&\int\int  DI(x,y) \left(\psi^{[j]}(x)\right)^2 
  \left(\psi^{[k]}(y)\right)^2
  d\mu(x)  d\mu(y)
  \label{propsign}
\end{eqnarray}
These are the most common  $DI(x,y)$ choices:
\begin{itemize}
\item
  Probability difference  between
  ``$j$ coming after $k$'' and ``$j$ coming before $k$'' events.
  Can be obtained from (\ref{propsign})
  with $DI(x,y)=\mathrm{sign}(x-y)$.
  It can be calculated analytically for the measures
  (\ref{muflaguerre}) and (\ref{muslegendre}). See java classes {\{KkQVMLegendreShifted, KkQVMLaguerre, KkQVMMonomials\}.\{\_getK2,\_getEDPsi\}} from Appendix \ref{appendix1} for
   implementation of probability difference function and infinitesimal time shift  operator.

\item Total volume traded
  \begin{eqnarray}
    V^{[j]}&=&\frac{\Braket{\psi^{[j]}_I|V|\psi^{[j]}_I}}{\Braket{\psi^{[j]}_I|\psi^{[j]}_I}} 
    \label{Vj}\\
    d_{jk}&=&V^{[j]}-V^{[k]}
    \label{dvol}
  \end{eqnarray}
  Corresponds to (\ref{propsign}) with $DI(x,y)=V(x)-V(y)$.
  The state with a greater volume can be considered
  as coming after the state with lower volume.

\item  Difference in projection to $\psi_0(x)$ from  (\ref{psix0}):
  \begin{eqnarray}
    d_{jk}&=&\left(\psi^{[j]}_I(x_0)\right)^2-\left(\psi^{[k]}_I(x_0)\right)^2
    \label{projdiff}
  \end{eqnarray}
  Corresponds to (\ref{propsign}) with $DI(x,y)=D_x-D_y$,
  with $D_x$ and $D_y$--  infinitesimal time shift  operators
  on $x$ and $y$.
  The state with a greater projection to $\psi_0(x)$
  is considered to be the one coming after the state with
  lower projection. The distance (\ref{projdiff}) is degenerated:
  it is equal to 0 for any two $\psi(x)$ for which $0=\psi(x_0)$.
  Also note, that  $\psi^{[j]}_I(x_0)=\Braket{ \psi^{[j]}_I|\psi_0} \psi_0(x_0)$, i.e. the
  $\psi^{[j]}_I(x_0)$ differ from the $\Braket{ \psi^{[j]}_I|\psi_0}$
  on a constant. 
 
\item
  One can variate the (\ref{Vj}) with infinitesimal time shift
  of $\psi^{[j]}_I$,
  applying (\ref{DpsiLag}) or (\ref{DpsiLegendre})
  operator to receive (after normalization)
  a time--distance like this:
  \begin{eqnarray}
    \delta V^{[j]}&=&\Braket{\psi^{[j]}_I|V_{x_0}-V|\psi^{[j]}_I}\left(\psi^{[j]}_I(x_0)\right)^2-\lambda^{[j]}_I \\
    d^{[j]}&=&\left(\psi^{[j]}_I(x_0)\right)^2\frac{\Braket{\psi^{[j]}_I|V_{x_0}-V|\psi^{[j]}_I}}{\lambda^{[j]}_I}-1
    \label{dj}\\
    d_{jk}&=&d^{[j]}-d^{[k]}
  \end{eqnarray}
  The (\ref{dj}) is a ``second order distance''.
  In contrast with the volume (\ref{Vj}),
  the (\ref{dj}) describe the difference in flows
  of volume since $\psi^{[j]}_I$ till ``now''
  per time $\left(\psi^{[j]}_I(x_0)\right)^2$
  and the rate $\lambda^{[j]}_I$.
\end{itemize}

\section{\label{twovarcorrela}$L^4\widetilde{\rho}(p,r)$: Value Correlation of Variables.}
For two variables $p$ and $r$, with some positive
measure $\Braket{p^mr^q}=\int p^m(t)r^q(t) d\mu$ on them,
regular $L^2covariation$
and a new one $L^4covariation$
can be obtained by differentiation (\ref{l2corr})  and (\ref{q2corrA}):
\begin{eqnarray}
  \Braket{(p-\overline{p})^2}&\to&\min\\
  \Braket{(r-\overline{r})^2}&\to&\min\\ 
  L^2covariation&=&\frac{1}{4}
\frac{\partial}{\partial\overline{p}} \frac{\partial}{\partial\overline{r}}
\Braket{(p-\overline{p})^2(r-\overline{r})^2} \label{l2corr} \\
L^2\rho(p,r)&=&
\frac{\Braket{(p-\overline{p})(r-\overline{r})}}
     {\sqrt{\Braket{(p-\overline{p})^2}\Braket{(r-\overline{r})^2}}}
\label{L2corrdef}
     \\
  \Braket{(p-p_1)^2(p-p_2)^2}&\to&\min \label{pmincov} \\
  \Braket{(r-r_1)^2(r-r_2)^2}&\to&\min \label{qmincov} \\ 
  L^4covariation&=&\frac{1}{16}
  \frac{\partial}{\partial p_1}
  \frac{\partial}{\partial p_2}
  \frac{\partial}{\partial r_1}
  \frac{\partial}{\partial r_2}
  \Braket{(p-p_1)^2(p-p_2)^2(r-r_1)^2(r-r_2)^2}
  \label{q2corrA} \\
  L^4\rho(p,r)&=&
  \frac{\Braket{(p-p_1)(p-p_2)(r-r_1)(r-r_2)}}
       {\sqrt{
           \Braket{(p-p_1)^2(p-p_2)^2}\Braket{(r-r_1)^2(r-r_2)^2}
       }}
       \label{q2corrAcorr}
\end{eqnarray}
where $p_{\{1,2\}}$ and $r_{\{1,2\}}$ are quadrature nodes
obtained from (\ref{pmincov}) and (\ref{qmincov})
minimization, exactly as we did in Eq. (\ref{p4min}) above.
The $L^4covariation$ (\ref{q2corrA})
(and (\ref{q2corrAcorr}) correlation)
covariate $p$ and $r$, but use higher order moments;
for $p=r$ it gives regular relations: $L^4volatility=L^4covariation$,
$L^4\rho(r,r)=1$ and $L^4\rho(r,const)=0$.

A much more interesting case is to consider the matrix $L^4covariation_{p_j,r_k}$,
that  covariate $j$--th level of $p$
with $k$--th level of $r$; (here $j,k=1,2$ and $s=\{p,r\}$).
Consider Lagrange interpolating polynomials $l^{(s)}_k$
built on quadrature nodes,
(they are proportional to (\ref{evp3quadr}) eigenfunctions):
\begin{eqnarray}
  l^{(s)}_{\{1,2\}}(s)&=&\frac{s-s_{\{2,1\}}}{s_{\{1,2\}}-s_{\{2,1\}}} \\
  l^{(s)}_{\{1,2\}}(s_{\{1,2\}})&=&1\\
  l^{(s)}_{\{1,2\}}(s_{\{2,1\}})&=&0 \\
  w^{(s)}_{\{1,2\}}&=&\Braket{l^{(s)}_{\{1,2\}}}=\Braket{\left(l^{(s)}_{\{1,2\}}\right)^2} \\
  \Braket{1}&=&w^{(s)}_{1}+w^{(s)}_{2} =\int d\mu \label{totmu}\\
  L^4covariation_{p_j,r_k}&=& \Braket{l^{(p)}_{j} l^{(r)}_{k}}=
  \int l^{(p)}_{j}(p(t)) l^{(r)}_{k}(r(t)) d\mu
  \label{l4conindex}
\end{eqnarray}
The $2\times 2$ covariation matrix (\ref{l4conindex})
can be interpreted as a joint distribution matrix of $p$ and $r$
variables. Corresponding to quadrature nodes
Lagrange interpolating polynomials $l^{(s)}_k$
are a useful tool to built such a matrix,
because their inner product can be obtained
for the measures of interest.
The (\ref{l4conindex}) covariance
definitions have integrals over time,
that can be calculated directly from distribution moments,
it can be obtained from observation sample
in a way similar to (\ref{sampleLag}) or (\ref{sampleLeg}).
For $p=r$ the matrix is diagonal:
$
  L^4covariation_{s_j,s_k}=\tiny
  \left(
  \begin{array}{ll}
    w^{(s)}_{1} &  0 \\
    0 & w^{(s)}_{2}
  \end{array}
  \right)
$.

$L^4covariation_{p_j,r_k}$ matrix components
have the dimension  of the measure $\Braket{1}$ from (\ref{totmu})
and can be easily written for two--point Gauss quadratures built
on $p$ and $r$:
\begin{equation}
L^4covariation_{p_j,r_k}=
  \frac{1}{(p_1-p_2)(r_1-r_2)}
  \left(
  \begin{array}{ll}
    \Braket{(p-p_2)(r-r_2)} &
      -\Braket{(p-p_2)(r-r_1)} \\
      -\Braket{(p-p_1)(r-r_2)} &
      \Braket{(p-p_1)(r-r_1)}
  \end{array}
  \right)
  \label{l4defmatr}
\end{equation}
quadrature weights $w^{(s)}_{\{1,2\}}$can be expressed through $L^4covariation_{p_j,r_k}$ elements sum:
\begin{subequations}
  \label{weights:wh}
\begin{eqnarray}
   w^{(p)}_{\{1,2\}}&=&L^4covariation_{p_{\{1,2\}},r_1}+L^4covariation_{p_{\{1,2\}},r_2} \\
   w^{(r)}_{\{1,2\}}&=&L^4covariation_{p_{1},r_{\{1,2\}}}+L^4covariation_{p_{2},r_{\{1,2\}}}
\end{eqnarray}
\end{subequations}
From (\ref{l4defmatr})
immediately follow that the sum of all four elements
of $L^4covariation_{p_j,r_k}$ matrix is equal to $\Braket{1}$ .
To obtain dimensionless ``correlation''--like
matrix the (\ref{l4defmatr}) can be divided
by $\Braket{1}$ from (\ref{totmu}),
the difference between diagonal and off-diagonal elements
of this ``correlation''--like matrix 
can be called $L^4\widetilde{\rho}(p,r)$ correlation:
\begin{eqnarray}
   L^4\widetilde{\rho}(p,r)&=&\frac{\sum\limits_{j,k=1}^{2}
    (-1)^{j-k} \,
     L^4covariation_{p_j,r_k}}{\sum\limits_{j,k=1}^{2}L^4covariation_{p_j,r_k}}
   \label{l3rhoprsum}
   \\
   L^4\widetilde{\rho}(p,r)&=&
   \frac{
   \overline{pr}-\overline{p}\,\overline{r}+
   \left(\frac{p_1+p_2}{2}-\overline{p}\right)\left(\frac{r_1+r_2}{2}-\overline{r}\right)}
   {0.25(p_1-p_2)(r_1-r_2)}
   \label{l3rhopr}
\end{eqnarray}
that is different from regular definition by
the term
$\left(\frac{p_1+p_2}{2}-\overline{p}\right)\left(\frac{r_1+r_2}{2}-\overline{r}\right)$
describing skewness correlation.
The (\ref{l3rhopr}) means, that if two distributions
have the skewness of the same sign,
their ``true'' correlation is actually higher, than the one,
calculated from the lower order moments as $\overline{pr}-\overline{p}\,\overline{r}$.
The (\ref{l3rhopr}) formula for $L^4\widetilde{\rho}(p,r)$
is obtained directly from joint distribution matrix (\ref{l4defmatr})
and has a meaning of values correlation:
the $L^4covariation_{p_j,r_k}$ element of (\ref{l4conindex})
matrix is the probability that
$p=p_j$ and $r=r_k$.
The conditions $L^4\widetilde{\rho}(r,r)=1$
and $L^4\widetilde{\rho}(r,const)=0$ also holds,
same as for $L^4\rho(p,r)$ from (\ref{q2corrAcorr}).
We want to emphasize, that,
in applications, the most intriguing feature
is not a new formula (\ref{l3rhopr}) or (\ref{q2corrAcorr})
for correlation, but an ability to obtain 
$(p,r)$ joint distribution matrix (\ref{l4defmatr})
from sampled moments of two distributions.

Quadrature nodes $p_{\{1,2\}}$ and $r_{\{1,2\}}$ 
are calculated from the moments (\ref{pimdefcor}) and (\ref{rhomdef})
respectively 
applying either formula (\ref{evp3nodes}) above or
the ones from  Appendix C of Ref. \cite{2015arXiv151005510G}
(or the formulas from Appendix \ref{quadrParam} of this paper with $dI=0$,
what give $P^{f}$--independent answers).
For $\Braket{pr}$ term in (\ref{l4defmatr})
one more moment (cross--moment) $(\pi\rho)_1$ from (\ref{pirho}) is required
in addition to regular $\pi_m$ and $\rho_m$ ($m=0,1,2,3$):
\begin{subequations}
  \label{moments:wh}
\begin{eqnarray}
  \pi_m&=&\Braket{p^m} \label{pimdefcor} \\
  \rho_m&=&\Braket{r^m}
  \label{rhomdef} \\
  (\pi\rho)_1&=&\Braket{pr}
  \label{pirho}
\end{eqnarray}
\end{subequations}
(to calculate (\ref{l4defmatr}) matrix it requires total 8 moment,
see the file \texttt{\seqsplit{com/polytechnik/utils/ValueCorrelation.java}} for
implementation example of numerical calculation of value correlation).
The (\ref{moments:wh})
definitions 
can be be generalized
to matrix averages (see Appendix E of Ref.\cite{2015arXiv151005510G}),
that corresponds to mixed state in quantum mechanics,
a generalization from pure states of $\Braket{\psi|p^mr^qI|\psi}$ form.

\section{\label{twoPvarcorrela}$\widetilde{\rho}(f,g)$: Probability Correlation of Variables.}
Obtained from sampled moments joint distribution estimator (\ref{l4defmatr})
of previous appendix is an important step in correlation estimation.
However, it still has a number of limitations to be applied to practical data.
\begin{enumerate}
\item It requires  two quadratures (on $p$ and $r$) to be built,
  this requires the moments (\ref{moments:wh})
  to be calculated from the data. Assume $r$ is execution flow $r=I=dv/dt$
  of some security, then, for example, $\Braket{r^2}$
  is problematic to calculate: it is not possible to calculate
  it directly from sample and (\ref{fg}) approach
  does not always give a good result.
\item The cross--moment $(\pi\rho)_1$ from (\ref{pirho})
  is often problematic to calculate.
\item Some of (\ref{moments:wh}) moments can diverge or even do not exist,
  their numerical estimation often becomes
  a kind of numerical regularization exercise.
\end{enumerate}
If we generalize ``correlation concept'',
then the approach to joint distribution matrix
estimation can be extended to using the
moments calculated in arbitrary basis, not only for the one
with 
basis functions argument as an observable,
the case considered in Appendix \ref{twovarcorrela}.
Assume we have two variables $f$ and $g$ (e.g. execution flow of
two securities), some basis $Q_m(x)$ for $m=0,1,2$ ($x$ can be e.g. time or price;
$Q_m(x)$ is a polynomial of $m$--th order),
inner product $\Braket{Q_j(x)|s|Q_k(x)}$ (where $s=\{f,g,const\}$
and  $j,k=0,1$) is defined in some way,
such that the inner product can be calculated directly from sample.
As we discussed in \cite{2016arXiv161107386V}
any observable variable sample can be converted to a matrix,
then generalized eigenvalue problems define the spectrum
of the observable. For $f$ and $g$ this would be the equations
(similar to Eq. (\ref{GEVI}) with $n=2$):
\begin{eqnarray}
  \sum\limits_{k=0}^{1}\Braket{Q_j|f|Q_k} \alpha^{f;[i]}_k &=&
  \lambda^{[i]}_f\sum\limits_{k=0}^{1} \Braket{Q_j|Q_k} \alpha^{f;[i]}_k
  \label{evf} \\
  \sum\limits_{k=0}^{1}\Braket{Q_j|g|Q_k} \alpha^{g;[i]}_k &=&
  \lambda^{[i]}_g\sum\limits_{k=0}^{1} \Braket{Q_j|Q_k} \alpha^{g;[i]}_k
  \label{evg}
\end{eqnarray}
For $n=2$
generalized eigenvalue problem
$\Ket{A|\psi}=\lambda\Ket{B|\psi}$
is reduced to solving quadratic on $\lambda$ equation: $0=\det \| A-\lambda B\|$, same as with Eq. (\ref{evp3quadr}):
\begin{eqnarray}
  \left(
  \begin{array}{ll}
    \Braket{Q_0|s|Q_0} &  \Braket{Q_0|s|Q_1} \\
   \Braket{Q_1|s|Q_0} &  \Braket{Q_1|s|Q_1}
  \end{array}
  \right)
  \left(
  \begin{array}{l}
    \alpha_{0}^{s;[i]}\\
    \alpha_{1}^{s;[i]}
  \end{array}
    \right)
    &=&
    \lambda_{s}^{[i]}
    \left(
    \begin{array}{ll}
      \Braket{Q_0|Q_0} &  \Braket{Q_0|Q_1} \\
   \Braket{Q_1|Q_0} &  \Braket{Q_1|Q_1}
  \end{array}
  \right)
  \left(
  \begin{array}{l}
    \alpha_{0}^{s;[i]}\\
    \alpha_{1}^{s;[i]}
  \end{array}
  \right)
  \label{d2EVproblem} \\
  \Ket{\psi^{[i]}_s}\,\, \mathrm{state}&:&\psi^{[i]}_s(x)=\alpha_{0}^{s;[i]}Q_0(x)+\alpha_{1}^{s;[i]}Q_1(x)
\end{eqnarray}
Found $\frac{\Braket{s \psi^2(x)}}{\Braket{\psi^2(x)}}\to \{\min ; \max \}$
solutions are chosen to have 
 normalized $\Ket{\psi^{[i]}_{s}}$ eigenvectors: 
 $\delta_{im}=\sum\limits_{j,k=0}^{1} \alpha^{s;[i]}_j \Braket{Q_j|Q_k} \alpha^{s;[m]}_k$ ;
$\lambda_{s}^{[i]}=\Braket{\psi^{[i]}_{s}|s|\psi^{[i]}_{s}}$,
and
ordered eigenvalues  $\lambda^{[0]}_{\{f,g\}} \le  \lambda^{[1]}_{\{f,g\}}$.
The square of eigenvectors scalar product
define $2\times 2$ matrix $P correlation_{\lambda^{[i]}_f,\lambda^{[m]}_g}$,
the elements of which are
the probabilities of how low/high $f$ is correlated to low/high $g$:
\begin{eqnarray}
  P correlation_{\lambda^{[i]}_f,\lambda^{[m]}_g}&=&
  \left(\sum\limits_{j,k=0}^{1}\alpha^{f;[i]}_j \Braket{Q_j|Q_k} \alpha^{g;[m]}_k\right)^2
  \label{welementsdef}\\
  \widetilde{\rho}(f,g)&=&\frac{\sum\limits_{i,m=0}^{1}
    (-1)^{i-m} \,
    P correlation_{\lambda^{[i]}_f,\lambda^{[m]}_g}}{\sum\limits_{i,m=0}^{1}P correlation_{\lambda^{[i]}_f,\lambda^{[m]}_g}}
  \label{wcorr}
\end{eqnarray}
The $\widetilde{\rho}(f,g)$ modified correlation is 
the difference between diagonal and off-diagonal elements
of $P correlation_{\lambda^{[i]}_f,\lambda^{[m]}_g}$ matrix.
This is similar 
to (\ref{l3rhoprsum}) of previous section,
but now the $P correlation_{\lambda^{[i]}_f,\lambda^{[m]}_g}$
matrix is built solely out from $\Braket{Q_j|s|Q_k}$ moments,
that can be defined in arbitrary basis.
An important difference between (\ref{welementsdef}) and (\ref{l4conindex})
matrices is that the (\ref{l4conindex}) elements are
scalar product of eigenvectors,
but (\ref{welementsdef}) elements are \textsl{squared} scalar product
of eigenvectors;
the elements of both matrices have a meaning of probability,
but the probability is defined differently.
The (\ref{welementsdef}), as squared scalar product of eigenvectors,
is a correlation of probabilities.
The $P correlation_{\lambda^{[i]}_f,\lambda^{[m]}_g}$
is a \textsl{probability of probability}\footnote{
  In quantum mechanics a scalar product
  of two wavefunctions can be interpreted
  as ``two wavefunctions correlation''.
  Taking it squared obtain the probability
  of probability correlation.
  If the wavefunctions are of the states $f$ having
  specific value $\lambda^{[i]}_f$ (\ref{evf}) and
  $g$ having
  specific value $\lambda^{[m]}_g$ (\ref{evg}),
  then squared scalar product of 
  corresponding eigenvectors can be similarly interpreted
  as a probability of probability 
  of $f=\lambda^{[i]}_f$ and $g=\lambda^{[m]}_g$.
  This interpretation also corresponds to (\ref{prowB:wh}) normalizing.
  }
that $f$ has a value $\lambda^{[i]}_f$
and $g$ has a value $\lambda^{[m]}_g$,
what is different from the $L^4covariation_{p_j,r_k}$, Eq. (\ref{l4conindex}),
that is a \textsl{probability} of $p=p_j$ and $r=r_k$.
Instead of (\ref{weights:wh}) we now have:
\begin{subequations}
  \label{prowB:wh}
\begin{eqnarray}
  1&=&P correlation_{\lambda^{[\{0,1\}]}_f,\lambda^{[0]}_g}+P correlation_{\lambda^{[\{0,1\}]}_f,\lambda^{[1]}_g} \\
  1&=&P correlation_{\lambda^{[0]}_f,\lambda^{[\{0,1\}]}_g}+P correlation_{\lambda^{[1]}_f,\lambda^{[\{0,1\}]}_g}
\end{eqnarray}
\end{subequations}
the sum of the elements in any row or column of $P correlation_{\lambda^{[i]}_f,\lambda^{[m]}_g}$
matrix is equal to $1$.
If $Q_0(x)=const$ (typical situation),
then, similar to (\ref{skewness}) definition,
a skewness--like (like a difference between median and average)
characteristics $\widetilde{\Gamma}$
of random variable $s=\{f,g\}$ can be introduced:
\begin{eqnarray}
\widetilde{\Gamma}&=&
\frac{2\overline{s}-\lambda^{[0]}_s-\lambda^{[1]}_s}{\lambda^{[0]}_s-\lambda^{[1]}_s}
\label{skewnessLike} \\
\overline{s}&=&\Braket{sQ_0}\Big/\Braket{Q_0} \\
\widetilde{\rho}(f,g) &=&
\frac{
\Braket{\psi^{[0]}_g|f|\psi^{[0]}_g} -\Braket{\psi^{[1]}_g|f|\psi^{[1]}_g}}
     {\lambda^{[0]}_f-\lambda^{[1]}_f}
     \label{wcorr:el} \\
     &=&
\frac{
\Braket{\psi^{[0]}_g|f|\psi^{[0]}_g} -\Braket{\psi^{[1]}_g|f|\psi^{[1]}_g}}
     {\Braket{\psi^{[0]}_f|f|\psi^{[0]}_f} -\Braket{\psi^{[1]}_f|f|\psi^{[1]}_f}}
     \label{wcorr:elA}
\end{eqnarray}
This skewness definition (\ref{skewnessLike}) has a meaning of $\psi(x)=const$
state $\Ket{\psi_C}$
expansion weights asymmetry on the states: $\Ket{\psi^{[0]}_s}$, corresponding to minimal $s=\lambda^{[0]}_s$,
and $\Ket{\psi^{[1]}_s}$, corresponding to maximal $s=\lambda^{[1]}_s$;
$\widetilde{\Gamma}=\Braket{\psi_C|\psi^{[0]}_s}^2-\Braket{\psi_C|\psi^{[1]}_s}^2$.
The (\ref{wcorr}) probability correlation $\widetilde{\rho}(f,g)$ can be also
written in a similar ``derivative--like'' form (\ref{wcorr:el}): the difference between $f$
in the state  $\Ket{\psi^{[0]}_g}$ of minimal $g$, 
and $f$ in the state $\Ket{\psi^{[1]}_g}$ of maximal $g$, 
 divided by minimal and maximal $f$ difference.
For probability correlation
classical condition $\widetilde{\rho}(f,f)=1$
holds,
for $f=g$ the (\ref{welementsdef}) matrix is diagonal:
$
  P correlation_{\lambda^{[i]}_f,\lambda^{[m]}_f}=\tiny
  \left(
  \begin{array}{ll}
    1 &  0 \\
    0 & 1
  \end{array}
  \right)
$.
But another classical condition \textsl{does not hold}:
$\widetilde{\rho}(f,const)\ne 0$,
if $g=const$ then eigenvalues problem (\ref{evg})
is degenerated and, without an extra condition on eigenvectors,
the value of probability correlation (\ref{wcorr})
can be arbitrary, depending on specific $g$-- eigenvectors choice.

The distinction between ``value'' and ``probability'' correlations
is an important topic of modern research in both
computer science and market dynamics.
The problems of Distribution Regression Problem\cite{dietterich1997solving,zhou2004multi}
(a number of observations of type ``bag of instances to a value''
are used to build a mapping: probability distribution to value)
and Distribution to Distribution Regression Problem
(a number of observations of type ``bag of instances to a bag of other instances''
are used to build a mapping: probability distribution to probability distribution)
are the most known generalization of regular Regression Problem
(a number of observations of type ``value to a value''
are used to build a mapping: value to value)
have been addressed from a number of points.
Our contribution to it is based on an application of
Christoffel function\cite{2015arXiv151107085G}, and
Radon--Nikodym derivatives\cite{2015arXiv151109058G}.
The difficulties in probability estimation using
real life data
have been emphasized\cite{taleb2014precautionary},
but very different mathematical technique have been used
for probability estimation.
The (\ref{wcorr}) answer is,
to the best of our knowledge,
the first probability correlation answer,
that is calculated from the moments of sampled data.
To calculate (\ref{welementsdef})
matrix it requires
$m=0,1,2$ moments: $\Braket{\{f,g,const\}Q_m(x)}$; total  9 moment,
see the file \texttt{\seqsplit{com/polytechnik/utils/ProbabilityCorrelation.java}} for
implementation example of numerical calculation of
probability correlation $\widetilde{\rho}(f,g)$ from (\ref{wcorr}),
also see the file \texttt{\seqsplit{com/polytechnik/utils/Skewness.java:getGSkewness}}
for calculation $\widetilde{\Gamma}$ from (\ref{skewnessLike}).
A remarkable feature of these answers is that
they use only first order moments on $f$ and $g$
and higher order moments on $Q_m(x)$.
This separation of observable
variables and basis functions
allows the approach to be applied to
$f$ and $g$ having non--Gaussian distributions,
even those with, say, infinite $\Braket{f^2}$ or $\Braket{g^2}$,
a distinguishable feature of Radon--Nikodym approach\cite{2016arXiv161107386V}.

\section{\label{quadrParam}Price distribution estimation
  with unknown future price $P^{f}$ as a parameter}
\begin{figure}
  \includegraphics[width=16cm]{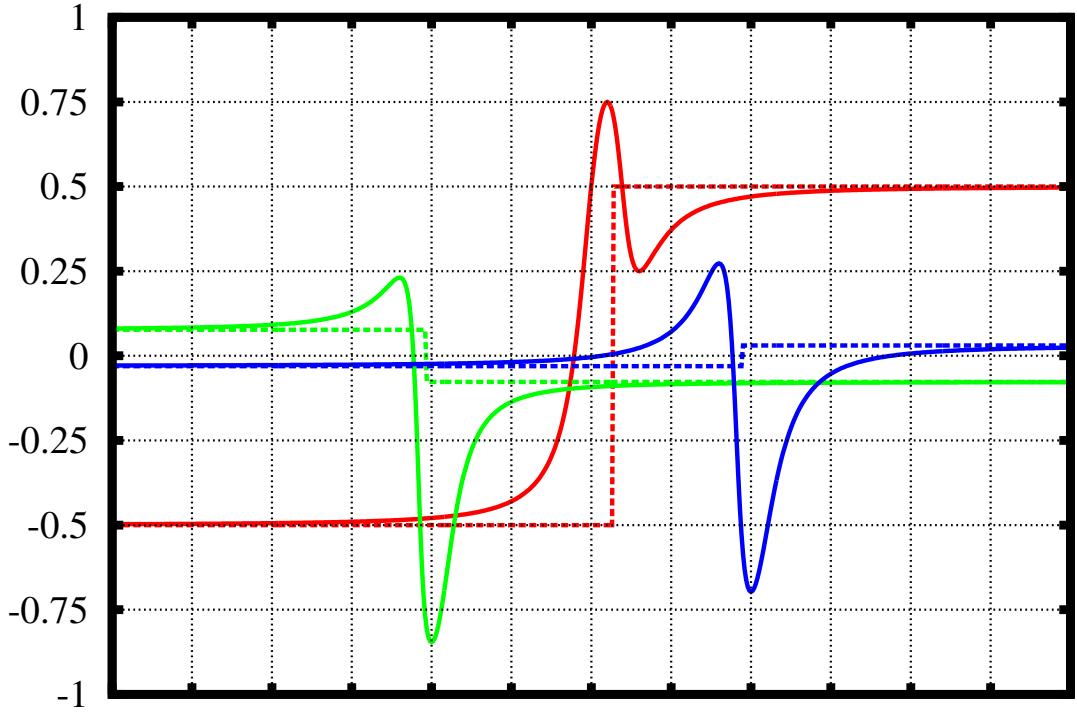}
\caption{\label{GPexample}
 Several examples of $\Gamma(P^{f})$ dependence for different $\pi_m$ and $di$.
  The $\Gamma(P^{f})$ has  maximum and minimum at unperturbed ($di=0$)
  quadrature nodes; the $P^{f}\to\pm\infty$ asymptotic is (\ref{spkewnessPfasymp}).
  Dashed line is $\Gamma(P^{f})$ skewness for the measure with single support
  at average value (single node quadrature).
}
\end{figure}

In Section \ref{Thr} we solved the problem
of price distribution estimation given $\pi_m$ moments (\ref{pimdef}).
However,  future price $P^{f}$ is required
to calculate future moments $\pi_m^{f}$;
``the last price as $P^{f}$ estimator (\ref{Pfmestlast})''
is a very crude approximation,
thus it is better to consider $P^{f}$ as a parameter
(This consideration is a special case of varying measures
orthogonal polynomials\cite{totik}. In this work,
instead of typicaly considered a sequence of measures,
a measure, depending on $P^{f}$  as a parameter, is considered.)
For a given $\Ket{\psi}$
the $\|p^mI^{f}\|$ operator from (\ref{pImfutureOp})
with an impact from the future term 
give future moments $\pi_m^{f}$:
\begin{eqnarray}
  \pi_m^{f}&=&\pi_m+\left(P^{f}\right)^m dI \Braket{\psi|\psi_0}^2 \label{pimFdef}
\end{eqnarray}
that are different from past moments $\pi_m=\Braket{\psi|p^mI|\psi}$ from
(\ref{pimdef}) by impact from the future term: $\left(P^{f}\right)^m dI \Braket{\psi|\psi_0}^2$.
The value of $P^{f}$ is unknown, however one
can repeat all the calculations of Section \ref{Thr}
above, using $P^{f}$ as  a parameter.
After simple algebra (see \texttt{\seqsplit{DiffSkewness.java}} from Appendix
\ref{appendix1} below  for numerical
implementation)
$P^{f}$--dependent
$\Gamma$ from (\ref{skewness}),
quadrature nodes $p_{\{1,2\}}(P^{f})$, weights $w_{\{1,2\}}(P^{f})$,
and monic second order orthogonal polynomial (\ref{opol2})
($P^{f}$ --dependent orthogonal system)
for the  measure with (\ref{pimFdef}) moments
are:
\begin{eqnarray}
  di&=&dI \Braket{\psi|\psi_0}^2  \label{didef} \\
  b&=&\frac{di}{\pi_0+di} \\
a_m&=&\frac{\pi_m}{\pi_0+di} \\
A(P^{f})&=&(a_3a_1-a_2^2)+(a_3b)P^{f}-2(a_2b)\left(P^{f}\right)^2
+(a_1b)\left(P^{f}\right)^3 \\
B(P^{f})&=&(a_2a_1-a_3)+(a_2b)P^{f}+(a_1b)\left(P^{f}\right)^2-(1-b)b\left(P^{f}\right)^3 \label{BPf} \\
D(P^{f})&=&(a_2-a_1^2)-2(a_1b)P^{f}+(1-b)b\left(P^{f}\right)^2 \label{DPf}\\
\Gamma(P^{f})&=&\frac{-B(P^{f})-2(a_1+bP^{f})D(P^{f})}
      {\sqrt{B^2(P^{f})-4A(P^{f})D(P^{f})} }
      \label{spkewnessPf} \\
      \Gamma(P^{f}\to\pm\infty)&=&    \pm \frac{\pi_0-di}{\pi_0+di}
      \label{spkewnessPfasymp} \\
p_{\{1,2\}}(P^{f})&=& \frac{-B(P^{f}) \mp \sqrt{B^2(P^{f})-4A(P^{f})D(P^{f})} }
{2D(P^{f})} \label{p12pertr} \\
w_{\{1,2\}}(P^{f})&=&\frac{\pi_0+di}{1+\left.{\left[p_{\{1,2\}}(P^{f})-\overline{p}(P^{f})\right]^2}\right/{D(P^{f})}} \label{w12pertr} \\
\overline{p}(P^{f})&=&
\frac{p_1(P^{f})w_{1}(P^{f})+p_2(P^{f})w_{2}(P^{f})}{w_{1}(P^{f})+w_{2}(P^{f})}
=a_1+bP^{f} \label{paverPf} \\
p_{mid}(P^{f})&=&\frac{p_1(P^{f})+p_2(P^{f})}{2}=-0.5\frac{B(P^{f})}{D(P^{f})}
\label{pmedestBA}  \\
p_{2}(P^{f})-p_{1}(P^{f})&=&\frac{\sqrt{B^2(P^{f})-4A(P^{f})D(P^{f})}}{D(P^{f})}
\label{p21diff} \\
P_2(p,P^{f})&=& (p-p_{1}(P^{f}))(p-p_{2}(P^{f}))=p^2+\frac{B(P^{f})}{D(P^{f})}p+\frac{A(P^{f})}{D(P^{f})}
\label{opol2} \\
E(P^{f})&=& (a_3a_1-a_2^2)+(a_2a_1-a_3(1-b))P^{f}+
    (a_2(1-b)-a_1^2)\left(P^{f}\right)^2 \nonumber \\
\overline{(p-p_1(P^{f}))^2(p-p_2(P^{f}))^2}&=&a_4+\frac{
  a_3B(P^{f})+a_2A(P^{f})+(P^{f})^2bE(P^{f})
}
{D(P^{f})} \\
\overline{(p-\overline{p}(P^{f}))^2}&=& D(P^{f})
\end{eqnarray}
The (\ref{spkewnessPf}) is a ratio
of third order polynomial in numerator
and square root of sixth order polynomial in denominator.
The $\Gamma(P^{f})$ is a function 
with  $\frac{w_1-w_2+di}{w_1+w_2+di}$ maximum at $P^{f}=p_1$
and $\frac{w_1-w_2-di}{w_1+w_2+di}$ minimum at $P^{f}=p_2$,
$p_1\le p_2$, where $p_{\{1,2\}}$ and $w_{\{1,2\}}$
are quadrature nodes and weights
of two--point Gauss quadrature built on $\pi_m$
moments (with $di=0$, unperturbed quadrature: $w_1+w_2=\pi_0$). 
The $\Gamma(P^{f})$ have  (\ref{spkewnessPfasymp}) asymptotic for $P^{f}\to\pm\infty$.
In Fig. \ref{GPexample} several examples for $\Gamma(P^{f})$
are presented, maximum, minimum and asymptotic
are clearly observed.

 When
$\pi_m$ moments are of single support point  distribution
the (\ref{spkewnessPf}) take a very simple form:
Gauss quadrature built on $\pi^{f}_m$ moments (\ref{pimFdef})
has the nodes: the support point and $P^{f}$;
quadrature weights are: $\pi_0$ and $di$;
the $\Gamma(P^{f})$ is
a step--function with (\ref{spkewnessPfasymp}) values,
changing the value at support point;
$L^4volatility$ from (\ref{p4min}) is zero.
In Fig. \ref{GPexample} this situation:
two support points: unperturbed average (with the weight $w_1+w_2$)
and $P^{f}$ (with the weight $di$)
is presented as dashed lines.

\begin{figure}
  \includegraphics[width=16cm]{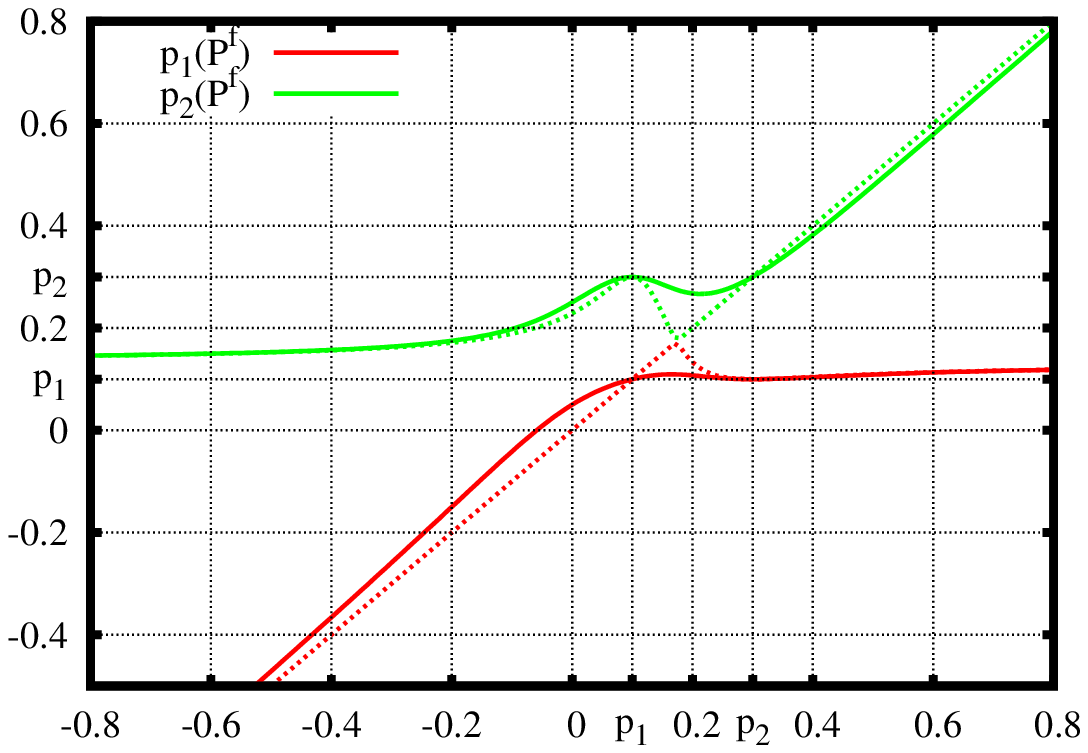}
  \includegraphics[width=16cm]{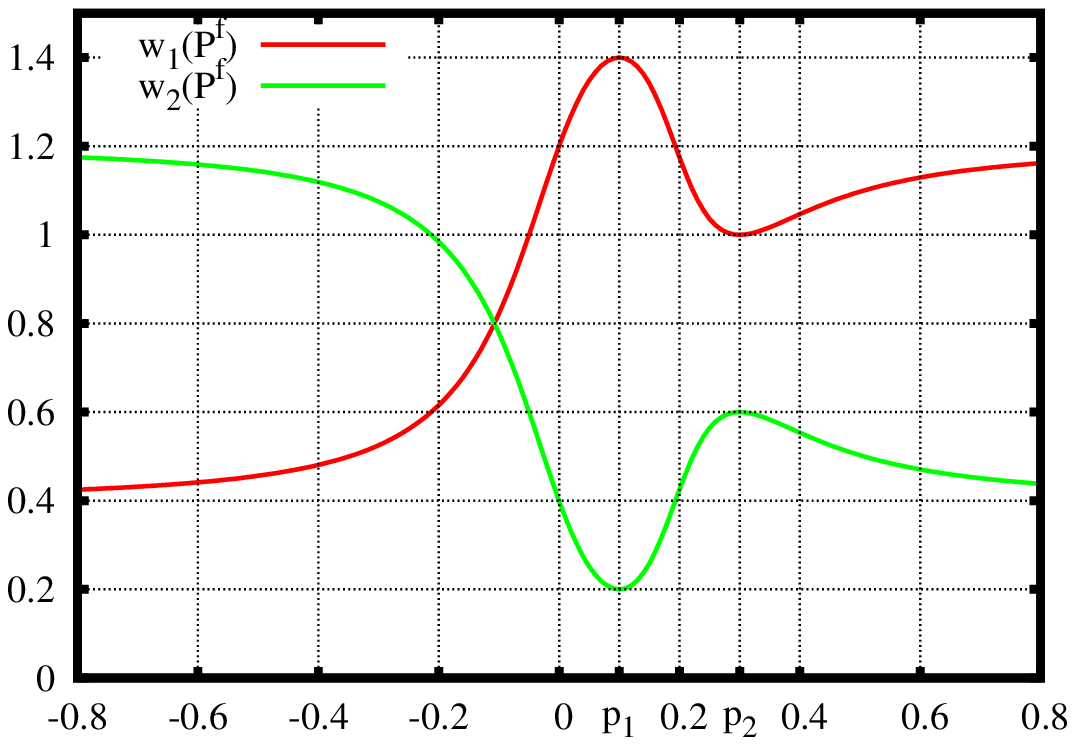}
  \caption{\label{PmPexample}
    An example of distribution (with $p_1=0.1$, $w_1=1$, $p_2=0.3$,
    $w_2=0.2$).
    Top: The dependence (\ref{p12pertr}) of $p_{1}(P^{f})$  (red)
    and $p_{2}(P^{f})$ (green) for  $di=0.4$ (solid lines)
    and $di\to\infty$ asymptotic (dashed lines).
    Bottom: The dependence (\ref{w12pertr}) of $w_{1}(P^{f})$  (red)
    and $w_{2}(P^{f})$ (green) for  $di=0.4$
}
\end{figure}
The
$p_{\{1,2\}}(P^{f})$ and $w_{\{1,2\}}(P^{f})$ (perturbed quadrature nodes and weights)
from (\ref{p12pertr}) and (\ref{w12pertr})
are often of interest.
In Fig. \ref{PmPexample} we present an example.
The weight $w_{\{1,2\}}(P^{f})$ has $w_{\{1,2\}}+di$ maximum  at $P^{f}=p_{\{1,2\}}$
and $w_{\{1,2\}}$ minimum at $P^{f}=p_{\{2,1\}}$.
The $p_{1}(P^{f})$ is a function with minimum (equal to unperturbed $p_1$)   at $P^{f}=p_2$
and $p_{2}(P^{f})$ is a function with maximimin (equal to unperturbed $p_2$)  at $P^{f}=p_1$,
(parabolic behavior of $p_{\{1,2\}}(p_{\{2,1\}}+\Delta p)$
for small $\Delta p$;
also note that $p_{\{1,2\}}(p_{\{1,2\}})=p_{\{1,2\}}(p_{\{2,1\}})=p_{\{1,2\}}$).
The behavior of $p_{\{1,2\}}(P^{f})$ for a constant $di$
and $di\to\infty$ asymptotic is shown in 
Fig. \ref{PmPexample} as solid and dashed  lines respectively.
In  applications the (\ref{pmedestBA}) midpoint (a function with min, max,
having $p_{mid}(p_1)=p_{mid}(p_2)=p_{mid}(\overline{p})$);
the (\ref{paverPf}) average (a linear function with $b$ slope)
can be also of interest.

A very important characteristic is ``volatility''--like
characteristic (\ref{p21diff}),
the difference between perturbed quadrature nodes:
$p_{2}(P^{f})-p_{1}(P^{f})$.
It is always positive, has the dimension of price
and can be used in place of standard deviation.
This difference reach the same value $p_2-p_1$ for $P^{f}$ equal to unperturbed quadrature nodes $p_{\{1,2\}}$
and has $|P^{f}-\overline{p}|$ asymptotic for  $P^{f}\to\pm\infty$.

\section{\label{PnLTrading}P\&L Trading Strategy and Frontrun Asymmetry}
\begin{figure}
  \includegraphics[width=8cm]{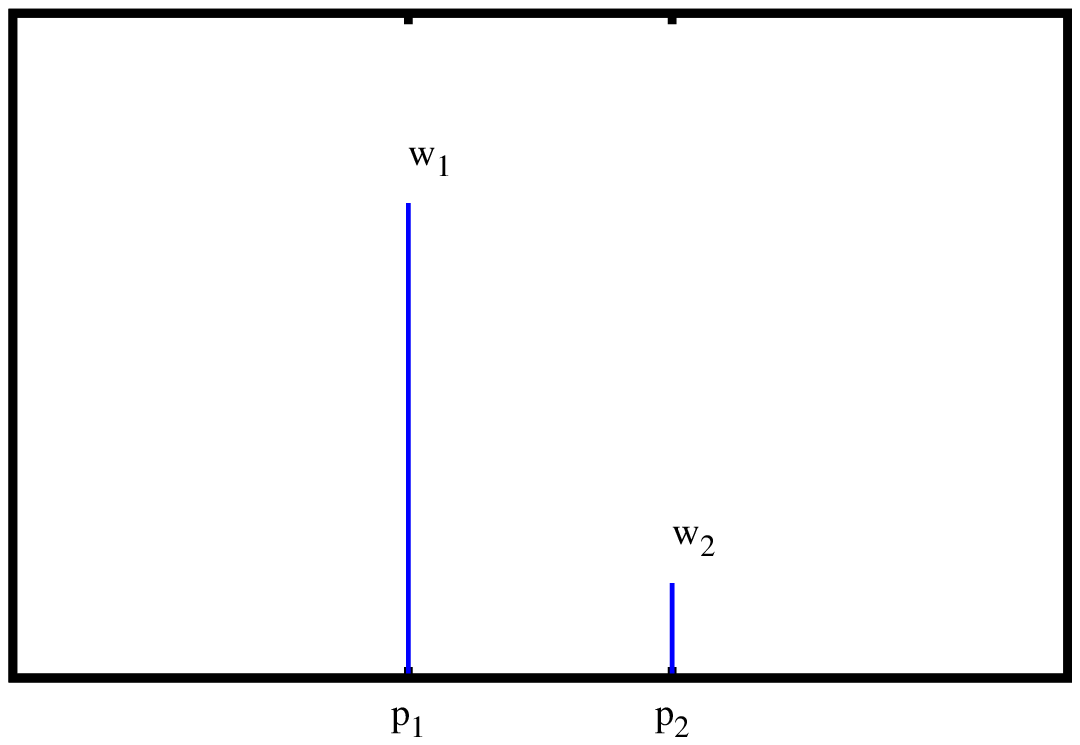}
  \includegraphics[width=8cm]{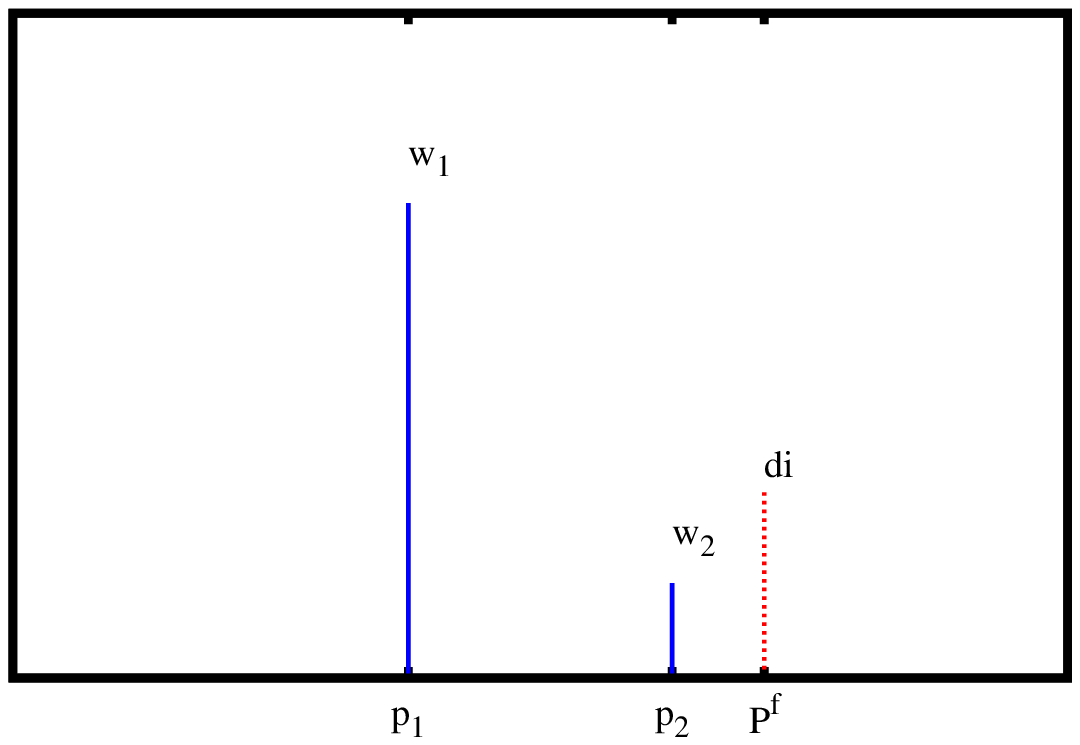} \\
  \includegraphics[width=8cm]{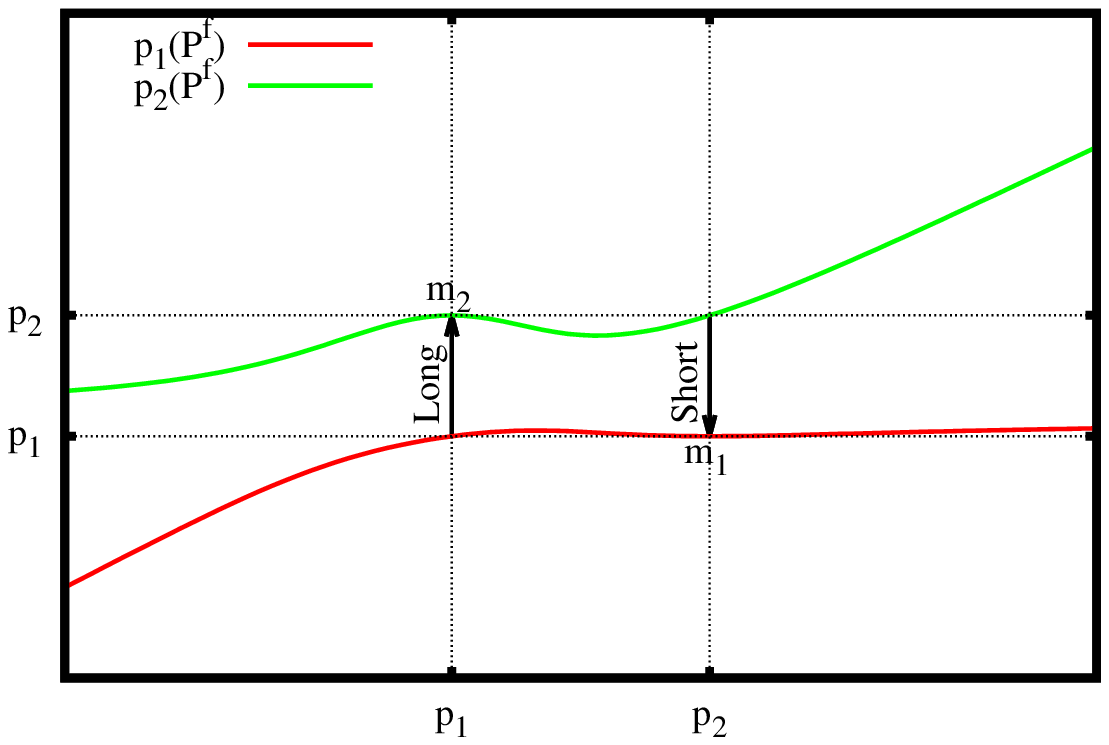}
  \caption{\label{frontrunF}
    Top Left: past information available for $\Ket{\psi}$ state:
    $w_1$ at $p_1$ and $w_2$ at $p_2$,
    where $p_{\{1,2\}}$ and $w_{\{1,2\}}$ are unperturbed
    quadrature nodes and weights built on past  moments (\ref{pimdef});
    the median is $w_1$, because $w_1>w_2$. 
    Top Right: past and future information for $\Ket{\psi}$ state,
    in addition to the data from ``the past'' the following is also available:
    known (\ref{didef}) impact from the future $di$ 
    at unknown future price $P^{f}$.
    Bottom: ``Band structure'' of Long/Short frontrunning alternatives.
    The asymmetry is determined by ``effective mass'' difference (\ref{directionalD}).
}
\end{figure}
In Section \ref{movavermomentsSK} we considered a simple frontrun
strategy and have shown that the median should be used as a threshold.
It is of great interest
to consider such a strategy in general case.
Important feature of trading distributions is that it is
a discrete one (price levels are discrete).
Moreover, ``real'' distribution
can be interpolated by Gauss quadrature
and discrete weights of the quadrature
can be considered as interpolating distribution.

Consider a very simple example:
let trading take place at price $p_1$ with volume $w_1$
and at price $p_2$ with volume $w_2$, Fig. \ref{frontrunF}
(we assume $p_1<p_2$, and the $w_{\{1,2\}}$ is the number of \textsl{matched}
buyers \& sellers at price $p_{\{1,2\}}$).
This distribution has the median equal to $p_{1}$ or $p_{2}$,
depending what weight $w_{1}$ or $w_{2}$ is a greater one.
As in Section \ref{movavermomentsSK},
were a speculator knows future trading profile,
buying below median and selling above the median,
the maximal P\&L a speculator can obtain is:
\begin{eqnarray}
  \mathrm{P\&L_{\max}}&=& (p_2-p_1)\min(w_1,w_2)
  \label{PnL2Lev}
\end{eqnarray}
At $p_1$ he should frontrun the buyers bidding at $p_1+\delta$
and at $p_2$ he should frontrun the sellers offering at $p_2-\delta$,
maximal volume $\min(w_1,w_2)$
come from the fact that at the level of highest
weight (equal to the median) the speculator have to partially trade both long and short
to avoid position accumulation at the end of investment horizon.
(If $p_1$ and $p_2$ are considered as \textsl{unmatched} levels of
limit order book --- this two--level example
is a classical demonstration of market--making,
but the whole point of this paper is a
transition from unmatched volume (supply/demand)
to describing
matched data execution flow $I=dV/dt$ (both past and future (\ref{dI})).
This allows us to avoid using (\textsl{unmeasurable} from the data) supply and demand
and, instead, to work with (\textsl{directly measurable} from the data)
execution flow fluctuations).
Similary, for $n$--point distribution
(either actual or the weights of Gauss quadrature,
built out of $0\dots 2n-1$ distribution moments),
one need to find quadrature nodes, the median,
then frontrun the buyers below median,
frontrun the sellers above median;
at the median  partially trade both long and short
to avoid position accumulation.
The P\&L calculations is very similar to Quantile regression
problem \cite{wiki:quantileregression},
but we will not discuss this relation here.
In this paper we are going to limit ourselves
to two--nodes distributions only, then all the calculations
can be performed without full blown Linear Programming theory.

In real life we do not know complete future trading profile.
We know  impact from the future $di$ from (\ref{didef}),
but at unknown future price $P^f$, see Fig. \ref{frontrunF} right.
As with any two--level Hamiltonian
arbitrary state can be expanded as a superposition of two--level states.
If $P^f$ was traded at $p_1$ (frontrun buyers), then
$w_1\to w_1+di$, $w_2\to w_2$. If $P^f$ was traded at $p_2$ (frontrun sellers),
then
$w_1\to w_1$, $w_2\to w_2+di$. (In both cases $p_{\{1,2\}}$ do not change.).
These two alternatives (frontrun buyers/frontrun sellers)
give identical price change,
and, if $di\le w_{\{1,2\}}$, also give identical maximal P\&L.
Otherwise a term $\min(di,w_{\{1,2\}})$
similar to the one in (\ref{PnL2Lev}) arise.

To obtain directional information
we need a criteria to distinguish the two alternatives.
They can be distinguished considering variations of $P^f$.
Assume execution flow to occur not at specific single price $P^f$,
but within some price interval $P^f\pm\Delta p$.
Note that according to time--price symmetry argument\cite{2015arXiv151005510G}
first order derivative cannot privide dynamics information,
thus the P\&L should be invariant with respect to $\Delta p\to -\Delta p$.
Consider the P\&L corresponding to impact from the future execution flow $di$,
with $P^f$, distributed within the
$p_{\{1,2\}}\pm\Delta p$ interval.
Then
\begin{eqnarray}
  \mathrm{P\&L_{fr\,long}}(P^{f})&=& (p_2(P^{f})-P^{f})\min(di,w_{2})
  \label{PnLfrlong} \\
  \mathrm{P\&L_{fr\,short}}(P^{f})&=& (P^{f}-p_1(P^{f}))\min(di,w_{1})
  \label{PnLfrshort} \\
  \Delta\mathrm{P\&L_{fr}}&=& \mathrm{P\&L_{fr\,long}}(p_2\pm\Delta p)
  -\mathrm{P\&L_{fr\,short}}(p_2\pm\Delta p)
  \label{pnlvar}
\end{eqnarray}
The $p_{\{1,2\}}(P^{f})$ is a function with min/max at $P^{f}=p_{\{2,1\}}$,
see Appendix \ref{quadrParam} Fig. \ref{PmPexample}.
Long/short assymetry  (\ref{pnlvar}),
can be considered as directional asymmetry and 
for infinitesimal  $\Delta p$ second order term is:
\begin{eqnarray}
  \Delta\mathrm{P\&L_{fr}}&\approx&
  \frac{(\Delta p)^2}{2}\left[\left.\frac{\partial^2 \mathrm{P\&L_{fr\,long}}(P^{f})}{\partial (P^{f})^2}
  \right|_{P^{f}=p_1}-
  \left.\frac{\partial^2 \mathrm{P\&L_{fr\,short}}(P^{f})}{\partial (P^{f})^2}
  \right|_{P^{f}=p_2}\right] \\
  &=&  \frac{(\Delta p)^2}{2}\left[\left.\min(di,w_{2})\frac{\partial^2 p_2(P^{f})}{\partial (P^{f})^2}
  \right|_{P^{f}=p_1}+
  \left.\min(di,w_{1})\frac{\partial^2 p_1(P^{f})}{\partial (P^{f})^2}
  \right|_{P^{f}=p_2}\right]
  \label{smmband}
\end{eqnarray}
The $p_{\{1,2\}}(P^{f})$ are similar to solid state physics ``band structure''.
It is convinient to introduce an ``effective mass'' near zone edge:
\begin{eqnarray}
  \frac{1}{m_1}&=&\left.\frac{\partial^2 p_1(P^{f})}{\partial (P^{f})^2}
  \right|_{P^{f}=p_2} \\
  \frac{1}{m_2}&=&\left.\frac{\partial^2 p_2(P^{f})}{\partial (P^{f})^2}
  \right|_{P^{f}=p_1} \\
  {\cal D}&=&
  \frac{1}{m_1}+\frac{1}{m_2}
  \label{directionalD}
\end{eqnarray}
We have $m_1>0$  and $m_2<0$, as for electrons and holes in a semiconductor,
see Fig. \ref{frontrunF} for this ``transition'' analogy.
The ${\cal D}$, directional assymetry of
distribution,
is related to distribution skewness (\ref{skewness})
and, in some situations, can be used as a directional indicator.

\section{\label{FuturePsiA} Future Wavefunction Without $I_0^{f}$.}
In the section \ref{FuturePsiOp} we have determined (\ref{iofuture})
future $I_0^{f}$ and made an attempt to convert this
information to price information using the dynamic equation of Ref. \cite{2015arXiv151005510G}. A question arise what kind of answer
can be obtained {\bf without} information about $I_0^{f}$?
It is clear, that in this case only perturbation theory
on ${dI}/{I_0^{f}}$
can be developed. Because
$dI\ge 0$ (\ref{dIge0}) some information can still be obtained,
even in case of unknown  $I_0^{f}$ value.

\begin{figure}
  \includegraphics[width=16cm]{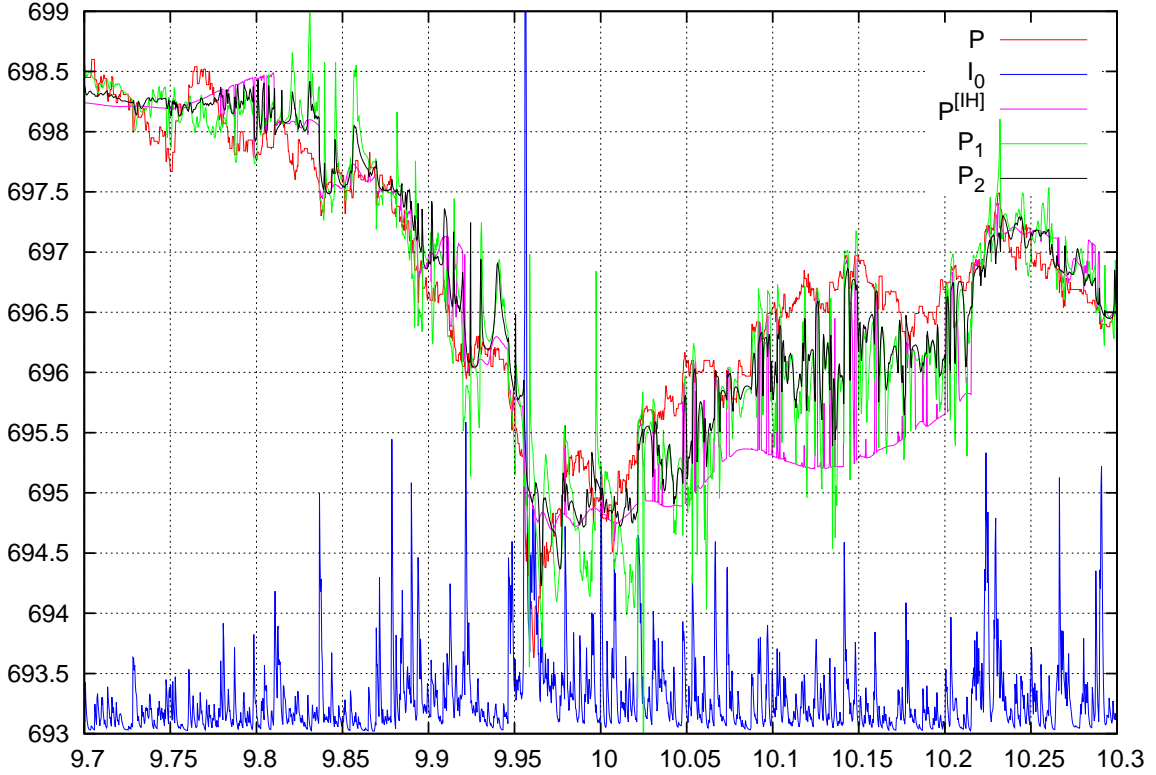}
\caption{\label{PIHP1P2}
  The AAPL stock price on September, 20, 2012.
  Calculated in Shifted Legendre basis with $n=7$ and $\tau$=128sec.
  The $P^{[IH]}$, $P_1$  and $P_2$ 
  are calculated according to (\ref{PIH}), (\ref{P1}) and (\ref{P2})
  respectively.
}
\end{figure}

Consider some wavefunction $\psi(x)$ and corresponding
execution flow $I_{\psi}$,
calculated as in (\ref{faver}). Consider simple variation $\delta \psi(x)$.
Then second order Rayleigh quotient perturbation is:
\begin{eqnarray}
  I_{\psi+\delta\psi}&=&\frac{\Braket{\psi+\delta\psi|I|\psi+\delta\psi} }
  {\Braket{\psi+\delta\psi|\psi+\delta\psi}} = D0+D1+D2 +\dots \label{psivar} \\
  D0&=&\frac{\Braket{\psi|I|\psi}}{\Braket{\psi|\psi}} \\
  D1&=&2\left(\frac{\Braket{\psi|I|\delta\psi}}{\Braket{\psi|\psi}}-
  D0\frac{\Braket{\psi|\delta\psi}}{\Braket{\psi|\psi}}\right) \label{rqD1} \\
  D2&=& \frac{\Braket{\delta \psi|I|\delta\psi}}{\Braket{\psi|\psi}}-
  D0\frac{\Braket{\delta \psi|\delta\psi}}{\Braket{\psi|\psi}}
  -2\frac{\Braket{\psi|\delta\psi}}{\Braket{\psi|\psi}}D1 \label{rqD2}
\end{eqnarray}
A rather complex perturbation theory  on $\Ket{\delta \psi}$
can be developed in a style 
of our earlier work \cite{malyshkin1999features} in a very different field
of multiple-scattering,
but  we limit here all the considerations to
first order $I$ variation only on $\delta \psi$
states, orthogonal to $\psi$, i.e. $\Braket{\delta \psi|\psi}=0$.
Then
\begin{eqnarray}  
  I_{\psi+\delta\psi}&\approx& \frac{\Braket{\psi|I|\psi}}{\Braket{\psi|\psi}} +\delta I \\
  \delta I&=&2\frac{\Braket{\psi|I|\delta\psi}}{\Braket{\psi|\psi}} = 2\Braket{b|\delta \psi}
  \label{ivar}\\
  \Ket{b} &=& \Ket{I|\psi}
\end{eqnarray}
Thus $\delta I$ (\ref{ivar}) is represented as a scalar product 
of   $\Ket{b}$ and $\Ket{\delta \psi}$ vectors.
What variation $\delta \psi$ to provide
maximal  $\delta I$? The one, different from $\Ket{b}$ only on a constant $\beta$,
i.e.
\begin{eqnarray}
  \Ket{\phi} &=& \Ket{I|\psi} - \Braket{\psi|I|\psi} \Ket{\psi} \label{phivar} \\
  \Ket{\delta \psi} &=& \Ket{\phi} \beta 
  \label{apsibvar}
\end{eqnarray}
The states (\ref{apsibvar}) provide maximal variation $\delta I$.
The term $\Braket{\psi|I|\psi} \Ket{\psi}$ is subtracted in (\ref{phivar}) to
have $\Braket{\delta \psi|\psi}=0$.
Put $\Ket{\psi}=\Ket{\psi_0}$
from Eq. (\ref{psix0}) to (\ref{phivar}),
this immediately lead to $\phi(x_0)=0$,
and consider $I$ as a function of $\beta$
\begin{eqnarray}
  \Ket{\phi} &=& \Ket{I|\psi_0} - \Braket{\psi_0|I|\psi_0} \Ket{\psi_0} \label{phi0} \\
  I(\beta)&=& \frac{\Braket{\psi_0 +\beta \phi|I|\psi_0 +\beta \phi}}
  {\Braket{\psi_0 +\beta \phi|\psi_0 +\beta \phi}} \nonumber \\
  &\approx& I_0 +  2\beta\Braket{\phi|I|\psi_0}+\dots
  \label{betaexp} 
\end{eqnarray}
As we noted in section \ref{openQuestions}
when $\Ket{\psi_0}$ is an eigenfunction of (\ref{GEVI})
(or $I=const$ and the problem (\ref{GEVI}) is degenerated),
 then theory fails (now for the reason of $\Braket{\phi|\phi}=0$ no first order perturbation theory possible).
Otherwise, because $\Braket{\phi|I|\psi_0}=\Braket{\phi|\phi} > 0$ we always have $\beta > 0$
and in the first order perturbation two answers, let us call them,
$P_1$ and $P_2$
in a weak hope to get a poor--man $P^{[IH]}$:
\begin{eqnarray}
  P_1&=&\frac{\Braket{\phi|pI|\psi_0}}{\Braket{\phi|I|\psi_0}} \label{P1} \\
  P_2&=&\frac{\Braket{\phi|pI|\phi}}{\Braket{\phi|I|\phi}} \label{P2} \\
  r&=&\frac{\sqrt{\Braket{\phi|\phi}}}{\Braket{\psi_0|I|\psi_0}}=
  \frac{\sqrt{\Braket{\psi_0|I | I|\psi_0}-\Braket{\psi_0|I|\psi_0}^2}}{\Braket{\psi_0|I|\psi_0}}
  \label{rf}
\end{eqnarray}
These answers, while being very crude estimates
in practice, may be still useful (especially $P_2$ from (\ref{P2})) in applications
for their simplicity. The $r$ (standard deviation --like estimate of $I$ on $\Ket{\psi_0}$ state) from (\ref{rf})
can serve as an estimate of how close is $\Ket{\psi_0}$ to $\|I\|$ eigenfunction.
The major drawback of all these first order perturbation answers is that
they are not as good in automatic selection of proper time--scale,
as eigenvalues problem.
In Fig. \ref{PIHP1P2} the $P^{[IH]}$, $P_1$  and $P_2$
are presented (calculated according to (\ref{PIH}), (\ref{P1}) and (\ref{P2})
respectively).
One can see that the $P_2$ has a similar to $P^{[IH]}$ behavior,
especially it tracks well market direction change.
The $P_1$,
because it is not averaged with always positive weight,
is more volatile than $P_2$, but also can be of interest.
A very important feature of $P_1$ (\ref{P1}) and $P_2$ (\ref{P2})
is that they are obtained without solving
eigenvalues problem, but, nevertheless,
still provide some information, thus can be
considered as a poor man $P^{[IH]}$.

\section{\label{appendix1} Computer Code Implementation}
\subsection{\label{installprep} Installation and Data Preparation}
\begin{itemize}
\item Install java 1.8 or later.
\item
  Download
  from \cite{polynomialcode}
NASDAQ ITCH data file
  \texttt{\seqsplit{S092012-v41.txt.gz}}, and
  the archive
\href{http://www.ioffe.ru/LNEPS/malyshkin/AMuseOfCashFlowAndLiquidityDeficit.zip}{\texttt{\seqsplit{AMuseOfCashFlowAndLiquidityDeficit.zip}}}
  with the source code.
\item Decompress and recompile the program:
\begin{verbatim}
unzip AMuseOfCashFlowAndLiquidityDeficit.zip
javac -g com/polytechnik/*/*java
\end{verbatim}

\item Extract triples (time, execution price, shares traded)
from NASDAQ ITCH data file:
\begin{verbatim}
java com/polytechnik/itch/DumpData2Trader \
      S092012-v41.txt.gz AAPL >aapl.csv
\end{verbatim}
Execution data and limit order book edges are now saved to tab--separated file \verb+aapl.csv+
of 15 columns and 634205 lines.
The columns of interest are:
\begin{itemize}
\item \verb+currenttime+ Time in nanoseconds since midnight.
\item \verb+exe_price_last+ Last Price.
\item \verb+exe_shares+ Shares traded.
\end{itemize}

\item Run the command to test the program
\begin{verbatim}
java com/polytechnik/algorithms/CallAMuseOfCashFlowAndLiquidityDeficit \
      --musein_cols=15:1:4:5 \
      --musein_file=aapl.csv \
      --museout_file=museout.dat \
      --n=7 \
      --tau=128 \
      --measure=ImpactQVMMuseLegendreShifted
\end{verbatim}
Program parameters are:
\begin{itemize}
\item[] \verb+--musein_file=aapl.csv+ : Input tab--separated file
  with (time, execution price, shares traded) triples timeserie.
\item[] \verb+--musein_cols=15:1:4:5+ : Out of total 15 columns of
  \verb+aapl.csv+ file, take column \#1 as time (nanoseconds since midnight),
  \#4 (execution price), and \#5 (shares traded), column index is base 0.
\item[] \verb+--museout_file=museout.dat+ : Output file name is set to \texttt{\seqsplit{museout.dat}}.
\item[] \verb+--n=7+ : Basis dimension. Typical values are: 2 (for testing a concept),
or some value about $[4\dots 12]$ for more advance use.
\item[]  \verb+--tau=128+ : Exponent time (in seconds) for the measure used.
\item[] \verb+--measure=ImpactQVMMuseLegendreShifted+
  The measure. The values 
  \texttt{\seqsplit{ImpactQVMMuseLaguerre,ImpactQVMMuseLegendreShifted,ImpactQVMMuse\_pi}}
  correspond the measures (\ref{sampleLag}),  (\ref{sampleLeg}), (\ref{samplePrice}) respectively.
  The results of \texttt{\seqsplit{ImpactQVMMuseMonomials}} (uses $Q_k(x)=x^k$)
  should be identical to \texttt{\seqsplit{ImpactQVMMuseLaguerre}} (uses $Q_k(x)=L_k(-x)$),
  as the measure is the same and all the calculations are $Q_k(x)$--basis invariant (but numerical stability is worse for \texttt{\seqsplit{ImpactQVMMuseMonomials}}).
\end{itemize}
\item The results are saved in the output file \verb+museout.dat+.
\item There is a short ``bundled'' data file \texttt{\seqsplit{dataexamples/aapl\_old.csv.gz}} of 9 colums and 28492 lines, that contains only executions (no limit order book events).
  It can be used for testing insead of \texttt{\seqsplit{aapl.csv}}
  obtained from   \texttt{\seqsplit{S092012-v41.txt.gz}}:
\begin{verbatim}
java com/polytechnik/algorithms/CallAMuseOfCashFlowAndLiquidityDeficit \
      --musein_cols=9:1:2:3 \
      --musein_file=dataexamples/aapl_old.csv.gz
      --museout_file=museout.dat \
      --n=7 \
      --tau=128 \
      --measure=ImpactQVMMuseLegendreShifted
\end{verbatim}
\end{itemize}

\subsection{\label{musecode} CallAMuseOfCashFlowAndLiquidityDeficit.java}
Output file is tab--separated file
with columns corresponding the calculations of this paper.
Most output data is saved in the objects of \texttt{Skewness} type
(skewness and generalized skewness)
and \texttt{\seqsplit{EVXData}} type (generalized eigenvalue problem $\Ket{I|\psi}=\lambda\Ket{\psi}$)
created by the \texttt{\seqsplit{ImpactQVMMuse}}.
Field number (and name) are printed in the first line of output file,
so they can be processed by any common plotting software (such as gnuplot or matlab).
Below are the description of most noticeable fields:
\begin{itemize}
\item \verb+T+ Time in nanoseconds since midnight (copied from input).
\item \verb+shares+ Shares traded (copied from input).
\item \verb+P_last+ Execution price (copied from input).
\item \verb+I.*+ Correspond to $\Ket{I|\psi}=\lambda\Ket{\psi}$ eigenvalues solution
  with the given \verb+--n=+. The \verb+I.Gamma0+ is $\widetilde{\Gamma^0}$ (\ref{skewnesslikeS0})
  of past sample. The \verb+I.sL+, \verb+I.sH+, and \verb+I.s0+ correspond to min/max eigenvalues,
  and $\Braket{\psi_0|I|\psi_0}$.
  The \texttt{\seqsplit{I.wL}} and \texttt{\seqsplit{I.wH}} are squared in the output.
\item \verb+P.*+ Correspond to $\Ket{pI|\psi}=\lambda\Ket{I|\psi}$ eigenvalues solution
  with the given \verb+--n=+. This eigenproblem for price is presented just for completeness.
\item \verb+SK_P_IH.*+ Skewness on $\max I$ state from Section \ref{PsiFutureI}, with $dI=0$.
  The \verb+SK_P_IH.xa+ is equal to $P^{[IH]}$ (\ref{PIH}).
\item \verb+pnlss.*+ fields correspond to $n=2$ (regardless of
  the given \verb+--n=+ value, use Laguerre basis to have $p^k$ and $t^k$ basis
  similar behavior without \texttt{\seqsplit{--tau=}} adjustment),
  calculations of Section \ref{Pcorrdirinfo}.
  Regular price skewness (\ref{skewness}) along with the generalized skewness (\ref{skewnesslikeS})
  for $I$ and $P$ are presented. Regular exponential moving average
  $\overline{p}_{\tau}=\Braket{Q_0 pI}/\Braket{Q_0I}$
  is equal to any of \texttt{\seqsplit{pnlss.\{SK\_P\_average,gSK\_P\_average\}.xa}},
  and $p_{\{1,2\}}$ nodes (\ref{evp3nodes}) are \verb+pnlss.SK_P_average.{x1,x2}+.
\item \texttt{\seqsplit{pnldidsk.*}} fields calculated by the
  \texttt{\seqsplit{PnLdIDSk}}
  class, most noticeable are: \texttt{\seqsplit{pnldidsk.SK\_spur\_\_nodI}} the skewness
  of Section \ref{MeasureSpur} density matrix states,
  the \texttt{\seqsplit{pnldidsk.SK\_spur\_\_nodI.xa}} is $\overline{p}^{spur}_{\tau}=Spur(\|pI\|)/Spur(\|I\|)$,
  moving average, calculated via operator spur (sum of diagonal elements).
  The  \texttt{\seqsplit{pnldidsk.Pf\_from\_pt\_true\_pi}} is (\ref{PfNative}).
\item \texttt{\seqsplit{pnlfutureSk.*}} correspond to Section \ref{FutureISK}
  calculations.
\end{itemize}
Current \texttt{\seqsplit{CallAMuseOfCashFlowAndLiquidityDeficit.java}}
output 77 fields. The code can be modified to adjust the output.
You may also use \texttt{\seqsplit{com/polytechnik/scripts/plot\_chart.pl}}
to select only specific fields, also you may run
\texttt{\seqsplit{com/polytechnik/trading/GenerateTrainingData.java}}
to produce more data in output.

\subsection{\label{codestr} Code Structure}
The codebase is huge. Most of the
code are my past fault attempts to find a market dynamics equation.
Once an idea is decided to be a fault --- all related code is moved to
the unit tests, thus increase the codebase\footnote{
  This section is adjusted from the
  \href{https://arxiv.org/abs/1709.06759v1}{earlier version}
  in order to reflect API changes in \cite{MalMuseScalp}.
  }.
To run all unit tests at once execute the command:
\begin{verbatim}
java com/polytechnik/trading/QVM
\end{verbatim}
It may take a while to finish all the unit tests (about 2 days to run,
the best usage I found for these
unit tests is
to catch Java HotSpot JIT compiler bugs $\stackrel{\cdot\cdot}{\frown}$).
But for the theory of this paper
the calculations are extremely fast and 
there are actually very
few classes of interest. Most noticeable of them are described below.
\begin{verbatim}
com/polytechnik/trading/QVMDataL.java
com/polytechnik/trading/QVMDataP.java
com/polytechnik/trading/QVMData.java
\end{verbatim}
These calculate the moments $\Braket{fQ_k}$ from a sequence of trades
using  Laguerre, Shifted Legendre, or monomials basis for $Q_k(x)$.
The calculations are optimized
to incrementally\footnote{Using the
$Q_n(ax+b)=\sum_{k=0}^{n}d^{(n)}_kQ_k(x)$
expansion,
  that is
Newton Binomial $(1+x)^n=\sum_{k=0}^{n}C_{n}^{k}x^k$
monomials basis generalization. For numerical implementation
see \texttt{\seqsplit{setNewtonBinomialLikeCoefs}}
method of classes extending the \texttt{\seqsplit{com/polytechnik/utils/BasisPolynomials.java}} class,
implementing the expansion using three term recurrence of basis
polynomials $Q_k(x)$, see Appendix A ``Non-monomials polynomial bases''
of Ref. \cite{2015arXiv151005510G}.}
update already calculated moments,
what make the calculations extremely fast,
thus applicable to a practical realtime HFT trading.
To access calculated distribution moments use
the classes implementing the \texttt{\seqsplit{DataInterfaceToMoments<T>}}:
\begin{verbatim}
com/polytechnik/trading/QVMDataLDirectAccess.java
com/polytechnik/trading/QVMDataPDirectAccess.java
com/polytechnik/trading/QVMDataDirectAccess.java
\end{verbatim}
To manipulate distribution moments obtained
in various $Q_k(x)$ bases
there are few classes  (they all extend the \verb+OrthogonalPolynomialsABasis+
and use a reference to \verb+BasisPolynomials+ to manipulate polynomials):
\begin{verbatim}
com/polytechnik/utils/OrthogonalPolynomialsLegendreShiftedBasis.java
com/polytechnik/utils/OrthogonalPolynomialsLegendreBasis.java
com/polytechnik/utils/OrthogonalPolynomialsLaguerreBasis.java
com/polytechnik/utils/OrthogonalPolynomialsChebyshevBasis.java
com/polytechnik/utils/OrthogonalPolynomialsHermiteEBasis.java
com/polytechnik/utils/OrthogonalPolynomialsMonomialsBasis.java
\end{verbatim}
Once the moments $\Braket{fQ_k}$ 
are calculated from a sequence of trades, the classes such as:
\begin{verbatim}
com/polytechnik/trading/MomentsData.java
com/polytechnik/trading/SMomentsData.java
\end{verbatim}
calculate and store the matrices: $\Braket{Q_j|Q_k}$, $\Braket{Q_j|I|Q_k}$,
$\Braket{Q_j|pI|Q_k}$, $\Braket{Q_j|dp/dt|Q_k}$ (and others,
the classes are different in attributes selection)
from the moments data using basis functions multiplication operator $c_l^{jk}$ :
\begin{eqnarray}
  Q_j(x)Q_k(x)&=&\sum_{l=0}^{j+k} c_l^{jk}Q_l(x)
  \label{cmul}
\end{eqnarray}
The $c_l^{jk}$ coefficients are available analytically for all practically interesting bases, see Appendix A of Ref. \cite{2015arXiv151005510G}
and references therein,
the calculations are implemented in the classes  above,  the ones extending
the \texttt{\seqsplit{OrthogonalPolynomialsABasis}} and \texttt{\seqsplit{BasisPolynomials}}.
The class: 
\begin{verbatim}
com/polytechnik/utils/EVXData.java
\end{verbatim}
given two matrices $\Braket{Q_j|Q_k}$ and $\Braket{Q_j|I|Q_k}$
solves generalized eigenvalue problem
$\Ket{I|\psi_{I}^{[i]}}=\lambda_{I}^{[i]}\Ket{\psi_{I}^{[i]}}$,
finds eigenvalues and eigenvectors, calculates $\Braket{\psi_0|\psi_{I}^{[i]}}$
projections, and $I_0=\Braket{\psi_0|I|\psi_0}$.
The class:
\begin{verbatim}
com/polytechnik/trading/PnLSimpleSkewness.java
\end{verbatim}
perform simple calculations of Sections \ref{movavermoments},
\ref{skewnessDemonstration}, and \ref{movavermomentsSK}
(for $n=2$ all the matrices are $2\times 2$).
This class calculates:
price regular skewness (skewness, quadrature nodes, and weights are calculated),
generalized skewness ($\widetilde{\Gamma}$ skewness, $\widetilde{\Gamma^0}$ skewness, $\lambda^{[\{0,1\}]}$, and weights),
and, out of curiosity, probability correlation $\widetilde{\rho}(p,I)$ of Appendix \ref{twoPvarcorrela}. The class:
\begin{verbatim}
com/polytechnik/trading/PnLdIDSk.java
\end{verbatim}
performs na\"{\i}ve dynamic impact calculations of Section \ref{naive}
along with some other skewness--related calculations
considered in Sections \ref{PsiFutureI} and \ref{MeasureSpur}. The class
\begin{verbatim}
com/polytechnik/trading/PnLFutureSk.java
\end{verbatim}
performs the calculations of Section \ref{FutureISK}.
It takes an instance of 
\verb+MomentsData+ and do the
following:
\begin{itemize}
\item Solve generalized eigenvalue problem (\ref{GEVI}),
  find $dI$ as (\ref{dI})  and $P^{[IH]}$ as (\ref{PIH}).
\item Construct $\|I^{f}\|$ operator (\ref{Ifuture}).
\item Solve generalized eigenvalue problem (\ref{GEVIf}).
\item Find past $\widetilde{\Gamma^0}$ and future $\widetilde{\Gamma^0}^f$
  skewness of $I$.
\item  Find  $P^{f}$ as (\ref{PfSkewF}).
\end{itemize}
This class demonstrate reference implementation of this paper theory:
\begin{verbatim}
com/polytechnik/algorithms/CallAMuseOfCashFlowAndLiquidityDeficit.java
\end{verbatim}
It read line-by-line
tab--separated timeserie file of triples (time, execution price, shares traded)
to update a sequence of executed trades.
For each new trade (new line read),
it calls\footnote{
  See the file  \texttt{\seqsplit{com/polytechnik/ImpactQVMMuse.java}},
  that, create \texttt{\seqsplit{MomentsData}} out of trade sequence,
  do the calculations, and save the results
  to (type:name) objects: \texttt{\seqsplit{PnLSimpleSkewness:pnlss}},
\texttt{\seqsplit{Skewness:SK\_IH}},
\texttt{\seqsplit{PnLdIDSk:pnldidsk}},
\texttt{\seqsplit{PnLFutureSk:pnlfutureSk}}, and 
along with (for demonstration)
separate calculation of $P^{[IH]}$ from (\ref{PIH})
and
$w^{[IL]}_I$, $w^{[IH]}_I$ projections (\ref{wL}), (\ref{wH})
 (using \texttt{\seqsplit{EVXData:I}}).
  }
\texttt{\seqsplit{com/polytechnik/ImpactQVMMuse<T>}}, 
that incrementally (optimization for speed)
calculates the moments, obtains the \texttt{\seqsplit{MomentsData}}
with $\Braket{Q_j|Q_k}$, $\Braket{Q_j|I|Q_k}$,
$\Braket{Q_j|pI|Q_k}$ matrices,
performs the calculations
and creates the \texttt{\seqsplit{ImpactQVMMuse}} object,
then, finally,
outputs the data out of the \texttt{\seqsplit{ImpactQVMMuse}} as described
in the  previous section.

\bibliography{LD}

\end{document}